\documentclass[12pt,a4paper,final]{iopart}

\usepackage{iopams}  
\usepackage{graphicx}
\usepackage{enumitem}
\usepackage[breaklinks=true,colorlinks=true,linkcolor=blue,urlcolor=blue,citecolor=blue]{hyperref}
\usepackage{subcaption}
\usepackage{wrapfig}
\usepackage{textcomp}
\usepackage[all]{xy} 

\newtheorem{thm}{Theorem}

\newtheorem{rmk}{Remark}

\usepackage{perpage} 
\MakePerPage{footnote}
\newcommand\bv[1]{\boldsymbol{#1}}

\begin{document}

\title{Crisis of the Chaotic Attractor of a Climate Model: A Transfer Operator Approach}

\author[cor1]{Alexis Tantet$^{1,2}$}
\ead{a.j.j.tantet@uu.nl}

\author{Valerio Lucarini$^{2, 3, 4}$}

\author{Frank Lunkeit$^{2}$}

\author{Henk A. Dijkstra$^{1}$}
\address{$^1$Institute of Marine and Atmospheric research Utrecht, Department of Physics and 
Astronomy,  University of Utrecht, Utrecht, The Netherlands}
\address{$^2$Meteorological Institute and Centre for Earth System Research and Sustainability (CEN), University of Hamburg, Hamburg, Germany}
\address{$^3$Department of Mathematics and Statistics, University of Reading, Reading, UK}
\address{$^4$Centre for the Mathematics of Planet Earth, University of Reading, Reading, UK}

\begin{abstract}

The destruction of a chaotic attractor leading to rough changes
in the dynamics of a dynamical system is studied.
Local bifurcations are known to be characterised by a single or a pair of characteristic exponents
crossing the imaginary axis.
As a result, the approach of such bifurcations in the presence of noise
can be inferred from the slowing down of the decay of correlations \cite{Held2004a}.
On the other hand, little is known about global bifurcations involving
high-dimensional attractors with several positive Lyapunov exponents.

It is known that the global stability of chaotic attractors may be characterised by the
spectral properties of the Koopman \cite{Mauroy2016} or the transfer operators
governing the evolution of statistical ensembles.
Accordingly, it has recently been shown \cite{Tantet2017} that a boundary crisis in the Lorenz flow
coincides with the approach to the unit circle of the eigenvalues of these operators
associated with motions about the attractor, the \emph{stable resonances}.
A second type of resonances, the \emph{unstable resonances},
is responsible for the decay of correlations and mixing on the attractor.
In the deterministic case, those cannot be expected to be affected by general boundary crises.

Here, however, we give an example of chaotic system 
in which slowing down of the decay of correlations of some observables
does occur at the approach of a boundary crisis.
The system considered is a high-dimensional, chaotic climate model of physical relevance.
Moreover, coarse-grained approximations of the transfer operators on a reduced space,
constructed from a long time series of the system, give evidence that this behaviour
is due to the approach of unstable resonances to the unit circle.
That the unstable resonances are affected by the crisis
can be physically understood from the fact that the process responsible for the instability,
the ice-albedo feedback, is also active on the attractor.
Implications regarding response theory and the design of early-warning signals are finally discussed.




\end{abstract}

\vspace{2pc}
\submitto{\NL}

\section{Introduction}

Much progress has been achieved during the last decades regarding bifurcation theory of low-dimensional dynamical systems and the emergence of aperiodic behaviour and sensitive dependence to initial conditions, or chaos, when a parameter of the system is changed.
Examples include changes  in the Rayleigh number in hydrodynamics \cite{Lorenz1963a} or in the baroclinic forcing responsible for midlatitude atmospheric variability \cite{Lucarini2007,Ghil2010}.
Much less is known, however, on the sudden appearance or destruction
of chaotic attractors during global bifurcations.
In particular, a \emph{boundary crisis} involves the destruction of the basin of attraction of an invariant set
due to the collision of this set with a saddle.
The review by \cite{Grebogi1983} provides a good starting point regarding the topological description of chaotic attractor crises and the statistics of chaotic transients.
However, little is known regarding the changes in the statistical properties
of high-dimensional chaotic systems undergoing a boundary crisis.

While it is a classical result that local bifurcations involving stationary points or periodic orbits
are respectively associated with eigenvalues of the Jacobian and Floquet multipliers
crossing the imaginary axis and with the critical slowing down of trajectories \cite{Guckenheimer1983,Kuznetsov1998}, what happens before a boundary
crisis during which a chaotic invariant set ceases to be attracting is less understood.
For example, Lyapunov exponents describing the divergence of nearby trajectories on the attractor
do not allow, in general, for the study of the global stability of chaotic invariant sets
(see \cite{Tantet2017}, for a discussion).
On the other hand, at the approach of a boundary crisis,
trajectories are expected to take more and more time to recover
from exogenous perturbations.
It is, however, unclear whether such a slowing down can be inferred
from the behaviour of unperturbed trajectories on the attractor alone.

For high-dimensional chaotic systems, statistical physics methods
developed for nonequilibrium systems are very fruitful \cite{Ruelle1999b, Gallavotti2014}.
Following the ground-breaking ideas of Boltzmann, a statistical steady-state is described
by a time invariant probability measure, say $\mu$.
While the most intuitive microscopic manifestation of chaos is the sensitive dependence on initial conditions
associated with positive Lyapunov exponents \cite{Oseledets1968,Eckmann1985},
a macroscopic manifestation of chaos is the decay with time of correlations between observables,
associated with the loss of memory to initial conditions of statistical ensembles (as they mix).
The evolution of statistical ensembles or probability densities
\emph{with respect to the invariant measure $\mu$}
is governed by linear operators, the \emph{transfer} or \emph{Perron-Frobenius operators} $\mathcal{P}_\mu$.
It is a classical result from ergodic theory \cite{Halmos1956,Arnold1968}
that correlations between observables vanish for time lags going to infinity only if there are
no eigenvalues of the transfer operator $\mathcal{P}_\mu$
in the complex unit circle other than the eigenvalue 1.
A more difficult problem, which is still a matter of investigation
\cite{Baladi2001,Young2013}, is to characterise the
rate at which correlations decay with time \cite{Pollicott1985,Ruelle1986}.
This rate depends on the position inside the unit disk of the complex plane
of the eigenvalues of transfer operators
acting on anisotropic Banach spaces adapted to the dynamics of
contraction and expansion of chaotic systems
\cite{Liverani1995a,Blank2001a,Gouezel2006,Butterley2007,Faure2014,Baladi2017}.
We will refer to these eigenvalues as the \emph{unstable resonances} \cite{Ruelle2009,Cessac2007}.

It is a fundamental property of nonequilibrium systems that \emph{physically relevant}
invariant measures \cite{Young2002} are supported by strange attractors of zero volume \cite{Eckmann1985}.
Information about ensembles initialised away from the attractor 
thus cannot be expected to be carried by the transfer operator $\mathcal{P}_\mu$
acting on densities with respect to the invariant measure $\mu$
(even if more general anisotropic spaces of distributions are considered).
Instead, one is led to study the spectral properties of transfer operators $\mathcal{P}_m$
acting on densities \emph{with respect to the Lebesgue measure} (i.e. possibly giving
weight to any region in state space, not only in the attractor, see \cite{Tantet2017}).
This has allowed for new characterisations of the global stability of dynamical systems
in terms of Lyapunov functions \cite{Vaidya2008} and in terms of the spectral properties
of the \emph{Koopman operator} \cite{Mauroy2016}, adjoint to the transfer operator $\mathcal{P}_m$.
Loosely speaking, while the unstable resonances describe the rate of mixing on the attractor,
eigenvalues of the transfer or Koopman operators acting everywhere in the state space
also describe the rate at which Lebesgue densities converge to the invariant measure.
Those eigenvalues which are not in the set of unstable resonances will be referred to as
the \emph{stable resonances} (see \cite{Tantet2017}).

Since the stable resonances characterise the rate at which distributions are
\emph{contracted towards the invariant measure},
the slowing down of this contraction
- expected to be observed at the nearing of a boundary crisis -
should be associated to the approach of the stable resonances to the unit circle.
In fact, exactly at the crisis, some resonances should
have reached the unit circle and prevent the response of the invariant measure to perturbations
to be smooth \cite{Ruelle2009, Lucarini2016}.
The approach of the stable resonances to the unit circle has been found
during a boundary crisis in the Lorenz flow \cite{Tantet2017}, while the unstable
resonances, and thus the decay of correlations, were not affected by the crisis.
Unfortunately, however, the stable resonances tend to be more difficult to study
than the unstable resonances and the associated decay of correlations,
as only information about the dynamics on the attractor (e.g. from a long time series)
is necessary to study the latter.

The purpose  of the present study is two-fold.
Using a high-dimensional dynamical system relevant for climate science,
we give numerical evidence that slowing down \emph{of the decay of correlations}
may occur during a boundary crisis (as opposed to the case of the Lorenz flow studied in \cite{Tantet2017}).
Under the assumption of the existence of a physical measure,
the latter can be calculated from correlation functions estimated from long time series.
Second, we show that information on the unstable resonances responsible
for this slowing down of the decay of correlations
may be obtained from projections of the transfer operator $\mathcal{P}_\mu$
(with respect to the invariant measure $\mu$)
onto coarse-grained spaces \cite{Chekroun2014,Chekroun2016}
and their estimation from long time series \cite{Tantet2016a}.
These projections allow one to go beyond the sole study of correlation functions
and to bring some of the operator techniques to high-dimensional systems.
The quality of these reductions depends of course on the
coarse-grained spaces onto which the operators are projected.
Note, moreover, that, as projections of the transfer operator $\mathcal{P}_\mu$
rather than $\mathcal{P}_m$, these \emph{reduced transfer operators}
yield in general no information about the stable resonances.
No analysis of the stable resonances is thus performed in this study.

The climate is a forced and dissipative nonequilibrium system featuring variability
on a vast range of spatial and temporal time scales \cite{Lucarini2014b}.  
The climate model considered is the Planet Simulator (PlaSim) \cite{Fraedrich2005a, Fraedrich2005b}.
This model displays chaotic behaviour \cite{Schalge2013} and
is complex enough to represent the main physical mechanisms of 
climate multistability, yet sufficiently fast to perform extensive
numerical simulations \cite{Lucarini2010a, Boschi2013}.
This multistability is due to the coexistence of two chaotic attractors associated
with a warm and a snow-covered state for a large range of values of the solar constant.
Moreover, a transition from the warm state to the snow-covered state occurs in this model,
as the solar constant is decreased,
which is due to the collision of a repeller (the edge state) with the attractor \cite{Lucarini2017}
(see also \cite{Bodai2015} for a simplified treatment of the problem).
Such a metastability is found for climate models across a hierarchical ladder of complexity, ranging 
from simple energy balance models  \cite{Budyko1969, Sellers1968, Ghil1976a} to fully coupled
global circulation models  \cite{Pierrehumbert2011a, Voigt2010}.
Moreover, the importance of such nonlinear processes to understand the response
of the climate system to anthropogenic forcing, namely the \emph{climate sensitivity},
has been studied in \cite{Zaliapin2010} using one such energy balance model.
The robustness of such features is tied to the presence of a powerful
positive feedback which controls the transitions 
between the two competing states, the so-called ice-albedo feedback.
Moreover, paleoclimatological studies support the fact that in the distant past (about 700 Mya) the 
Earth was in such a Snowball state, and it is still a matter of debate which mechanisms 
brought the planet in the present warm state. Note that the issue of multistability is 
relevant for the present ongoing investigations of habitability of exoplanets \cite{Lucarini2013a}. 

In the present study, we first demonstrate that slowing down of the decay of correlations
between physically relevant observables
occurs in PlaSim before the loss of stability of the chaotic attractor corresponding to a warm climate.
The decay of correlations is investigated from robust sample estimates of the correlation functions
using long time series and do not require information about the dynamics away from the attractor.
Moreover, we show that this slowing down of the decay of correlations
is described by the approach to the unit circle
of the leading eigenvalues of coarse-grained approximations of
the transfer operator $\mathcal{P}_\mu$ also constructed from time series.

These results suggest that, while the approach of the unstable resonances to the imaginary axis
at a boundary crisis cannot be expected to be generic
(a counter-example being investigated in \cite{Tantet2017},
for which only the stable resonances approach the imaginary axis),
it may, however, typically occur in high-dimensional systems
for which the instability mechanism responsible for the crisis is already affecting
the unperturbed dynamics on the attractor.
Such a statement should be taken as a motivation for further investigations
rather than as a rigorous result.

The paper is structured as follows.
Section 2  below briefly describes the climate model as well as the set up
of different simulations for varying values of the control parameter. 
The spectral theory of transfer operators and the link 
with slowing down of the decay of correlations as well as of
the convergence of probability densities to the invariant measure
is formally described in section 3. There, we take care to distinguish
between the different role played by the stable and the unstable resonances.
A reduction method used to extract information on the \emph{unstable resonances} of 
transfer operator $\mathcal{P}_\mu$ (with respect to the invariant measure $\mu$)
from time series is presented in section 4.
The results are presented in section 5, where slowing down in the decay of correlations is shown 
to occur in the model as the attractor crisis is approached.
It is also shown that this slowing down of the decay of correlations is associated with the
eigenvalues of the reduced transfer operators getting closer to the imaginary axis.
The results are summarised in section 6 where we discuss why the unstable resonances
are affected by this crisis, but not in general boundary crises.
Implications to response theory and the design of early-warning indicators of the crisis are also discussed.
A final appendix discusses the robustness of the numerical estimates of the spectrum
of the reduced transition matrices.

\section{Model and simulations}
\label{sec:model}

As mentioned above, the attractor crises corresponding to transitions between warm states 
and snowball climate states have been replicated with qualitatively similar feature in a variety of models 
of various degrees of complexity. Given the goals of this study, we need a climate model that is simple 
enough to allow for an  exploration of parameters in the vicinity of the attractor crisis and complex 
enough to feature essential characteristics of high-dimensional, dissipative, and chaotic systems, 
as the existence of a limited horizon of predictability due to the presence of instabilities in the flow. 
We have opted for using PlaSim, a climate model of so-called intermediate complexity. This model 
has already  been used for several theoretical climate studies and provides an efficient platform 
for investigating climate  transitions when changing boundary conditions and parameters.  

\subsection{The Planet Simulator (PlaSim)}
\label{sec:PlaSim}

The dynamical core of PlaSim is based on the Portable University Model of the Atmosphere PUMA  
\cite{Fraedrich1998a}. The atmospheric dynamics is modelled using the multi-layer primitive equations formulated 
for vorticity, divergence, temperature and the logarithm of surface pressure. Moisture is included by 
transport of water vapour (specific humidity). The governing  equations are solved using the spectral 
transform method \cite{Eliasen1970, Orszag1970} on a T21 grid (resulting in a horizontal resolution
of about 5 to 6 degrees in the midlatitudes) and with a semi-implicit time integration \cite{Hoskins1975}.
In the vertical, 5 non-equally spaced sigma (pressure 
divided by surface pressure) levels are used.
Counting for each prognostic variable of each layer a total of $21 \times 22 + 22$
real and imaginary coefficients
adds up to about $10^4$ degrees of freedom for the atmospheric component alone.

The parametrisation of unresolved processes 
consists of long- \cite{Sasamori1968} and short-  \cite{Lacis1974} wave radiation, interactive clouds 
\cite{Stephens1978, Stephens1984, Slingo1991}, moist  \cite{Kuo1965, Kuo1974} and dry convection, 
large-scale precipitation, boundary layer fluxes of latent and sensible heat and vertical and horizontal diffusion
\cite{Louis1979, Louis1981, Laursen1989, Roeckner1992}. The land surface scheme uses five diffusive 
layers for the temperature and a bucket model for the soil hydrology. The oceanic part is a 50 m 
mixed-layer (swamp) ocean, which includes a thermodynamic sea ice model  \cite{Semtner1976}.
Loosely speaking, this means that only the energy balance within a vertically diffusive water/ice/land column is modelled
but not the actual nonlinear dynamics of advection.

In addition, the horizontal transport of heat in the ocean can either be prescribed or parametrised by horizontal 
diffusion. In this case, we consider the simplified setting where the ocean gives no contribution to the 
large-scale heat transport through advection. While such a simplified setting is somewhat less realistic, the climate 
simulated by the model is definitely Earth-like, featuring qualitatively correct large scale features 
and turbulent atmospheric dynamics.
Finally, the model is forced by the orbital annual cycle and has atmospheric CO$_2$ concentrations
fixed at the value of $360$ppmv observed at the end of the twentieth century.

Beside standard output,  PlaSim provides  comprehensive 
diagnostics for the nonequilibrium thermodynamical properties of the climate system and in 
particular for  local and global energy and entropy budgets. PlaSim is freely available including 
a graphical user interface facilitating its use. PUMA and PlaSim have been applied to a variety 
of problems in  climate response theory \cite{Ragone2016a, Lucarini2017d}, entropy production 
\cite{Lucarini2014a, Fraedrich2008}, and in the dynamics of exoplanets  
\cite{Lucarini2013a}. 

In summary, rich chaotic dynamics is only possible in the atmospheric component of the model,
while the land surfaces, the oceanic mixed-layer and the sea ice,
interacting with the atmosphere at their interface and adding up to about $10^4$ degrees of freedom,
contribute to the energy balance of the model.

\subsection{Attractor crisis and instability mechanism}
\label{sec:crisisPlasim}

\begin{figure}[h!]
	\centering
	\includegraphics[width=7.5cm]{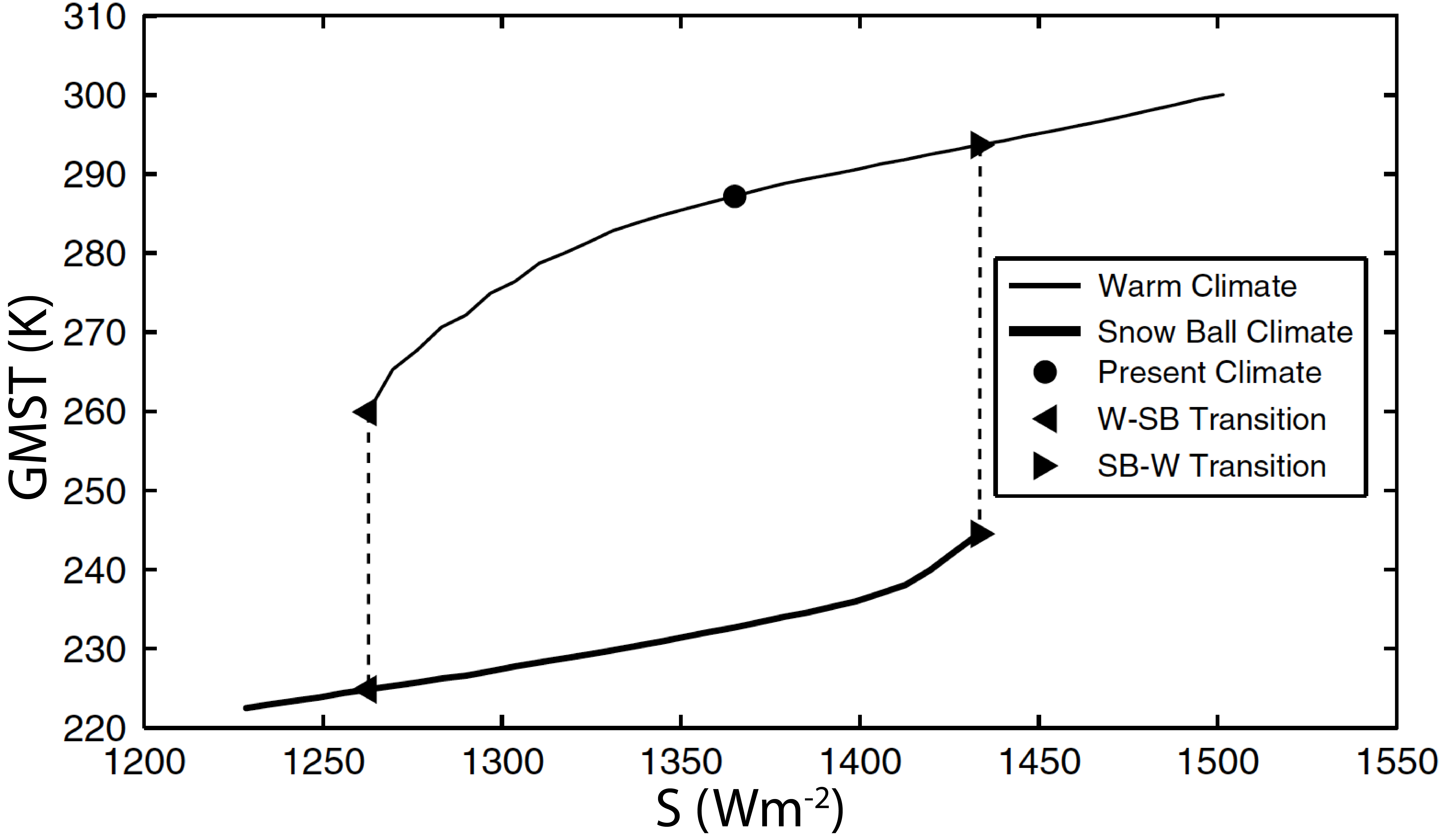}
	\caption{Hysteresis plot of the GMST (in Kelvin) for a varying solar constant $S$ (in $Wm^{-2}$) in PlaSim. Figure from \cite{Boschi2013}.}
	\label{fig:hysteresis}
\end{figure}

In \cite{Lucarini2010a} and  \cite{Boschi2013}, a parameter sweep in PlaSim is performed by changing the 
solar constant. The resulting changes in the Global Mean Surface Temperature (GMST) are represented 
in figure \ref{fig:hysteresis}, taken from \cite{Boschi2013}.
For a large range of values of the solar constant there are 
two distinct statistical steady states, which can be characterised by a large difference --- of the order of 
40-50 K --- in the value of the GMST. The upper branch in figure \ref{fig:hysteresis}  corresponds to  warm 
(W) climate conditions (akin to present-day state) and the lower cold branch  represents snowball (SB) states. 
The W branch is depicted with a thin line, whereas the SB branch is depicted with a thick line. In the 
SB states, the planet is characterised by a  greatly enhanced albedo due to a massive ice-cover. 

The actual integration of the model is performed by starting from present-day  climate conditions and 
decreasing the value of $S$, until a sharp W $\rightarrow$ SB transition is observed, corresponding 
to the attractor crisis, occurring at $S \approx 1265~W  m^{-2}$.  The critical  value of the solar 
constant needed to induce the onset of snowball conditions basically agrees with that found in 
\cite{Poulsen2004a} and in \cite{Voigt2010} (see also \cite{Wetherald1975}, Fig. 5). The SB 
state is realised for values of the solar constant up to $S \approx 1433~W  m^{-2}$, where the 
reverse SB $\rightarrow$ W transition occurs.  Hence, in the model there are two rather distinct 
climatic states for the present day  solar irradiance \cite{Voigt2010}. 

These W$\leftrightarrow$SB transitions are induced by the  ice-albedo positive feedback \cite{Budyko1969, Sellers1968} which 
is strong enough to destabilise both W and SB states. The main ingredient of this feedback is the increase in the 
albedo (i.e. the fraction of incoming solar radiation reflected by a surface) due to the presence of ice. When a negative (positive) perturbation in surface temperature results in temperatures below (above) the freezing point 
($T_f \approx 271.8 K$, for sea ice), ice is formed (melts). Because of the larger albedo of the ice compared 
to ocean or land surfaces, this increase (decrease) in the ice-cover extent leads to a decrease (increase) of the 
surface temperature, further favouring the formation (melting) of ice.
The dominant damping processes of this growth mechanism include
the local decrease of emitted long-wave  radiation with  decreasing temperatures and the meridional 
heat transport  from the equator to the poles, which tends to reduce the meridional temperature gradient.
However, as the solar constant is decreased to criticality, these 
negative feedbacks are not sufficient to damp the ice-albedo feedback  and the  transition occurs.

In more idealised models  \cite{Scott1999} or, in general, in non-chaotic models \cite{Lucarini2007a}, bistability 
is typically associated with the presence of two stable fixed points separated by a saddle-node, and the loss 
of a fixed point is related to a bifurcation determining the change in the sign of one eigenvalue of the system 
linearised around the fixed point. In the present case, instead, the two branches define the presence of two 
parametrically modulated (by the changes in the value of the solar constant) families of disjoint strange 
attractors, as in each climate state the dynamics of the system is definitely chaotic.
This is suggested by the fact that, for example, the system features variability on all time-scales  \cite{Schalge2013}.
The loss of stability occurring through 
the W $\rightarrow$ SB and SB $\rightarrow$ W transitions is related to the
loss of the attracting character of one of the two strange attractors,
as will be described in greater details in section \ref{sec:crisisMixing}.

\subsection{Set-up of long simulations for varying solar constant}
\label{sec:simulations}

While the studies in  \cite{Lucarini2010a, Boschi2013} were focused on the thermodynamic properties of the 
model solutions along the two branches of statistical steady-state,  we here focus on the changes in the 
characteristic evolution of statistics (cf. section \ref{sec:specTransfer}) along the warm branch 
as the solar constant $S$ is decreased to its critical value $S_c \approx 1265~W  m^{-2}$. The 
method presented in section \ref{sec:approxSpectrum} requires very long time series of observables. 
Hence, the model is ran for 10,000 years in the configuration presented section \ref{sec:PlaSim} and 
for 10 different fixed values of the solar constant, ranging from the critical value $S_c$ to $1360~
W  m^{-2}$ (approximately the present day value). 

The initial state of each simulation is taken in  the basin of attraction of the warm 
state. While this is difficult to guarantee, taking as initial state the annual 
mean of the 100th  year of a simulation for a solar constant as high as $1450~W  m^{-2}$
(i.e for  which only the warm state  exists) resulted in the convergence 
to the warm state  for  each of the 10 values of $S$. A spin-up of 200 years was more than 
enough for the  simulations to converge to the statistical steady-state (facilitated by the 
fact that the  ocean has no dynamics). Preparing for the analysis of section \ref{sec:results}, the 
spin-up period was removed from the time series.
These time series are thus guarantied to only carry information about the attractor dynamics.
The time series for each observable is subsequently sub-sampled from daily to annual averages, in order to focus on  interannual-to-multidecadal variability and to avoid having to deal with the  seasonal cycle.
In the following, we will consider these time series as trajectories of a time-one-year map
of a dynamical system.

\section{Transfer operators and correlations}
\label{sec:methodology}

In this section, we start by shortly reviewing the spectral theory of transfer operators for maps
and, in particular, the characterisation of important statistical properties,
such as the decay of correlations and the convergence to a statistical steady-state.
We analyse the climate model in the framework of discrete-time autonomous dynamical system.
The presence of a seasonal forcing in the model is, in fact not problematic
as only one year time steps will be considered in all subsequent applications.

\subsection{Transfer operator and ensemble propagation}
\label{sec:specTransfer}

We consider a diffeomorphism $\Phi:X \to X$ on a state space $X$ in  $\mathbb{R}^d$
generating a discrete-time trajectory
\begin{eqnarray}
	x_n = \Phi^n x_0, \quad n \ge 0,
	\label{eq:ODE}
\end{eqnarray}
for some initial state $x_0$ in X.

When the system is chaotic, it is fruitful to follow the evolution of 
probability densities and observables rather than that of solutions of (\ref{eq:ODE})
\footnote{This is a common practice in climate modelling
where ensemble simulations are ran for which members 
differ only in their slightly different initial conditions \cite{Leutbecher2008}.}.
For that purpose, we endow the state space $X$ with its Borel $\sigma$-algebra
$\mathcal{B}$ and some probability measure $\eta$ (to be specified) on $(X, \mathcal{B})$.
The transformation $\Phi$
\footnote{
The map $\Phi$ will be assumed to be nonsingular for $\eta$,
so that it maps sets of null measure into sets of null measure.}
induces a linear operator, the \emph{Koopman operator},
\begin{eqnarray}
	\mathcal{U}_\eta: g \to g \circ \Phi,
\end{eqnarray}
acting on any observable $g$
\footnote{Observables are real-valued functions of the state space
which give partial information on the state of the system,
such as an index of surface temperature or sea-ice cover in some region.}
in the space bounded measurable functions $L^q_\eta(X), 1 \le q < \infty$.
On the other hand, there exists a linear operator $\mathcal{P}_\eta$,
the \emph{transfer operator} or \emph{Perron-Frobenius operator}
on $L^p_\eta(X)$, with $1/p + 1/q = 1$, and such that the duality relation
\begin{eqnarray}
\int_X g(x)~\mathcal{P}_\eta f(x)~\eta(dx) = \int_X \mathcal{U}_\eta g(x)~f(x)~\eta(dx)
\label{eq:transfer}
\end{eqnarray}
holds for any observable $f$ in $L^p_\eta(X)$ and $g$ in $L^q_\eta(X)$.
It follows that $\mathcal{U}_\eta$ is the adjoint of $\mathcal{P}_\eta$.
%
Moreover, taking $g$ in (\ref{eq:transfer}) as the indicator function $\mathbf{1}_A$
of some set $A \in \mathcal{B}$, one has that
%
\begin{eqnarray}
	\int_A \mathcal{P}_\eta f(x) ~ \eta(dx)
	= \int_{\Phi^{-1}A} f(x) ~ \eta(dx).
	\label{eq:transferProba}
\end{eqnarray}
For a probability density $f$, i.e.~$f \ge 0$ and $\int_X f(x)  \eta(dx)= 1$,
(\ref{eq:transferProba}) expresses the fact that the probability to find a member $x$
sampled from an initial ensemble $f$ in a set $A$ after one application of the map $\Phi$
is none-other than the probability of this member to be initially in the preimage of this set by the flow.
%

The relationship between the nonlinear transformation $\Phi$ and these operators
provides the connection between the ergodic properties of dynamical systems
and the functional analysis of linear operators \cite{Yosida1980,Engel2001,Davies2007},
as was first recognised in \cite{Koopman1931} and \cite{Neumann1932}.
%

\subsection{Ergodicity, mixing and correlations}

A key concept in ergodic theory is that of an \emph{invariant measure} for the transformation $\Phi$,
that is, a probability measure $\mu$ such that
\begin{eqnarray*}
	\mu(\Phi^{-1} A) = \mu(A), \quad \mathrm{for~any~} A \in \mathcal{B}.
\end{eqnarray*}
In other words, a measure is invariant if, according to this measure, the probability
of a state to be in some set does not change as this state is propagated by the flow.
It follows then that the average with respect to the invariant measure $\mu$
of any integrable observable $g$ is also invariant with time, i.e.
\begin{eqnarray}
	\int_X g(\Phi^n x) \mu(d x) = \int_X g(x) \mu(d x) :=  \left<g\right>_\mu, \quad n \ge 0,
	\label{eq:invariantAverage}
\end{eqnarray}
so that the invariant measure gives a statistical steady-state.

A transformation $\Phi$ with an invariant measure $\mu$ has interesting statistical properties
when $\mu$ is \emph{ergodic}, that is, when the sets $A$ which are invariant,
i.e.~such that $\Phi^{-1} A = A$, are either of measure 0 or 1.
Then, by the celebrated \emph{pointwise ergodic theorem} of Birkhoff \cite[Chap.~7.3]{Lasota1994},
the average of any $\mu$-integrable observable $g$ is such that
\begin{eqnarray}
	 \left<g\right>_\mu = \lim_{N \to \infty} \frac{1}{N} \sum_{n = 0}^N g(\Phi^n x) := \bar{g},
	 \quad \mathrm{for ~} \mu\mathrm{-almost~every~} x.
	 \label{eq:Birkhoff}
\end{eqnarray}
Thus, when $\mu$ is ergodic, the time mean is independent of the initial state $x$
except for a set of null measure.
\begin{rmk}
\label{rmk:physical}
There may exist many ergodic measures.
This is the case for PlaSim in the configuration considered in section \ref{sec:model}
with the solar constant $S$ ranging from 1265 to 1430 $Wm^{-2}$, for which two attractors coexist.
Each attractor supports at least one invariant measure (\cite{Katok1996}, Chap.~4),
here associated with the W and the SB steady-states, respectively.
In this case, the equality (\ref{eq:Birkhoff}) between ensemble averages and time averages
may hold only for initial states belonging to the attractor supporting the measure,
while time averages for two initial conditions in different basins of attraction will not coincide in general.
More useful for experiments is then the eventual \emph{physical} property of the measure.
The latter ensures that the equality (\ref{eq:Birkhoff}) between ensemble averages and time averages
holds not only for initial states in a set of positive measure $\mu$,
but also for states in any set of positive Lebesgue measure $m$ in the basin of attraction
of a given attractor \cite{Young2002}.
\end{rmk}

A particularly important quantity in ergodic theory and climate science is the correlation function,
\begin{eqnarray} 
C_{f,g}(n) :=  \int_X  f(x) g(\Phi^n x)  \mu(dx) - \left<f\right>_\mu \left<g\right>_\mu, \quad n \ge 0,
\label{eq:defCorr} 
\end{eqnarray}
between any observables $f, g \in L^2_\mu(X)$.
It gives a measure of the relationship between two observables as one evolves with time.
A possible macroscopic manifestation of chaos is the decay of correlation functions with time, i.e. when
\begin{eqnarray*}
	\lim_{n \to \infty} C_{f,g}(n) = 0 \quad \mathrm{for~any~} f, g \in L^2_\mu(X).
\end{eqnarray*}
This decay is indeed equivalent to the mixing property
(see e.g.~\cite[Chap.~4]{Lasota1994} and \cite[Chap.~4]{Katok1996}),
\begin{eqnarray}
	\lim_{n\to\infty} \mu(A \cap \Phi^{-n} B) = \mu(A) \mu(B),
	\quad \mathrm{for~any~} A, B \in \mathcal{B}, \label{eq:defMixing}
\end{eqnarray}
In other words, the probability for some state to be in any set $B$ after some
$n$ iterations is independent of the probability of this state to be in any set $A$ initially.
Any information about the initial state of an ensemble is thus gradually forgotten.
Systems which are mixing are necessarily ergodic.
\begin{rmk}
\label{rmk:physicalCorr}
For a physical measure $\mu$ supported by an attractor $\Lambda$,
the correlation function can be estimated by the time mean
\begin{eqnarray} 
	C_{f,g}(n) = \lim_{N \rightarrow \infty} \frac{1}{N} \sum_{p = 0}^N
	(f(\Phi^p x) - \bar{f})(g(\Phi^{p+n}  x) - \bar{g}),
	\label{eq:corrTime} 
\end{eqnarray}
from a single trajectory initialised in the basin of attraction of $\Lambda$.
For this reason, we will see in section \ref{sec:approxSpectrum} that
complete information on the transfer operator $\mathcal{P}_\mu$
with respect to the invariant measure
can be obtained from observations on the corresponding attractor alone.
Unfortunately, for the forced and dissipative systems considered here,
volumes contract on average \cite[Chap.~2.8]{Gallavotti2014a} so that the invariant measure $\mu$
is supported by an attractor with zero Lebesgue measure $m$.
It follows that, as opposed to the transfer operator $\mathcal{P}_m$ with respect to the Lebesgue measure,
no information on the dynamics away from the attractor
is carried by $\mathcal{P}_\mu$.
\end{rmk}
%

\subsection{Spectral theory of ergodicity and mixing}
\label{sec:spectrumDecay}

One can see from the definition (\ref{eq:defCorr})
that the correlation function is fully determined by the
transfer (Koopman) operator $\mathcal{P}_\mu$ ($\mathcal{U}_\mu$) on $L^2_\mu(X)$,
i.e.~\emph{with respect to the invariant measure $\mu$}.
In fact, classical results relate the ergodic properties of
measure-preserving dynamical systems to the behaviour of these operators
\cite{Halmos1956,Arnold1968,Lasota1994}.
Note first, that the invariance of the measure $\mu$ together with
the invertibility of the transformation $\Phi$ ensure that the Koopman and transfer operators are 
unitary in $L^2_\mu(X)$.
As a consequence, when considering this functional space,
the spectrum of the operators is contained in the unit circle
$|z| = 1, z \in \mathbb{C}$ and $\mathcal{P}_\mu$ and $\mathcal{U}_\mu$
have the same eigenvalues and eigenfunctions.
Moreover, the following results hold.
\begin{thm}
If the measure-preserving dynamical system
$(X, \mathcal{B}, \Phi, \mu)$ is
\begin{enumerate}[label=(\roman*)]
\item ergodic, then $1$ is a simple eigenvalue of $\mathcal{P}_\mu$, and conversely.
\item mixing, then the only one eigenvalue is one.
\end{enumerate}
\end{thm}
The first item of this theorem can be understood from the fact that
ergodic systems have only one stationary density with respect to the invariant measure.
The second is due to the fact that eigenfunctions associated with eigenvalues
on the unit circle do not decay and thus prevent mixing for general observables.

The rate of decay of correlations, or mixing,
has been the subject of intense research these past decades.
Once again, the latter can be studied from the spectral properties
of the transfer operator $\mathcal{P}_\mu$ \cite{Liverani1995a}.
However, the eigenvalues responsible for such decay,
the \emph{unstable resonances},
lie inside the unit disk and are thus not accessible from operators on
regular functional spaces such as $L^p_\mu$.
Anisotropic Banach spaces of distributions capturing
the dynamics of contraction and expansion of the system and on which the operators
are contracting should instead be considered
\cite{Liverani1995a,Blank2001a,Gouezel2006,Butterley2007,Faure2014,Baladi2017}
\footnote{Contraction by operators can be interpreted in physical systems as determining entropy production.
Recognising this has been essential to understand how reversible microscopic evolutions can lead
to irreversible macroscopic properties of systems out of thermal equilibrium
(see the pioneering work \cite{Misra1979} and \cite{Gaspard1998,Garbaczewski2002}
for reviews).}.

Nonetheless, the coarse-graining induced by numerical algorithms
such as used in section \ref{sec:approxSpectrum}
gives access to the unstable resonances,
although a certain degree of hyperbolicity may be necessary for these resonances to
be stable to perturbations (see \cite{Keller1998,Gouezel2006} and \cite{Baladi1999,Froyland2007a},
respectively, for results in this direction).

The spectral properties of the operators with respect to an invariant measure $\mu$
thus allow for the study of the ergodic properties of this measure.
On the other hand, for the study of the global stability of this set,
operators acting on a neighbourhood of positive Lebesgue measure
of a chaotic set should be considered.

\subsection{Resonances and boundary crisis}
\label{sec:crisisMixing}

In the introduction, we have advocated on heuristic grounds
that the global stability of invariant sets could be studied from the
evolution of densities.
In this case, however, the transfer (Koopman) operators
$\mathcal{P}_m$ ($\mathcal{U}_m$) \emph{with respect to
the Lebesgue measure $m$} should be considered.
Indeed, this should allow one not only to study the mixing dynamics
of an invariant measure, but also the contraction to/escape from such a measure.
This has allowed for new developments in the theory 
of the global stability of dynamical systems
from Lyapunov functions related to the transfer operator $\mathcal{P}_m$ \cite{Vaidya2008}
and from the stable resonances calculated from the Koopman operator $\mathcal{U}_m$ \cite{Mauroy2016}.
In \cite{Mauroy2016}, the Koopman operator is defined on continuous functions.
Along the (contracting) direction of the leaves of the stable manifold,
continuous functions are smoothed under the (pullback) action of the Koopman operator.
In principle, the spectral properties of the transfer operators could allow to study
the global stability of an attractor, although appropriate spaces of distributions should be considered
(see section \ref{sec:spectrumDecay} and \cite{Tantet2017}, Appendix~A).

It follows that the stables resonances, given by the eigenvalues of the transfer
operator $\mathcal{P}_m$ with respect to the Lebesgue measure which
do not belong to the spectrum of the transfer operator $\mathcal{P}_\mu$
with respect to the invariant measure,
are expected to approach the unit circle during the nearing
of a boundary crisis whereby the attractor loses stability
(although exceptions may occur, see \cite{Baladi2007b}, Sect.~6).
This has recently been shown to be the case during a boundary crisis in the Lorenz flow \cite{Tantet2017}.
In this case, the unstable resonances were not affected by the crisis,
so that no changes could be observed in the decay of correlations.

There are situations, however, where the instability mechanism responsible for the crisis
is already affecting the dynamics of the system on the attractor.
This is the case in PlaSim, for which the ice-albedo feedback responsible for the
warm to snowball transition can be triggered by atmospheric variability
before the crisis.
Heuristically, this might be related to the fact that near the crisis the edge state
and the nearby climate state are dynamically extremely similar \cite{Lucarini2017}.
In such case, one would expect slowing down in the decay of correlations
to be visible without having to perturb the system away from the attractor.
In section \ref{sec:slowing}, we will show that this is indeed what happens in PlaSim.
In order to show that this slowing down of the decay of correlations
is indeed due to the approach of unstable resonances
to the unit circle, a reduction method allowing to give coarse-grained approximations
of the unstable resonances will first be presented in the following section \ref{sec:approxSpectrum}.

\section{Approximation of the unstable resonances on reduced spaces}
\label{sec:approxSpectrum}

The resonances are given by the eigenvalues of the transfer operator (and its adjoint)
governing the evolution of densities.
Analytical results on the properties of this operator
for chaotic systems are difficult to obtain (see e.g. \cite{Hasegawa1992b,Gaspard1992,Gaspard1995,Gaspard2001a}).
For high-dimensional systems, to our knowledge,
only qualitative results limited to uniformly hyperbolic systems
have been obtained regarding the distribution of the resonances
in the complex plane \cite{Gouezel2006,Butterley2007,Faure2014}.
Numerical methods are thus required to approximate the
resonances and associated eigenvectors.
Here, we explain how these operators can be approximated from
transition matrices estimated from time series.

For still relatively low-dimensional systems, but with chaotic dynamics,
one approach is to calculate the spectrum of transition matrices resulting
from the projection of the transfer operator on a finite family of basis functions.
The transition probabilities can then be estimated from many short time series
sampling the state space in order to approximate the transfer operator $\mathcal{P}_m$
acting on densities with respect to the Lebesgue measure (see section \ref{sec:methodology}).
This Galerkin truncation with estimation is referred to as \emph{Ulam's method} \cite{ulam1964collection,Dellnitz1999} in the literature.
The method is not limited to the use of characteristic functions (see \cite{Klus2015a} and references therein)
and has also been referred to as the \emph{Extended Dynamical Mode Decomposition} \cite{Williams2015}
and convergence results have recently been obtained \cite{Korda2017}.
An alternative is to apply the ergodic theorem to estimate the transition matrix from
a single long time series converged to the attractor, in which case only
the transfer operator $\mathcal{P}_\mu$ with respect to some physical measure $\mu$
is approximated.
These two methods have been applied to the Lorenz flow
in \cite{Lucarini2016} and \cite{Tantet2017} to respectively calculate the linear response
to forcing from transition matrices approximating the transfer operators and the
resonances across a boundary crisis.

High-dimensional systems, say of more than three dimensions, are not easily amenable to Ulam's method,
since the number of basis functions required to resolve properly a given scale
increases exponentially with the dimension \cite{Lucarini2016} as well as the number
of time series needed to sample the state space.
One is thus led to study the evolution of densities in a reduced space
and to estimate the projection of the transfer operator $\mathcal{P}_\mu$ on this space.
\begin{rmk}
	Whether performed analytically or numerically, one may wonder whether 
	the approximations of the eigenvectors on a finite number of basis functions
	carry relevant information on the decay of correlations described by eigendistributions.
	Indeed, these basis functions usually being regular,
	their linear combinations should not decay under the action of transfer operators
	associated with invertible maps (see section \ref{sec:methodology}).
	However, the projection of the image of densities by the transfer operator
	back to the finite dimensional space spanned by the basis functions introduces some irreversibility.
	As a consequence the eigenvalues and eigenvectors obtained from these truncations may actually
	converge to the unstable resonances and their associated eigendistributions.
	This is explained in detail in \cite{Hasegawa1992b} and \cite{Fishman2002}
	with simple yet highly illustrative examples.
\end{rmk}

\subsection{Reduced transfer operators}
\label{sec:reduced}

The reduction method followed in this study consists in three main steps:
(i) the projection of the transfer operator to a lower dimensional space
through the conditional expectation of the invariant measure with respect to this space ;
(ii) the discretisation of the resulting reduced operator on a finite family of basis functions ;
(iii) the estimation of the transition probabilities from a long time series.

Such a reduction has been introduced by \cite{Chekroun2014} to the climate field,
building up on the work of \cite{Shutte1999} in molecular dynamics.
For this purpose, a continuous mapping $h$ from
the state space $X$ to the lower dimensional space $Y$ is defined.
For a given measure $\mu$ on $\mathcal{B}$,
this mapping induces a Borel $\sigma$-algebra $\mathcal{C}$ of $Y$
and a \emph{marginal measure} $\rho$ such that
for any integrable function $f$ on the reduced space $Y$,
\begin{eqnarray*}
	\int f(h(x)) \mu(dx) = \int f(y) \rho(dy).
\end{eqnarray*}
Thus, the marginal measure $\rho$
allows one to calculate averages on the reduced space $Y$ according to $\mu$.
In addition, from the \emph{disintegration theorem} (\cite{Dudley2004}, Chap.~10),
there exists a unique \emph{conditional measure} $y \mapsto \mu_y$ such that,
for any $f$ in $L^1_\mu(X)$,
\begin{eqnarray*}
	\int_X f(x) \mu(dx) = \int_Y \int_{h^{-1}(y)} f(x) \mu_y(dx) \rho(dy).
\end{eqnarray*}

The conditional expectation $\mathbb{E}[ \cdot | \mathcal{C}]$ 
associated with the conditional measure $y \mapsto \mu_y$
defines an orthogonal projection from $L^1_\mu(X)$ to $L^1_\rho(Y)$ (\cite{Kallenberg2002}, Chap.~5).
This allows one to define a family of \emph{reduced transfer operators}
$\mathcal{P}_\rho^{(n)}, n \ge 0,$ on $L^1_\rho(Y)$,
which is as close as possible to the transfer operators $\mathcal{P}_\mu^n$
in the sense of the $L^1$-norm. This operator is explicitly given by
\begin{eqnarray}
	(\mathcal{P}_\rho^{(n)} f)(y)
	:= \mathbb{E}\left[\mathcal{P}_\mu^n (f \circ h) | y\right]
	= \int_{h^{-1}(y)} \mathcal{P}_\mu^n (f \circ h)(x) \mu_y(dx),
	\quad f \in L^1_\rho(Y).
	\label{eq:defRTO}
\end{eqnarray}
In other words, the action of $\mathcal{P}_\rho^{(n)}$
consists in lifting the function $f$ from the reduced function space $L^1_\rho(Y)$
to $L^1_\mu(Y)$, next in applying the true transfer operator $\mathcal{P}_\mu$ $n$-times to this lift
and then in projecting back to the reduced space by applying the conditional expectation.
This is summarised in the following diagram,
with $\mathcal{P}_\rho := \mathcal{P}_\rho^{(1)}$,
\begin{displaymath}
	\centering
	\xymatrix{
		f \circ h \ar[rr]^{\mathcal{P}_\mu} && \mathcal{P}_\mu (f \circ h) \ar[dd]^{E[ \cdot | \mathcal{C}]} \\
		&&\\
		f \ar[rr]_{\mathcal{P}_\rho}	\ar[uu]^{\circ h}	&&	\mathcal{P}_\rho f
	}
\end{displaymath}

As was shown in \cite{Chekroun2014}, Theorem A,
%
%
each operator $\mathcal{P}_\rho$ is a contraction on $L^1_\mu(X)$
such that, for any two sets $B$ and $C$ in the reduced space $Y$,
\begin{eqnarray}
	\frac{\left<\mathcal{P}_\rho \mathbf{1}_B, \mathbf{1}_C \right>_\rho}{\rho(B)}
	= \frac{\mu\left(h^{-1}(B) \cap \Phi^{-1} h^{-1}(C)\right)}{\mu(h^{-1}(B))}
	\label{eq:reducedTransition}
\end{eqnarray}
where
\begin{eqnarray*}
	\left<\mathcal{P}_\rho f, g \right>_\rho
	= \int_Y \mathcal{P}_\rho f(y) \enskip g(y) \rho(dy)
\end{eqnarray*}
denotes the scalar product between functions $f$ and $g$ in $L^1_\rho(Y)$ and $L^\infty_\rho(Y)$, respectively.
In other words, according to (\ref{eq:reducedTransition}), the reduced transfer operators
allow to calculate transition probabilities between coarse grained sets in the full state space $X$
as long as the latter can be obtained from the preimage by $h$ of any two sets in the reduced space $Y$.
In fact, it follows from the definition (\ref{eq:defRTO}) that
the correlation function between two arbitrary functions
$f$ and $g$ in the reduced function space $L^1_\rho(Y)$ (assumed with zero mean for convenience)
can be directly calculated from the family of reduced transfer operators, i.e.
\begin{eqnarray*}
	\left< \mathcal{P}_\rho^{(n)} f, g \right>_\rho
	= \left< \mathcal P_\mu^n (f \circ h), g \circ h \right>_\mu
	= C_{f \circ h, g \circ h}(n).
\end{eqnarray*}

It is then possible to extract information about the unstable resonances,
i.e.~the eigenvalues $\zeta_k$ of $\mathcal{P}_\mu$ (on some appropriate space of distributions)
describing the decay of correlations,
from the eigenvalues $\bar{\zeta}_k$ of the reduced transfer operator $\mathcal{P}_\rho$,
referred to as the \emph{reduced unstable resonances}\footnote{
Let us stress that no information about the stable resonances
can be extracted from the reduced transfer operator $\mathcal{P}_\rho$
as they are given as a projection of the transfer operator $\mathcal{P}_\mu$
with respect to the invariant measure $\mu$ rather than with respect to the Lebesgue measure $m$.}.

However, it is not in general true that
$\left< \mathcal{P}_\rho^n f, g \right>_\rho = C_{f \circ h, g \circ h}(n)$,
so that it is not possible to calculate the correlation function
between observables on the reduced space at a given time $n$
by iterating the reduced transfer operator for $n = 1$.
Indeed, the non-commutativity of the diagram associated with
the loss of information when going from the state space to the reduced space
introduces memory effects \cite{Tantet2015}.
The dynamics in the reduced space is thus non-Markovian, as can be understood
from the Mori-Zwanzig formalism \cite{Zwanzig2001, Chorin2009, Kondrashov2014}.
As a consequence, the family of reduced transfer operators $\mathcal{P}_\rho^{(n)}$ for $n \ge 0$,
do not in general constitute a semigroup and their eigenvalues $\bar{\zeta}_k(n)$
are not related by a spectral mapping formula.
Indeed, due to the semigroup property of the family of transfer operators $\mathcal{P}_\mu^n, n \ge 0$,
the eigenvalues $\zeta_k(n)$ of the $n$th powers of the transfer operator $\mathcal{P}_\mu$
are related by the spectral mapping formula \cite{Engel2001}
\begin{eqnarray}
	\lambda_k := \frac{1}{n} \log \zeta_k(n), \quad n \ge 0,
	\label{eq:reducedEigenvalues}
\end{eqnarray}
where $\log$ denotes the complex logarithm (assuming that we can take the principal part)
and $\lambda_k$ is a complex number independent of $n$.
The eigenvalues $\bar{\zeta}_k(n)$ of the reduction of the $n$th powers of the transfer
operator $\mathcal{P}_\mu$ do not in general satisfy such a relation.
Depending on the reduced space defined by $h$,
the eigenvalues of $\mathcal{P}_\rho$ will thus contain mixed information about
different eigenvalues of $\mathcal{P}_\mu$.

However, in some cases, it may still be that the leading reduced resonances satisfy the spectral mapping
formula (\ref{eq:reducedEigenvalues}) and appropriately describe the decay of correlations in the reduced space.
For example, in the limit of a strong time-scale separation between the dynamics
in the reduced space and its complement, it can be rigorously shown \cite{Crommelin2011}
thanks to homogenisation techniques \cite{Pavliotis2008}
that the reduced unstable resonances actually converge to some of the eigenvalues
of the true transfer operator, so that they do not depend on $n$.
Moreover, even if no time-scale separation exists,
the slowest decay rates $\bar{\lambda}_k(n) := \frac{1}{n} \log \bar{\zeta}_k(n)$ may happen
to depend only weakly on number of iterations of the time $n$, thus behaving "as Markovian" \cite{Tantet2015}.
Testing the dependence of the reduced unstable resonances on time thus
provides an important test of their robustness to memory effects in the reduced space (see \ref{ref:robustLag}).
Note that when presenting the numerical results in section \ref{sec:results},
the formula (\ref{eq:reducedEigenvalues}) will always be applied to represent the rates $\lambda_k$.
Thus the approach of the unstable resonances to the unit circle will correspond to the approach of these
rates to the imaginary axis.

%
%
\begin{rmk}
\label{rmk:choiceObs}
	In theory, the observation operator $h$ defining the reduced state space should be chosen
	so as for the eigenvectors associated with the leading unstable resonances
	to project significantly on the reduced space and leaving other eigenvectors in its complement.
	However, we do not know of any systematic method to choose the observation operator $h$,
	without the full knowledge of the true transfer operators.
	One approach would be to select observables such that their correlation functions, including the cross-correlations,
	decay slowly.
	Another approach, based on physical grounds, is to choose observables which are known to be involved
	in the physical processes of interest, such as the ice-albedo feedback.
	In this study, both techniques have been applied to choose the observables used in section \ref{sec:results}.
\end{rmk}

\subsection{Estimation of the reduced transfer operators from time series}

Once the reduced transfer operator has been defined,
their discrete approximation follows easily.
The reduced transfer operator is thus projected on a
truncated family $G = \{\chi_1, ..., \chi_n\}$ of orthogonal basis functions,
i.e. such that $\left<\chi_i, \chi_j \right>_\rho = 0$ when $i \ne j$.
This can be seen as a further reduction
from $L^1_\rho(Y)$ to $L^1_\rho(\mathrm{span~}G)$,
as in the previous section \ref{sec:reduced}.
Any vector of components $\mathbf{f} = (f_1, ..., f_m)$ in $\mathbb{R}^m$,
defines an observable
\begin{eqnarray*}
	f = \sum_{i = 1}^m f_i \frac{\chi_i}{\rho(\chi_i)} \quad \mathrm{such~that}
	\quad f_i = \left<f, \chi_i\right>_\rho
	\quad \mathrm{and} \quad \int_X f d\rho = \sum_{i = 1}^m f_i.
\end{eqnarray*}
For any such $f$, the component of $\mathcal{P}_\rho f$
on the basis function $\chi_j$ is then given by
\begin{eqnarray*}
	(\mathcal{P}_\rho f)_j = \left<\mathcal{P}_\rho f, \chi_j \right>_\rho
	&= \sum_{i = 1}^m f_i \frac{\left<\mathcal{P}_\rho \chi_i, \chi_j\right>_\rho}{\rho(\chi_i)}
	= \left(\bv{f} \bv{P}\right)_j,
\end{eqnarray*}
where we have defined the transition matrix $\bv{P}$
with elements the normalised correlations
\begin{eqnarray}
	\bv{P}_{ij}
	&:= \frac{\left<\mathcal{P}_\rho \chi_i, \chi_j\right>_\rho}{\rho(\chi_i)}.
	\label{eq:correlationMatrix}
\end{eqnarray}

In this study, we have chosen to only consider families of indicator functions
$G = \{\mathbf{1}_{B_1}, ..., \mathbf{1}_{B_n}\}$
on a grid of disjoint boxes $\{B_1, ...,B_n\}$ such that $\cup_{i = 1}^m B_i \subseteq \mathcal{C}$
(this choice providing us with sufficiently robust results, see \ref{sec:robust}).
In this case, the transition probabilities are simply given by
\begin{eqnarray}
	\bv{P}_{ij}
	&= \frac{\left<\mathcal{P}_\rho \mathbf{1}_{B_i}, \mathbf{1}_{B_j}\right>_\rho}{\rho(B_j)}
	= \frac{\mu\left(h^{-1}(B_i) \cap \Phi_{-1} h^{-1}(B_j)\right)}{\mu(h^{-1}(B_j))},
	\label{eq:transitionProb}
\end{eqnarray}
where the second equality follows from (\ref{eq:reducedTransition})
and relates the transition probabilities in the reduced space to the
transition probabilities between coarse grained sets in the full state space (\cite{Chekroun2014}, Theorem A).

Once the analysis of the reduced transfer operator $\mathcal{P}_\rho$
has been reduced to a discrete problem on a transition matrix $\bv{P}$,
it remains to calculate the transition probabilities of this matrix.
While in the original version of Ulam's method (see e.g.~\cite{Dellnitz1999})
this is done by an estimation from an ensemble of many short simulations
sampling a given volume in state space, this cannot be done
in the high-dimensional case considered here.

This is why, in practice, the reduction method presented in the previous section \ref{sec:reduced}
only allows for the reduction of the transfer operator $\mathcal{P}_\mu$ with respect to the invariant measure,
and thus for the approximation of the unstable resonances only.
On the other hand, assuming that the invariant measure $\mu$ is physical (see remark~\ref{rmk:physical}),
the transition probabilities in (\ref{eq:transitionProb}) can be estimated 
from a single long time series $\{x_s\}_{1 \le s \le T}$ converged to the attractor, according to the time means
\begin{eqnarray}
	\bv{P}_{ij}
	= \frac{\frac{1}{T} \sum_{s = 0}^T \chi_i(h(x_{s})) \chi_j (h(x_{s + t}))}{\frac{1}{T} \sum_0^T \chi_i (h(x_s))}
	= \frac{\#\{h(x_s) \in B_i, h(x_{s + t}) \in B_j\}}{\#\{h(x_s) \in B_i\}},
	\label{eq:MLE}
\end{eqnarray}
where $h(x_s)$ is the measurement taking a sample $x_s$ at time $s$
of the full state to the reduced space.
The second equality is valid only for the case of indicator functions and the right-hand side
is just the normalised count of the number of times the time series transits
from one box to another after one application of the map $\Phi$.
In practice, the time series is sampled at a finite rate for a finite time,
so that the time means yield the Maximum Likelihood Estimator (MLE, see e.g~\cite{Billingsley1961})
of the transition probabilities.

Once the transition matrices have been estimated,
the eigenvalues and the right and left eigenvectors of the transition matrix $\bv{P}$
can be calculated numerically (with an eigensolver such as ARPACK \cite{Lehoucq1997})
to get discrete approximations of the eigenvalues, eigenvectors
and adjoint eigenvectors of the reduced transfer operator $\mathcal{P}_\rho$
\footnote{See \cite{Crommelin2009} for useful estimates
of the convergence of the eigenvalues and eigenvectors of the transition matrices.}.
To summarise, the following numerically tractable problem is solved:
\begin{enumerate}
	\item Integrate a long time series $\{x_s\}_{1 \le s \le T}$ from a numerical model.
	\item Define an observation operator $h: X \to Y$ to a reduced space $Y$ of low dimension
	(typically one to three-dimensional, depending on the length of the available time series)
	inducing a family of reduced transfer operators $\mathcal{P}_\rho$, according to (\ref{eq:defRTO}).
	\item Select a finite number $m$ of basis function $\{\chi_1, ..., \chi_n\}$
	on which to discretise the reduced operators
	(once again, the number of basis functions is limited by the length of the time series).
	\item Estimate the transition probabilities (\ref{eq:MLE}) of the
	coarse grained reduced transfer operators $\mathcal{P}_\rho$.
	\item Calculate the eigenvalues $\bar{\zeta}_k$ of the transition matrix $\bv{P}$.
	\item Convert the $\bar{\zeta}_k$ to the complex rates
	$\bar{\lambda}_k$, according to (\ref{eq:reducedEigenvalues}).
	\item Check the robustness of the $\bar{\lambda}_k$ to the number of iterations $n$
	of the original transfer operator, the discretisation and the sampling (see \ref{sec:robust}).
\end{enumerate}

\section{Results}
\label{sec:results}

This section presents the main results of this article, namely that
(i) slowing down of the decay of correlations occurs before the boundary crisis in PlaSim ;
and (ii) such a slowing down is explained by the approach of the reduced unstable resonances to the unit circle.

\subsection{Changes in the statistical steady-state}
\label{sec:moments}

The long simulations described in section \ref{sec:simulations} are now used to study the statistical 
changes occurring along the warm branch of statistical steady-states as the solar constant is decreased towards 
its critical value $S_c$.
We start by discussing the statistics of a few observables to stress their key role in the ice-albedo feedback, the instability mechanism responsible for the attractor crisis (see section \ref{sec:crisisPlasim}).
Based on physical grounds (see remark~\ref{rmk:choiceObs}),
the following observables will be used
\footnote{These observables play the role of the function $g$ in section~\ref{sec:methodology}).}:
\begin{itemize}
	\item the fraction of Sea Ice Cover (SIC) in the Northern Hemisphere (NH),
	\item the Mean Surface Temperature (MST) averaged around the Equator (Eq, i.e. from $15^\circ S$ to $15^\circ N$),
\end{itemize}
The SIC is the primary variable involved in the ice-albedo feedback as an ice-covered ocean has a much bigger albedo than an ice-free ocean. Sea ice forms when the surface temperature is below the freezing point ($T_f \approx 271.8 K$) motivating the choice of a surface temperature indicator. Furthermore, the MST at the equator is an indicator of the amount of heat stored in the ocean at low-latitudes
%
and that can potentially be transported to high latitudes through horizontal diffusion in the Ocean or, indirectly, through the general circulation of the Atmosphere.
\begin{figure}[h!]
	\centering
        \begin{subfigure}[b]{5cm}
        (a)\\
		\includegraphics[width=\textwidth]{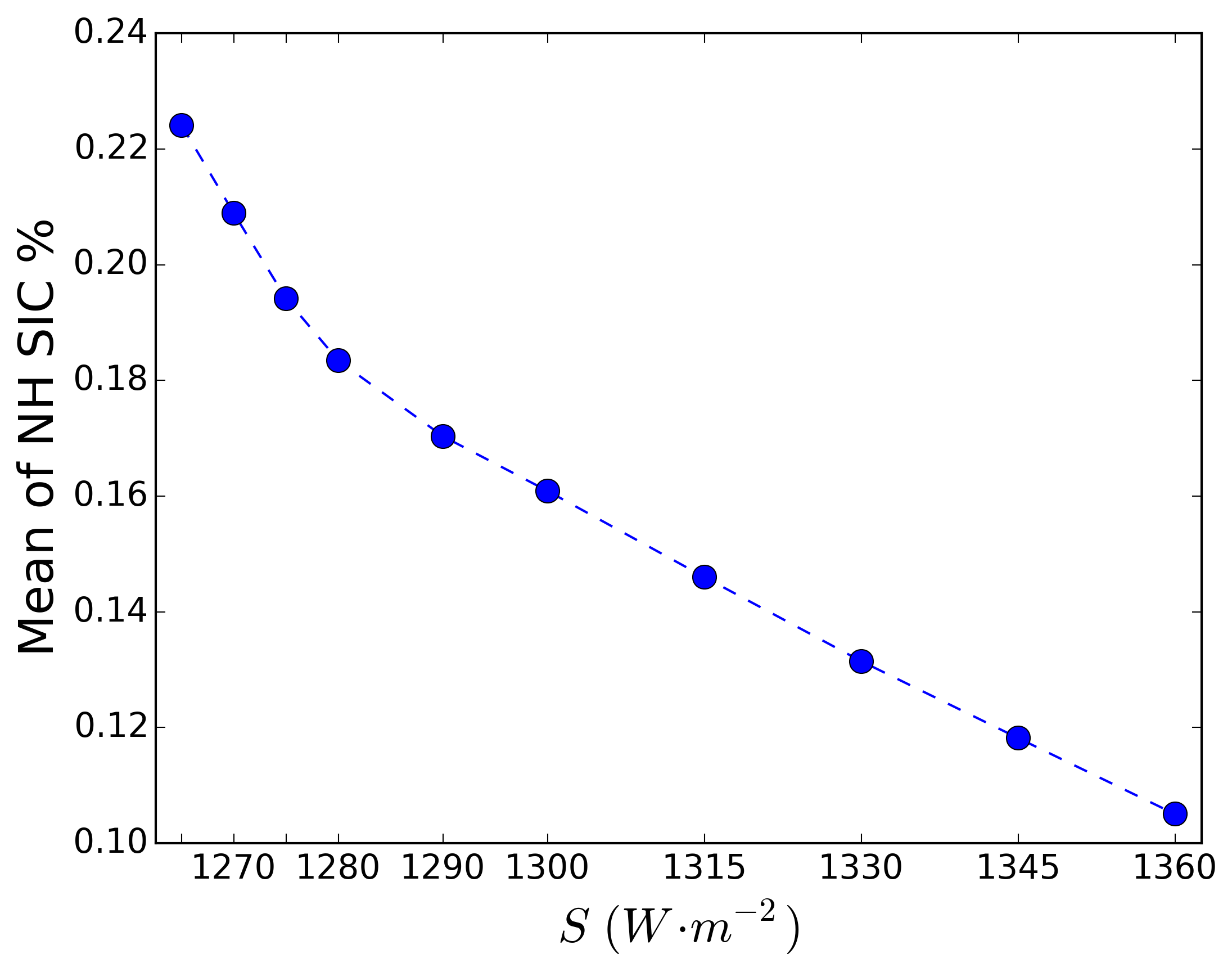}
	\end{subfigure}
        \begin{subfigure}[b]{5cm}
        (c)\\
		\includegraphics[width=\textwidth]{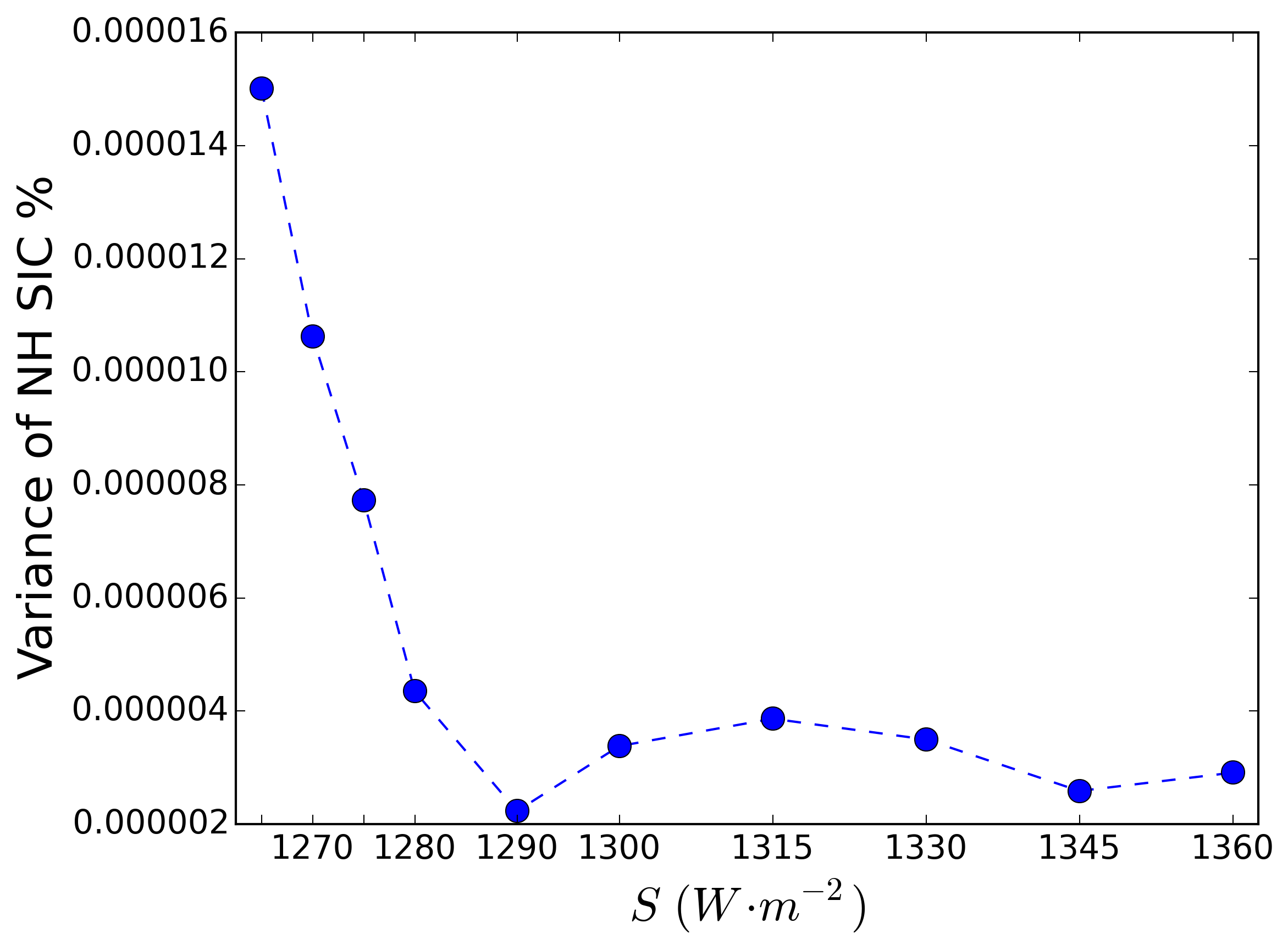}
	\end{subfigure}
        \begin{subfigure}[b]{5cm}
        (e)\\
		\includegraphics[width=\textwidth]{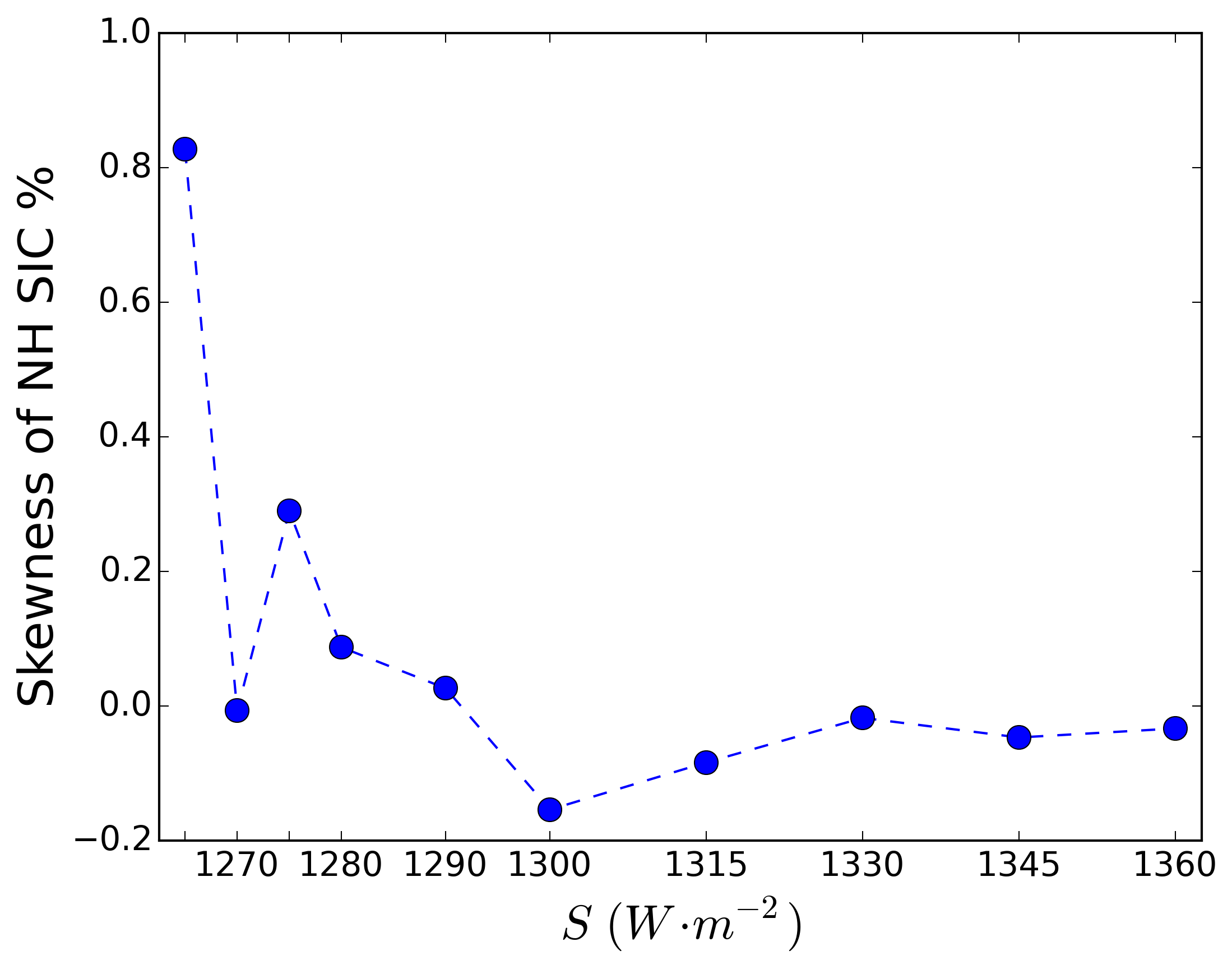}
	\end{subfigure}
        \begin{subfigure}[b]{5cm}
       (b)\\
		\includegraphics[width=\textwidth]{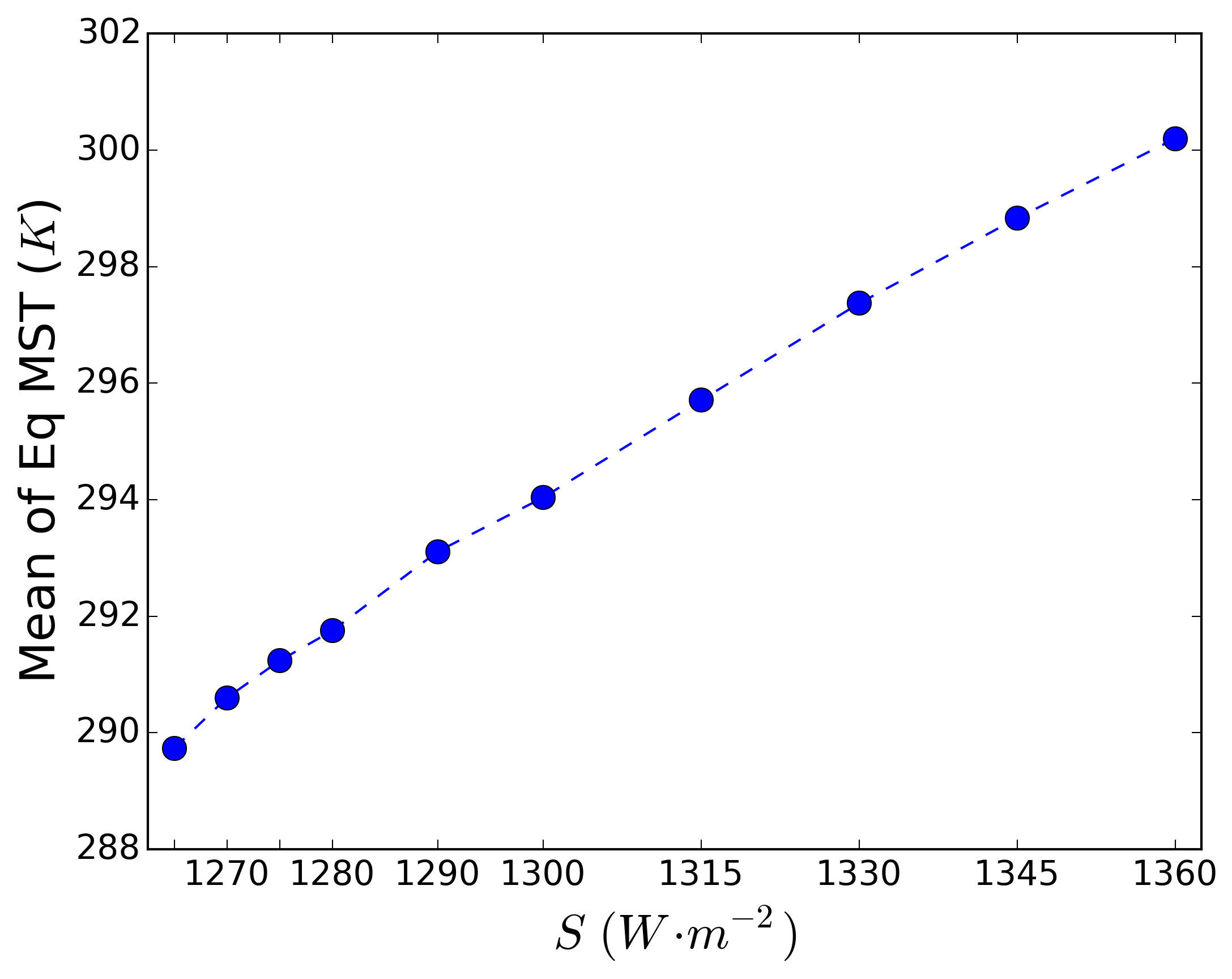}
	\end{subfigure}
        \begin{subfigure}[b]{5cm}
        (d)\\
		\includegraphics[width=\textwidth]{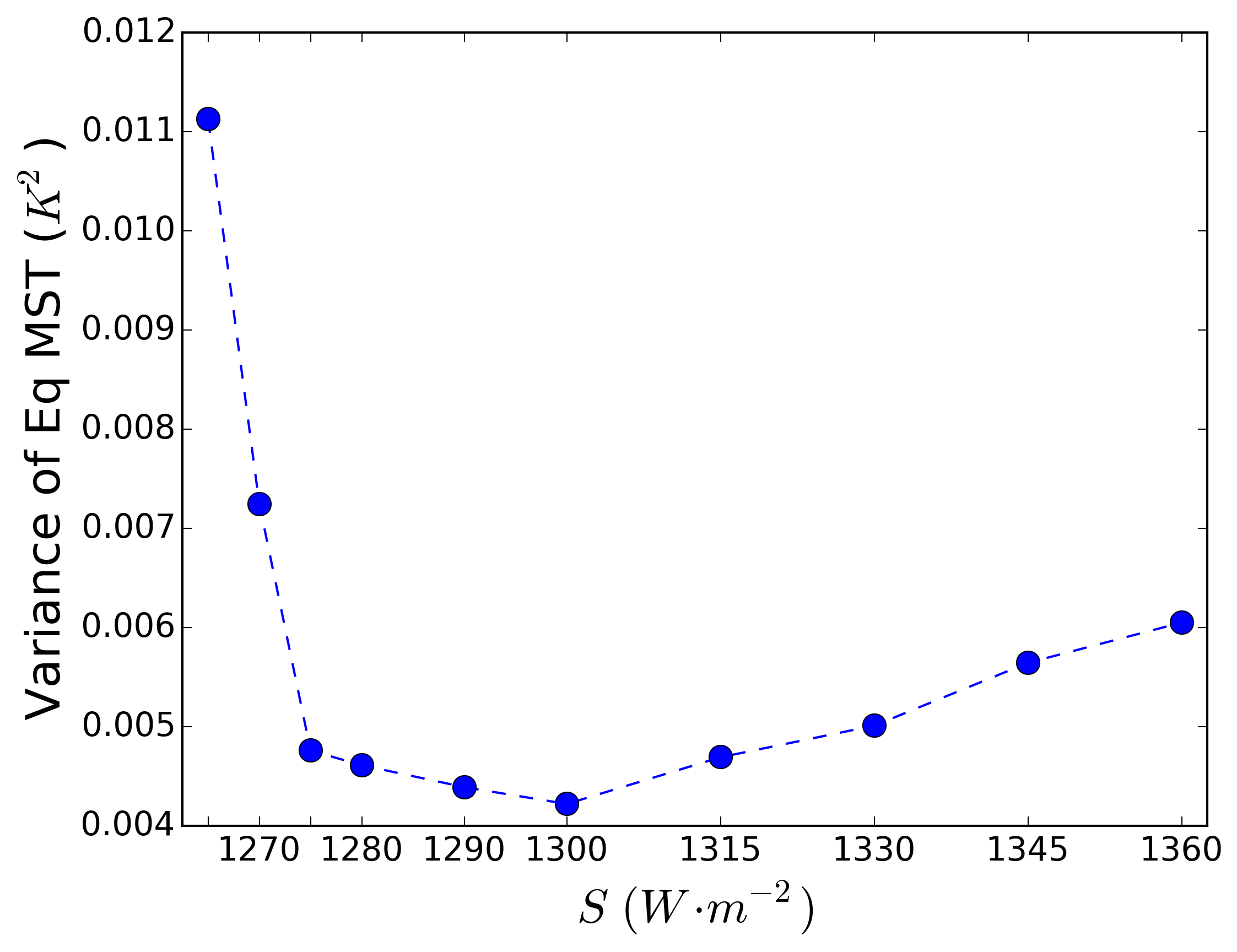}
	\end{subfigure}
        \begin{subfigure}[b]{5cm}
        (f)\\
		\includegraphics[width=\textwidth]{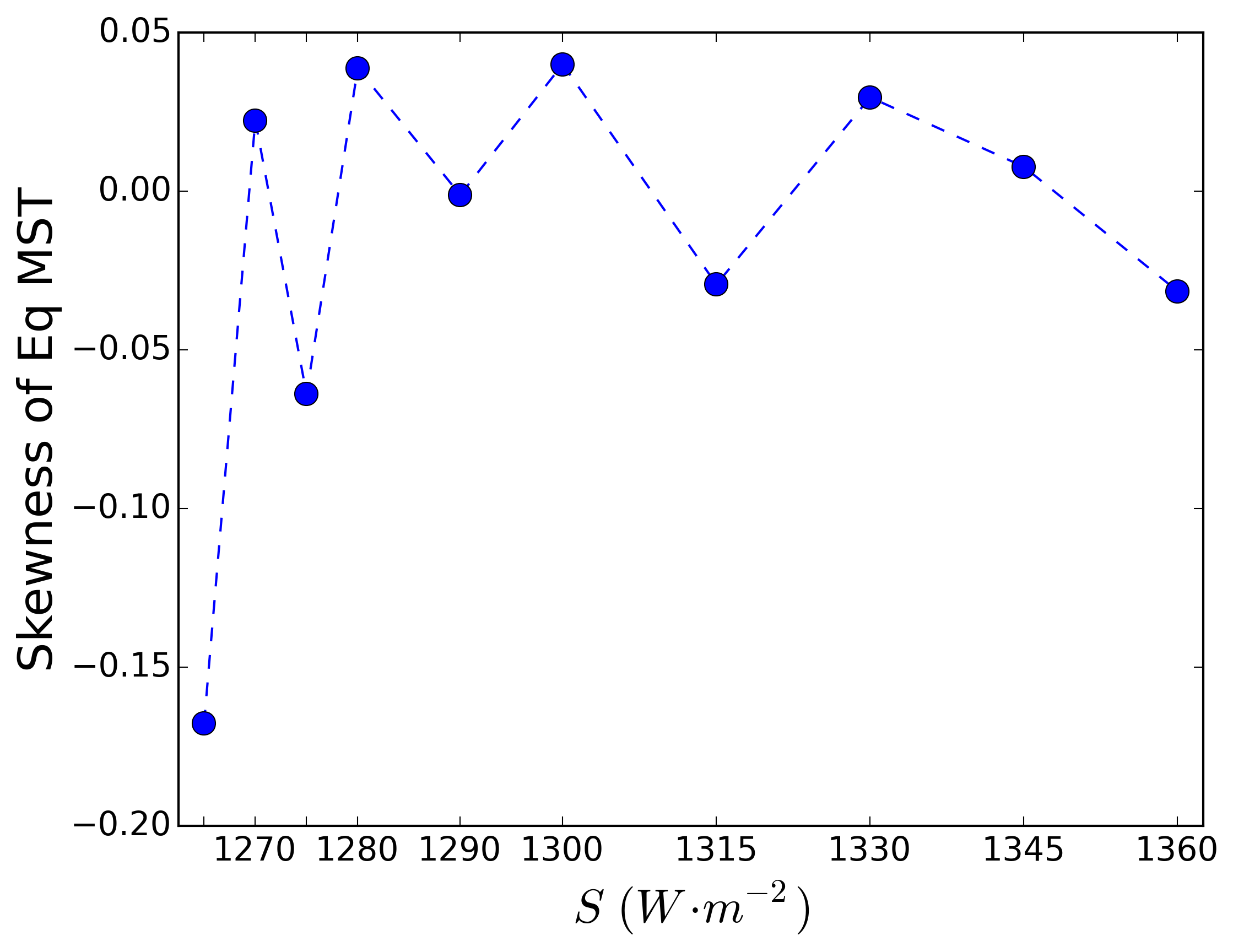}
	\end{subfigure}
	\caption{Sample mean (a-b), variance (c-d) and skewness (e-f) of yearly averages
	of the NH SIC fraction (top) and the Eq MST in Kelvin (bottom), versus the solar constant $S$ (Wm$^{-2}$).}
	\label{fig:moments}
\end{figure}

The sample mean (calculated from the long simulations presented in section \ref{sec:simulations}, according to (\ref{eq:Birkhoff}))
of these observables are represented in figure \ref{fig:moments}(a-b), 
allowing us to recap the changes in the climate steady-state  \cite{Lucarini2010a, Boschi2013}.
As the solar constant is decreased, 
less thermal Outgoing Longwave Radiation (OLR) is necessary to balance the Incoming Shortwave Radiation (ISR) from the sun, so that, as predicted by the Stefan-Boltzmann law of black bodies, the temperature of the Earth cools down, explaining the decrease in the Eq MST (fig. \ref{fig:moments}(b)). The cooling of the surface of the Earth induces an increase of the extent of the sea ice towards low latitudes, further weakening the amount of absorbed solar radiation and strengthening the cooling. These changes in the mean of these observables are smooth, almost linear. Only for a value of the solar constant smaller than $1280~Wm^{-2}$ does the increase in NH SIC strengthen.

However, the sample variance (fig. \ref{fig:moments}(c-d)) and skewness (fig. \ref{fig:moments}(e-f)) experience rougher changes with the solar constant. Indeed, the variance of the NH SIC and Eq MST increases dramatically for $S < 1300~W/m^2$. This increase indicates that the ice-albedo feedback is less and less damped as the criticality is approached. Thus, an anomaly in, for instance, the surface temperature can lead to an increase in the SIC which will be less easily damped by heat transport from a cooler equator.

\begin{figure}[h!]
	\centering
        \begin{subfigure}[b]{7.5cm}
        (a)\\
		\includegraphics[width=\textwidth]{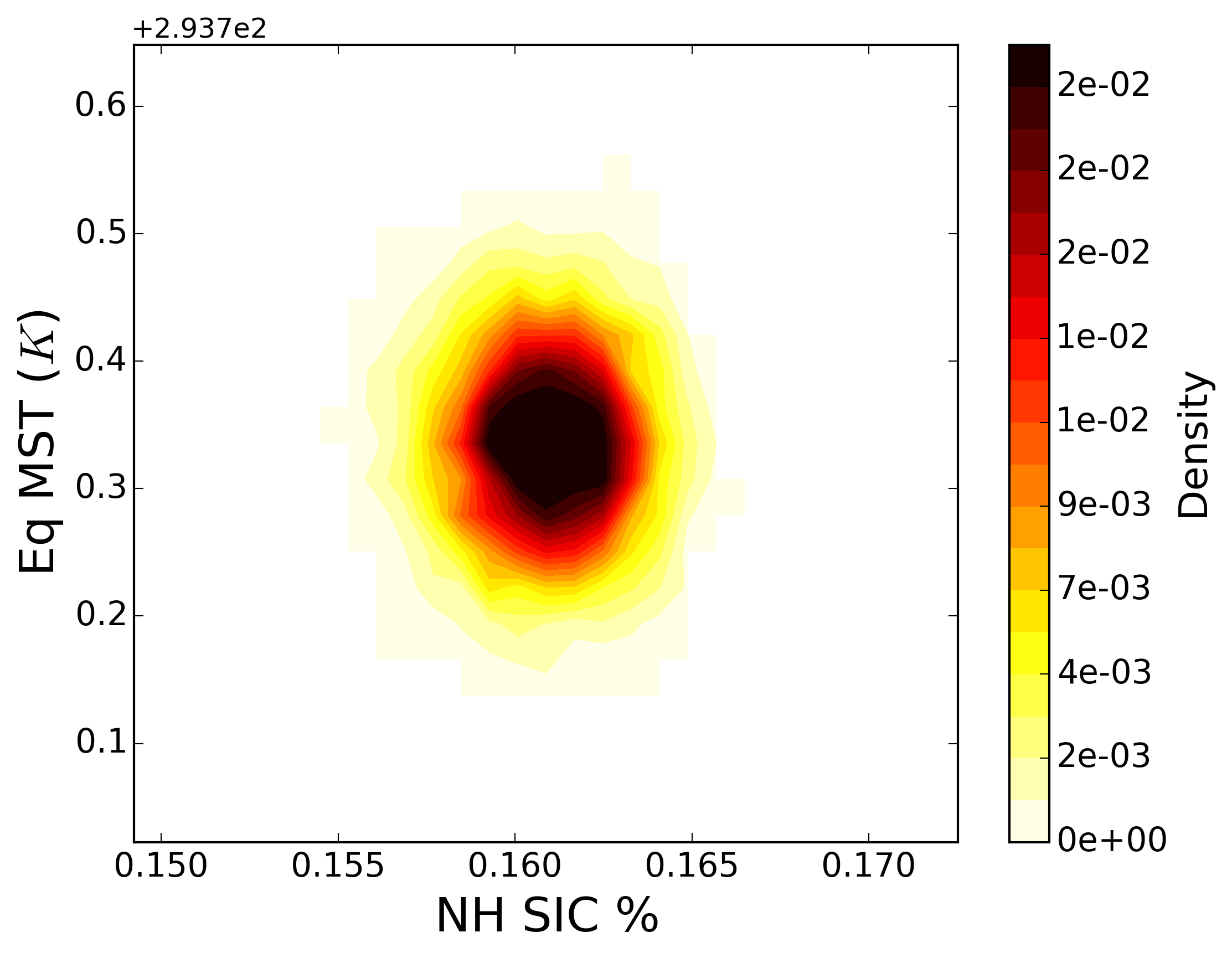}
	\end{subfigure}
        \begin{subfigure}[b]{7.5cm}
        (b)\\
		\includegraphics[width=\textwidth]{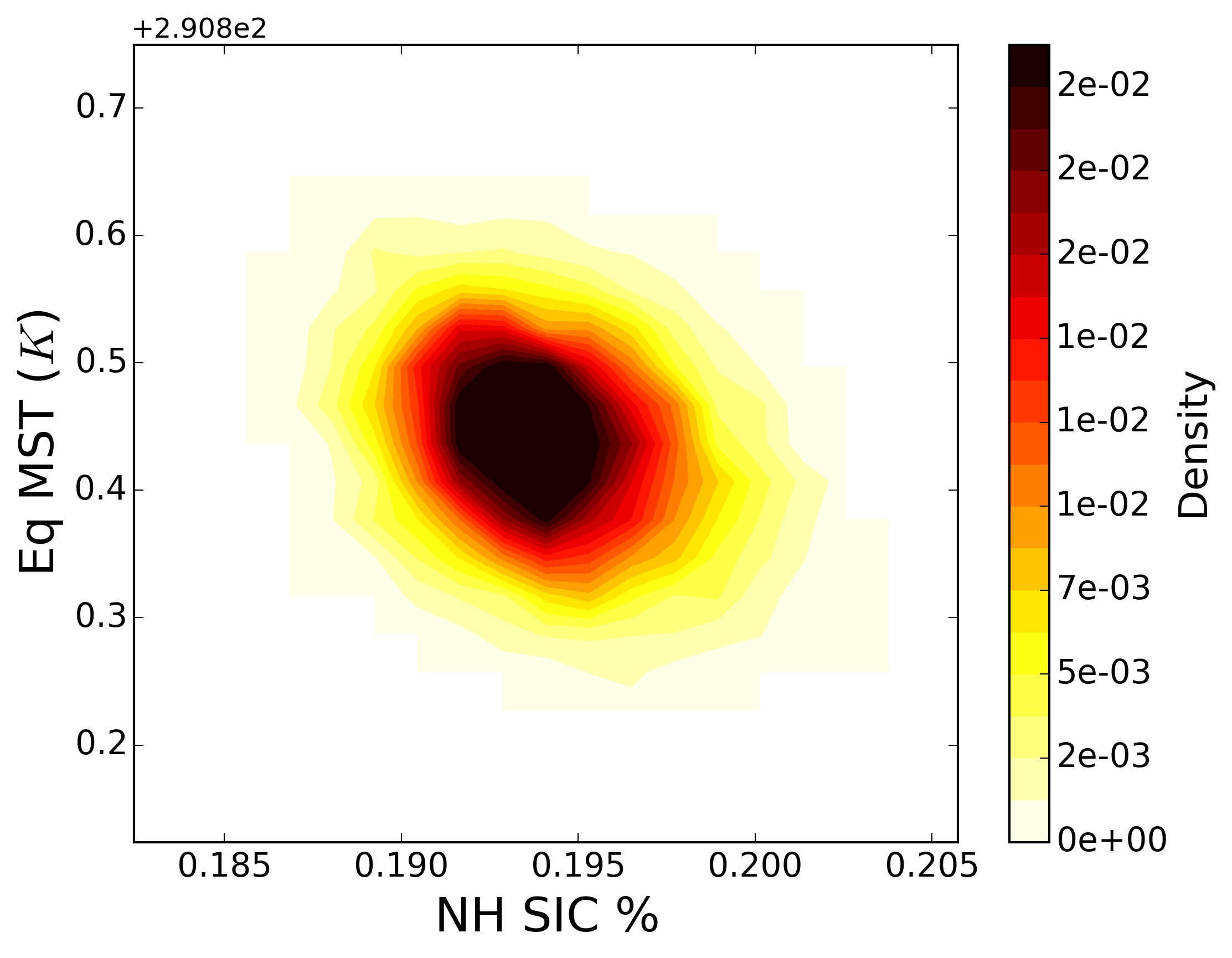}
	\end{subfigure}
        \begin{subfigure}[b]{7.5cm}
        (c)\\
		\includegraphics[width=\textwidth]{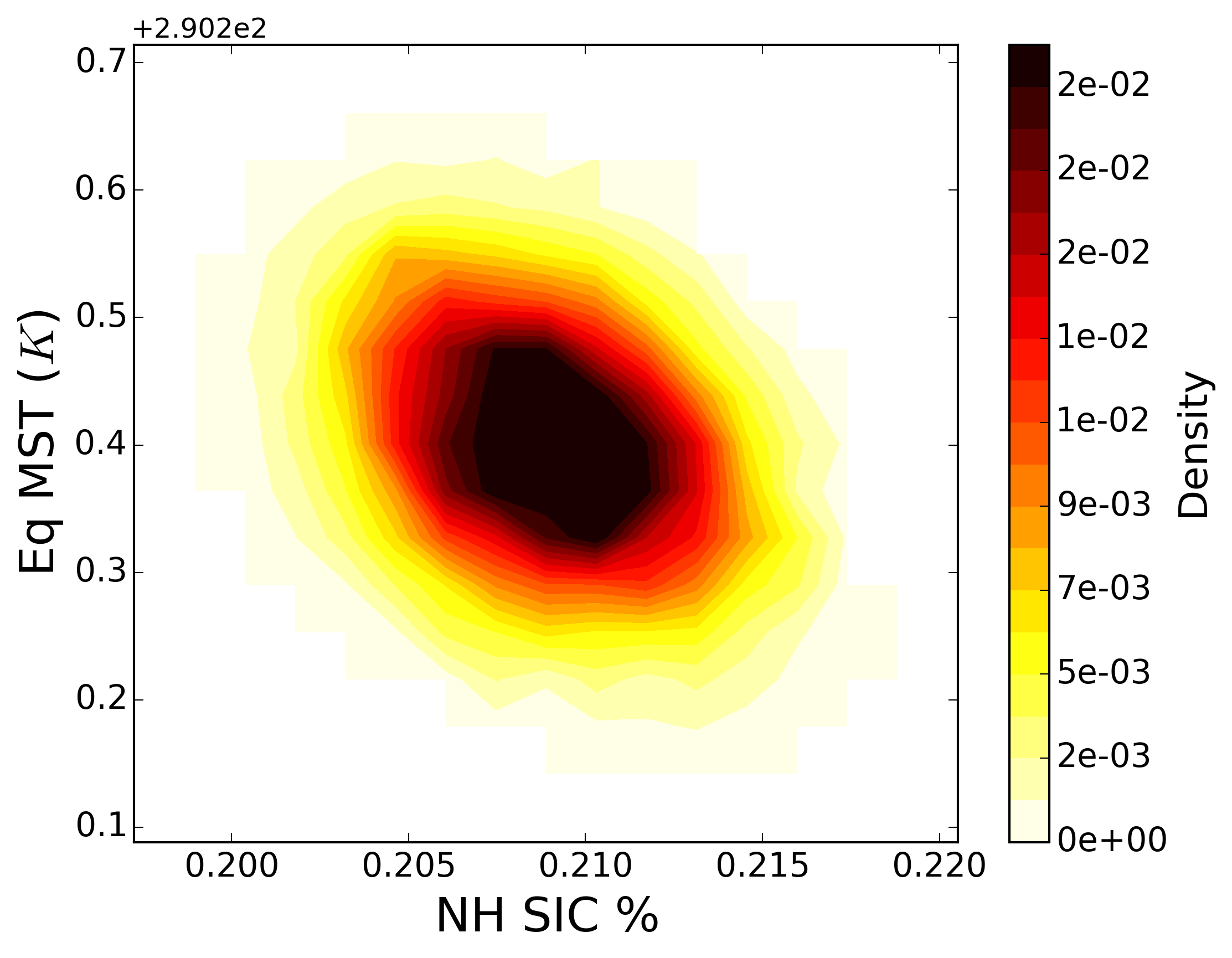}
	\end{subfigure}
        \begin{subfigure}[b]{7.5cm}
	(d)\\
		\includegraphics[width=\textwidth]{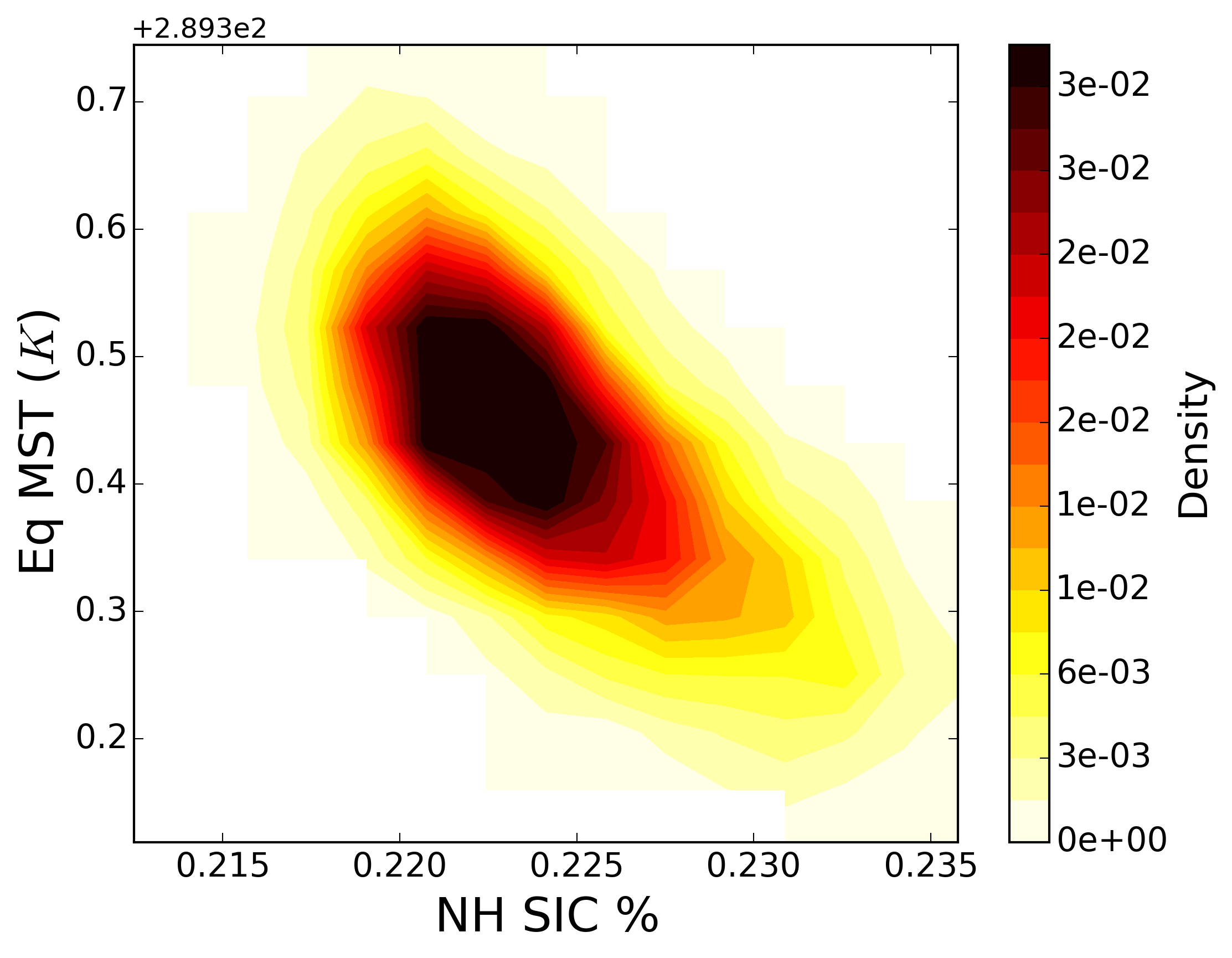}
	\end{subfigure}
	\caption{Joint PDFs of the NH SIC and Eq MST for (a) $S = 1300~W m^{-2}$, (b) $S = 1275~W m^{-2}$, (c) $S = 1270~W m^{-2}$ and (d) $S = 1265~W m^{-2}$, estimated from histograms on the same grid as used for the transition matrices of section \ref{sec:resultSpectrum}. While the axes are translated to follow the mean of the PDFs, their scaling is the same for each panel and is taken so as to span -3 to +3 standard deviations of the NH SIC and the Eq MST for $S = 1265~W m^{-2}$.
	Note that the offset of the Eq MST is reported (in K) on the upper right side of each panel.}
	\label{fig:PDFJOINT}
\end{figure}

To stress the statistical relationship between the NH SIC and the Eq MST,
their joint Probability Density Function (PDF)
\footnote{Similarly as for the transition probabilities (\ref{eq:MLE}),
the PDF is estimated from the number of realisations
falling in each box of the grid defined in section \ref{sec:resultSpectrum}.
In other words, it is a binned histogram which
gives a discrete approximation of the density of the marginal measure $\rho$
(see section \ref{sec:approxSpectrum}) for an observation operator $h$ with components the NH SIC and the Eq MST.}
is plotted figure \ref{fig:PDFJOINT}, for decreasing values of the solar constant.
Note that the offset of the Eq MST is reported (in K) on the upper right side of each panel.
Apart from the changes in the mean and the increase in the variance (the axis of each panel have the same scaling), these plots show that while for $S = 1300~W m^{-2}$ (panel (a)) the NH SIC and Eq MST are only weakly correlated and approximately normally distributed, their distribution becomes more and more tilted and skewed as the solar constant is decreased (panel (b-d)). This increase in the skewness of the NH SIC and Eq MST is also visible in figure \ref{fig:moments}(e-f) from which one can verify that positive values of the NH SIC and negative values of the Eq MST are favoured as the criticality is approached. This shape is in agreement with the fact that,
as the ice-albedo feedback is less and less damped,
larger excursions of the SIC to low latitudes and concomitant cooler temperatures are permitted.

\subsection{Slowing down of the decay of correlations}
\label{sec:slowing}

\begin{figure}[h!]
	\centering
	\begin{subfigure}[b]{7.5cm}
	(a)\\
		\includegraphics[width=\textwidth]{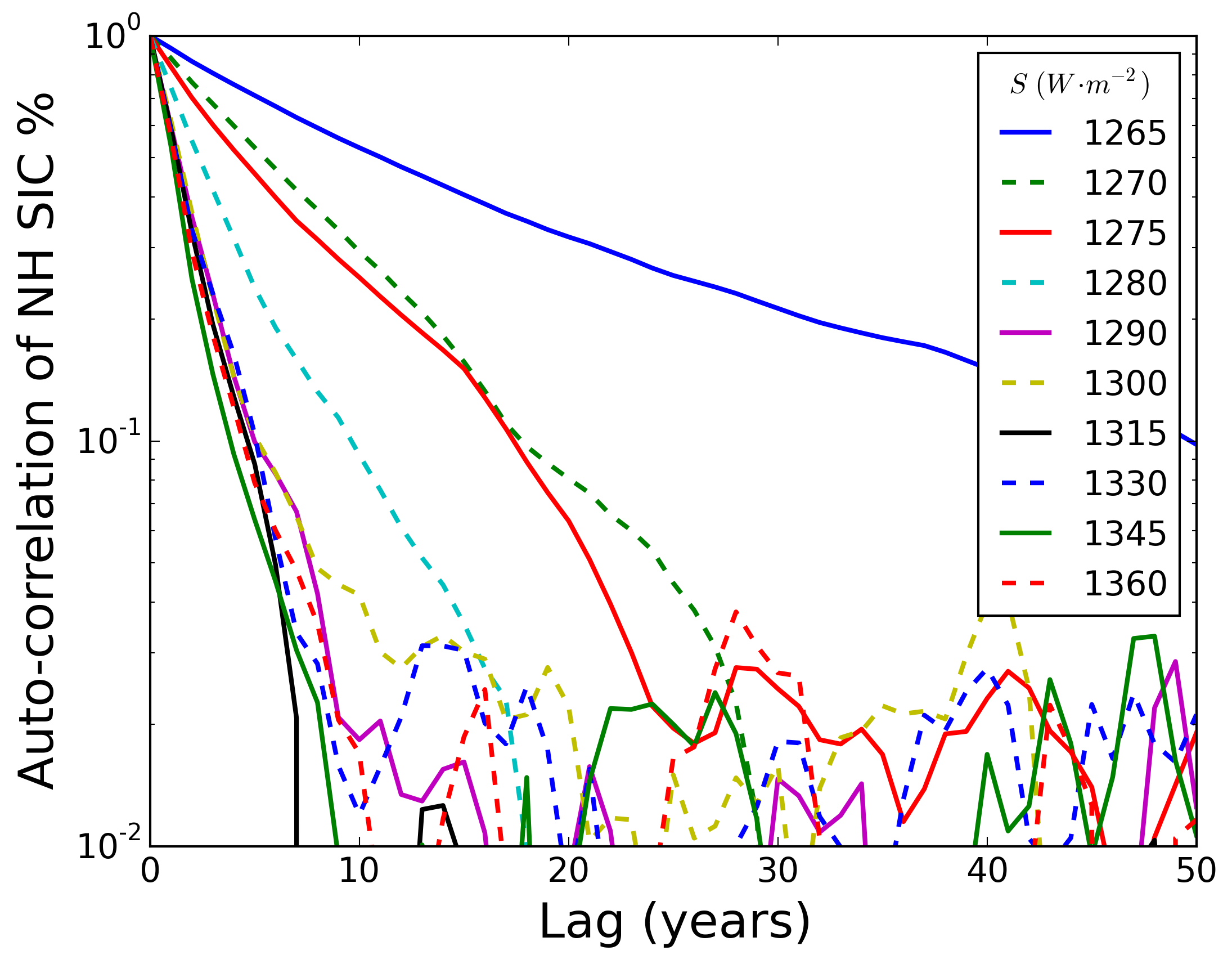}
	\end{subfigure}\\
	\begin{subfigure}[b]{7.5cm}
	(b)\\
		\includegraphics[width=\textwidth]{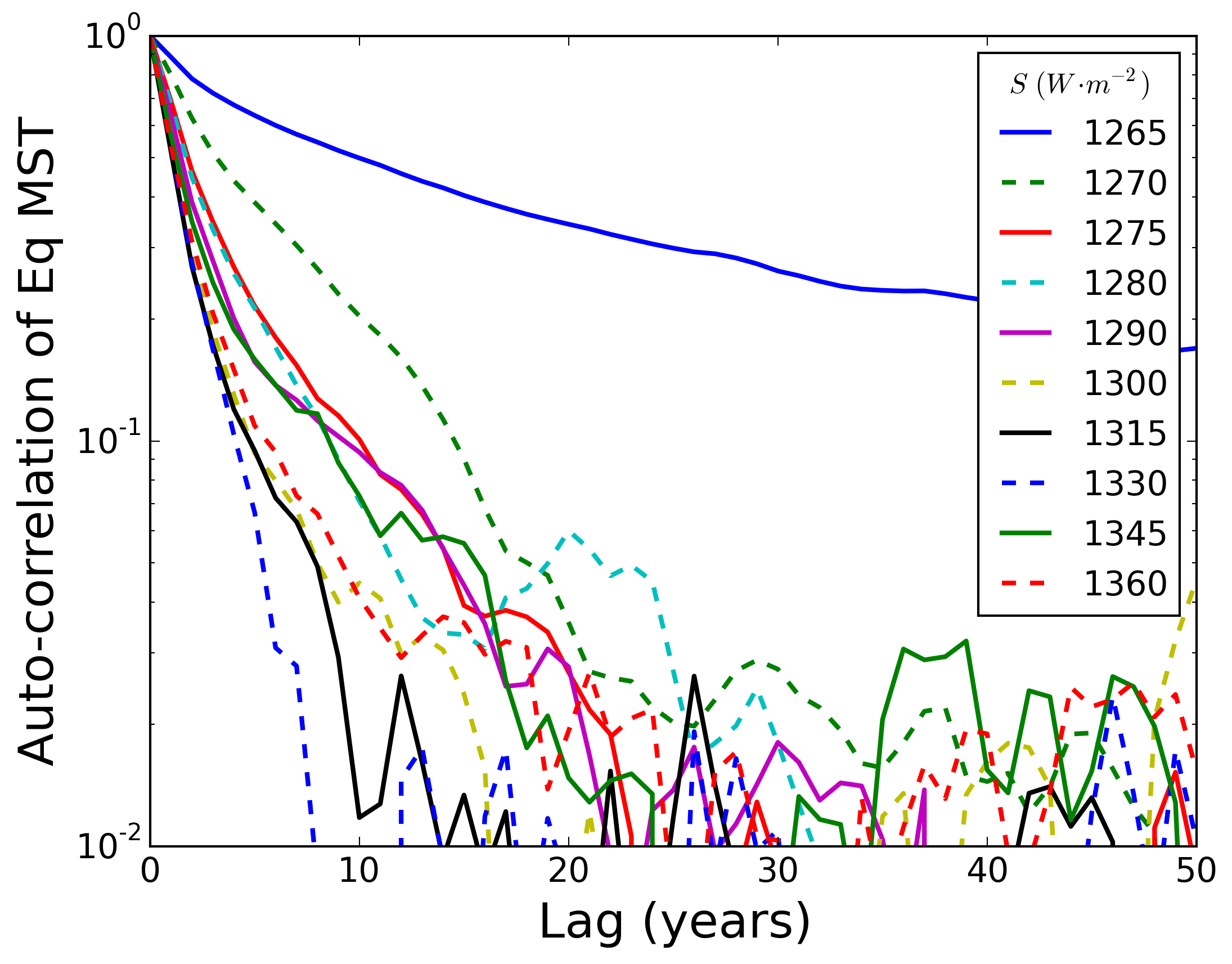}
	\end{subfigure}\\
	\begin{subfigure}[b]{7.5cm}
	(c)\\
		\includegraphics[width=\textwidth]{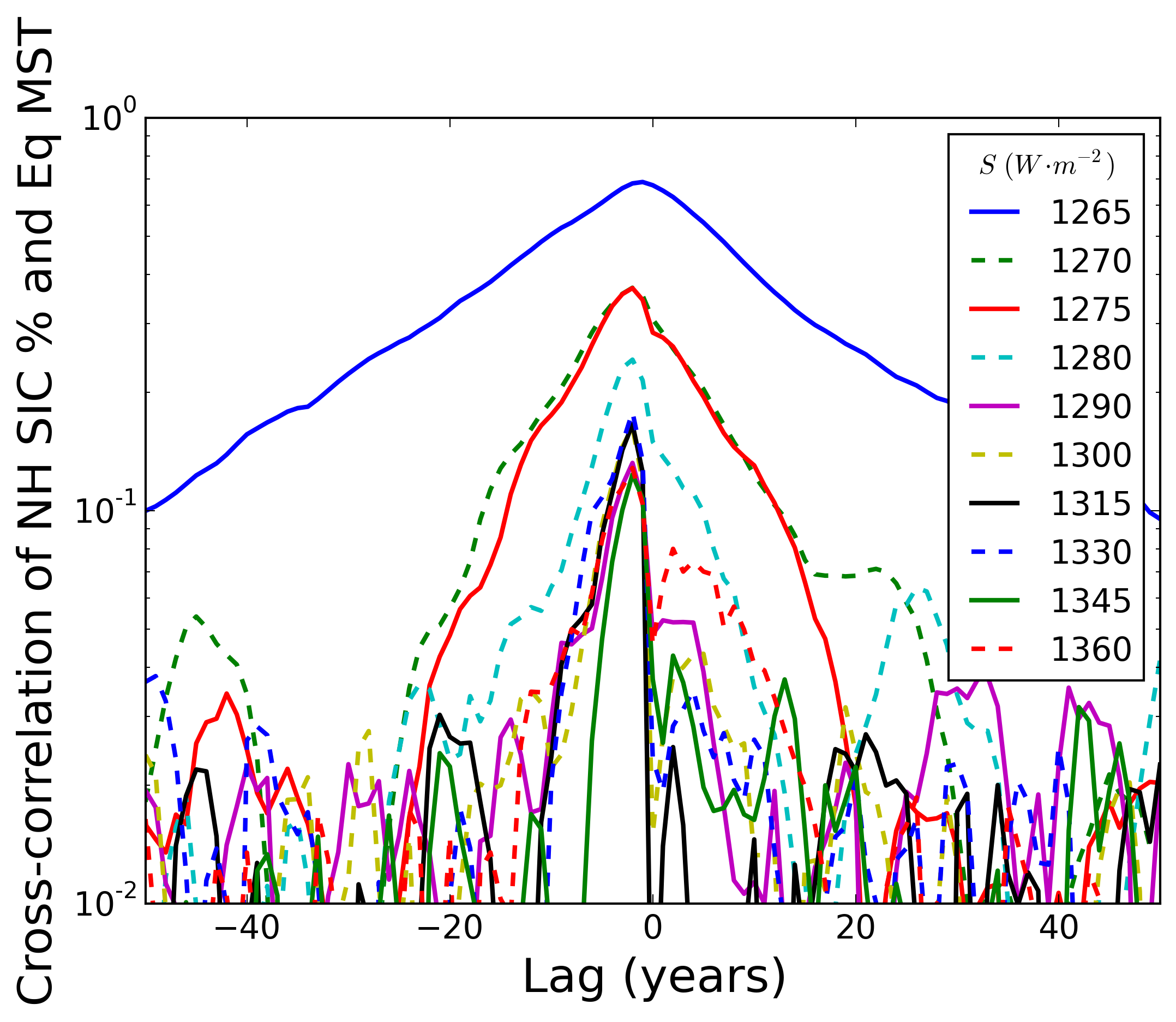}
	\end{subfigure}
	\caption{(a) ACF of the NH SIC, (b) ACF of the Eq MST and (c) CCF (in absolute value) of the NH SIC and Eq MST,
		for different values of the solar constant; note that the ordinate has a logarithmic scale.}
	\label{fig:ACF}
\end{figure}

Section \ref{sec:moments} revealed the changes in the moments of the steady-state along the warm 
branch, which could be linked to the physical mechanism of ice-albedo feedback. We now show that, 
consistent   with the increase in variance and skewness of the NH SIC and Eq MST,
the decay of their correlation functions also slows down.
For this purpose, the sample Auto-Correlation Function (ACF) and Cross-Correlation 
Function (CCF) of these observables are calculated from the simulations presented in section 
\ref{sec:simulations}, according to (\ref{eq:corrTime}).
The results are shown in figure \ref{fig:ACF} with a logarithmic scale for the $y$-axis
to check for an eventual exponential decay of correlations.
Note that, 
because the NH SIC and Eq MST are mostly anti-correlated, the absolute value of the CCF between 
the NH SIC and the Eq MST is plotted in figure \ref{fig:ACF}(c).
%

Let us first note that the fact that the correlation functions in figure \ref{fig:ACF}.a-c
show different decay rates depending on the choice of the observables is indicating that these observables 
project differently on the eigenvectors of the transfer operator.
For example, the fact that the auto-correlation functions for the NH SIC (fig. \ref{fig:ACF}.a) decay more slowly that the auto-correlation functions for the Eq MST (fig. \ref{fig:ACF}.b) suggests that the NH SIC projects
more strongly on the eigenvectors associated with unstable resonances
close to the imaginary axis than the Eq MST.
Furthermore, no oscillatory behaviour  is found  in the 
correlation functions, indicative that the leading reduced unstable resonances must be real. 
Note also the strong correlations between the NH SIC and the Eq MST figure \ref{fig:ACF}(c), in agreement with the PDFs of figure \ref{fig:PDFJOINT}.

Most importantly, the correlation functions 
decay more and more slowly as the solar constant is decreased towards its critical value,
although not in a regular fashion.
Slowing down of the decay of correlations is thus apparent before the boundary crisis 
in this model, even when the unperturbed evolution of the system alone is observed (i.e.~from 
time series converged to the attractor). This result can be explained in physical terms by the 
fact that, as the criticality is approached, an initial ensemble may experience more and more
persistent fluctuations associated with the strengthened ice-albedo feedback and thus
take a longer time to converge to the statistical steady-state.
 From a dynamical point of view, this effect is associated to the nearby presence of the edge state \cite{Lucarini2017}. From a macroscopic physical point of view, one can see the slowing down of the decay of correlations as resulting from the decreased efficiency of the large scale macroscopic thermodynamic fluxes in damping anomalies across the physical extent of the system \cite{Lucarini2014b,Lucarini2014a}.

\subsection{Changes in the eigenvalues of the reduced transfer operator}
\label{sec:resultSpectrum}

The theory presented in section \ref{sec:methodology} suggests that the slowing down
\emph{of the decay of correlations} observed before the attractor crisis can be explained by the approach of 
some \emph{unstable resonances} to the imaginary axis.
To test this theory here, we apply the numerical 
method presented in section \ref{sec:approxSpectrum} and estimate transition matrices from the 
time series of the NH SIC and the Eq MST, observables selected
for being most sensitive to the slowing down of the decay of correlations (see remark~\ref{rmk:choiceObs})
and for being most relevant in terms of ice-albedo feedback (see section \ref{sec:moments}).

Note first that the theory presented in section \ref{sec:methodology}
has been worked out for autonomous maps.
In the application considered here, the time series have been sub-sampled to one year time steps,
by taking yearly averages.
The transfer operators considered in this application are thus induced by
the time-one-year map corresponding to these yearly averaged time series.

Several observation operators $h$ will be considered.
Let us first take a one dimensional reduced state space with $h$ having only one component,
either defined as the NH SIC or as the Eq MST alone. 
A grid of $50$ boxes spanning the interval $[-5 \sigma, 5\sigma]$, where $\sigma$ is the standard deviations 
of each observable, is taken.
The width of the domain is chosen so as to avoid boundary effects which 
tend to result in a spectrum spuriously far from the imaginary axis.
The choice of the number of boxes is a trade-off between the resolution of short spatial and temporal scales and the quality in the estimates of the transition probabilities.
A lag of one year was chosen  (i.e.~at the sampling frequency of the time series).
It is shown in \ref{sec:robust} that 
our results are fairly robust to changes in the grid, the lag and to the sampling.
This does not mean that the true unstable resonances are being approximated
but rather that the eigenvalues of the reduced transfer operators have converged
and that memory effects in the reduced space are weak.

\begin{figure}[h!]
	\centering
        \begin{subfigure}[b]{7cm}
        (a)\\
		\includegraphics[width=\textwidth]{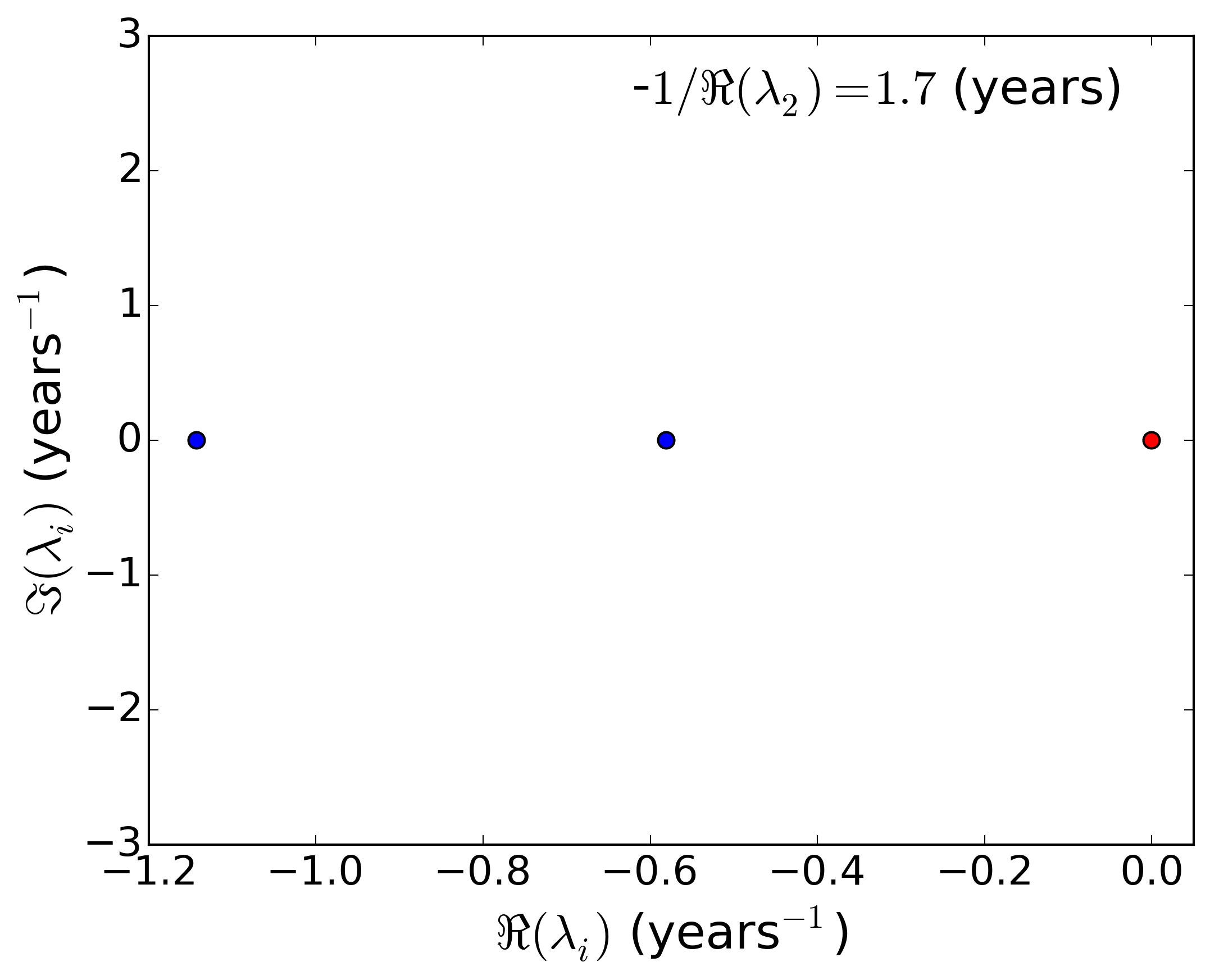}
	\end{subfigure}
        \begin{subfigure}[b]{7cm}
        (b)\\
		\includegraphics[width=\textwidth]{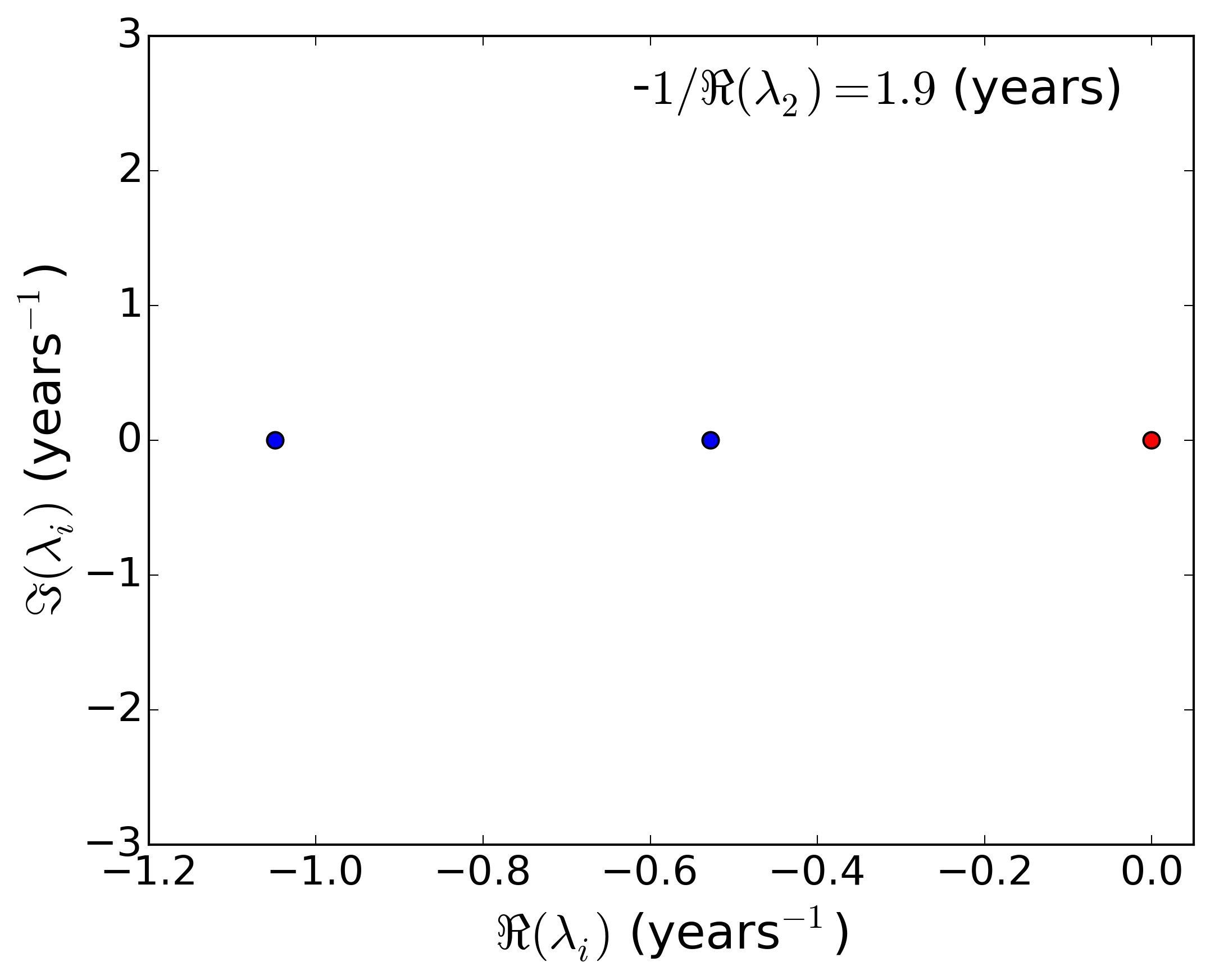}
	\end{subfigure}
        \begin{subfigure}[b]{7cm}
        (c)\\
		\includegraphics[width=\textwidth]{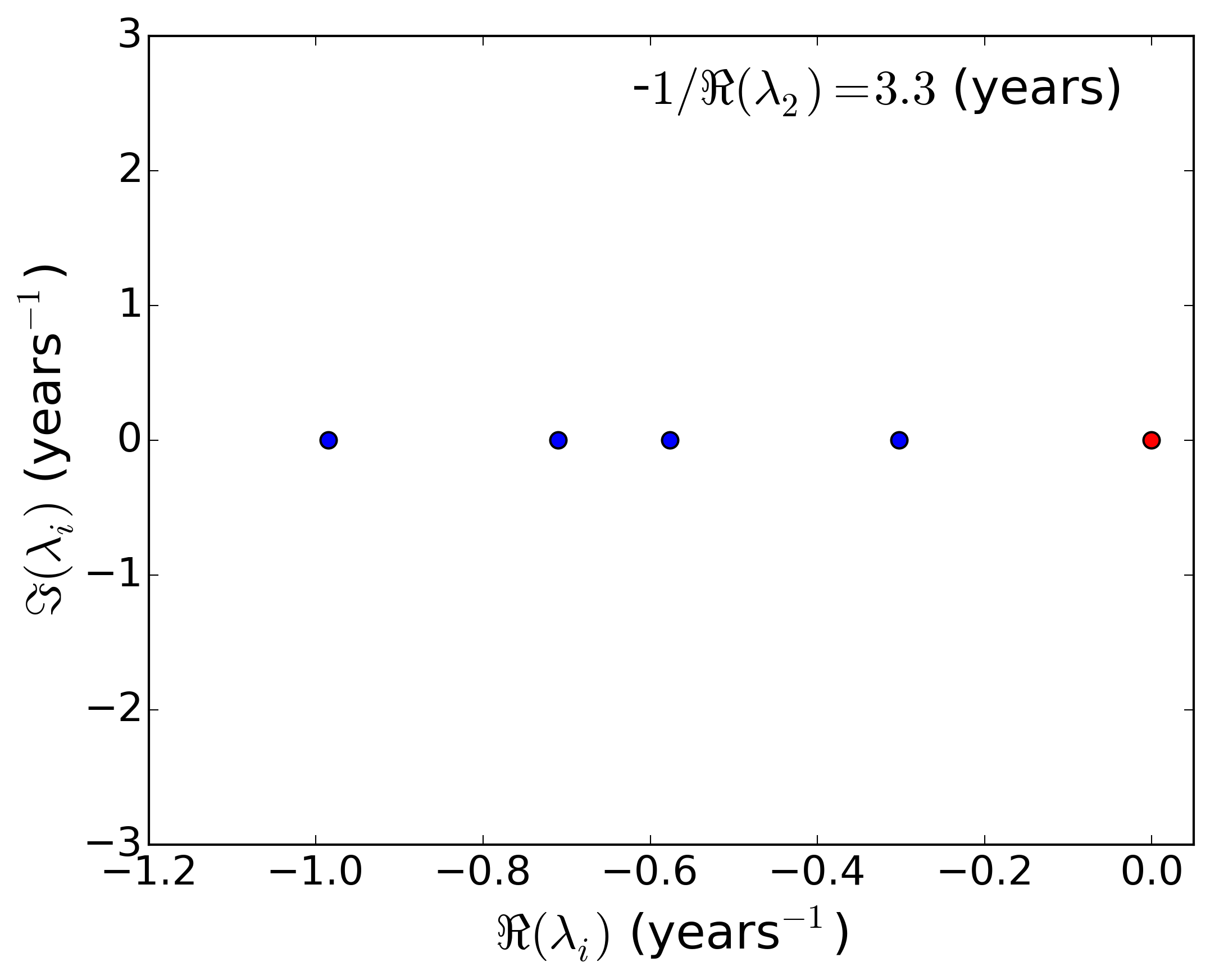}
	\end{subfigure}
        \begin{subfigure}[b]{7cm}
        (d)\\
		\includegraphics[width=\textwidth]{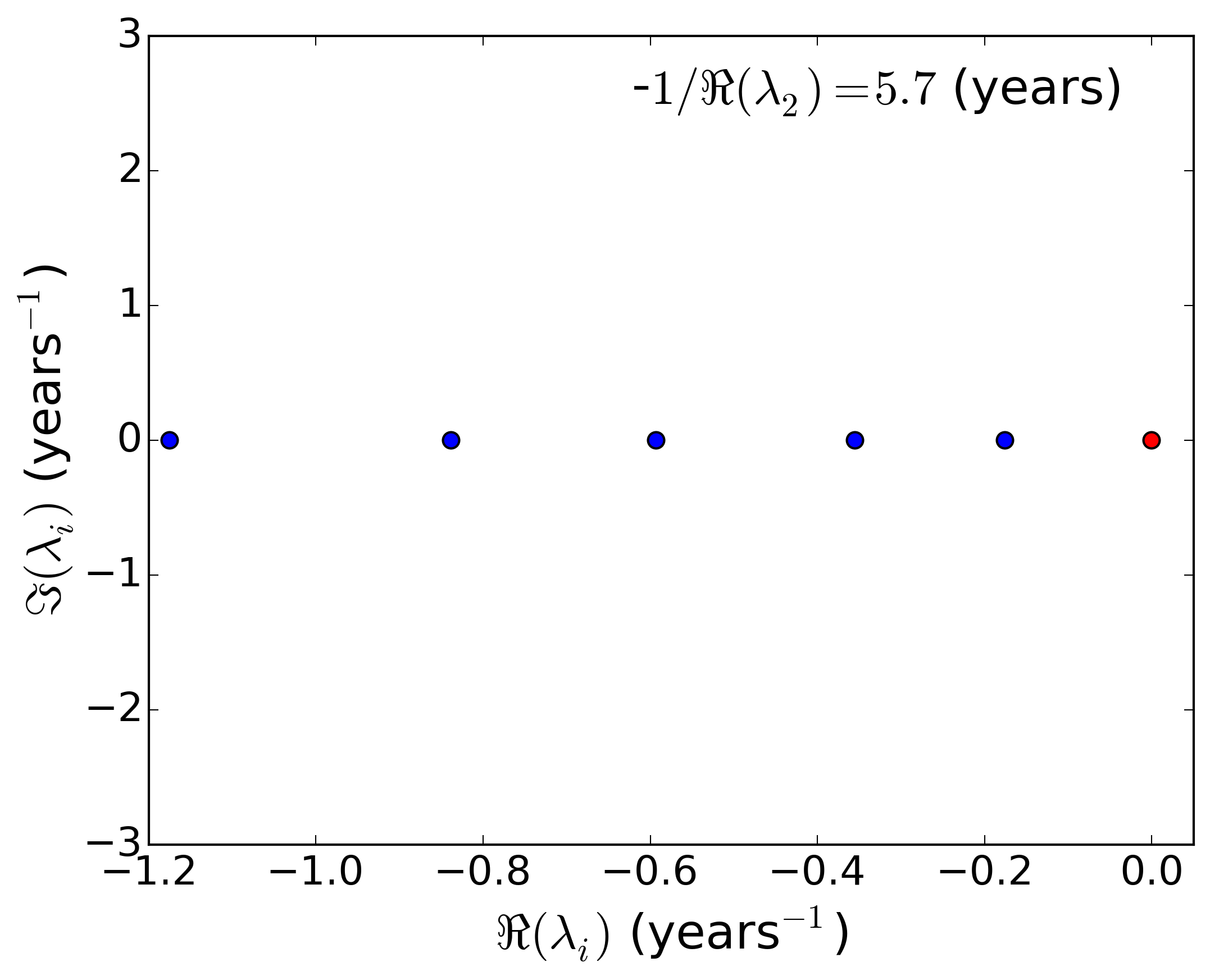}
	\end{subfigure}
        \begin{subfigure}[b]{7cm}
        (e)\\
		\includegraphics[width=\textwidth]{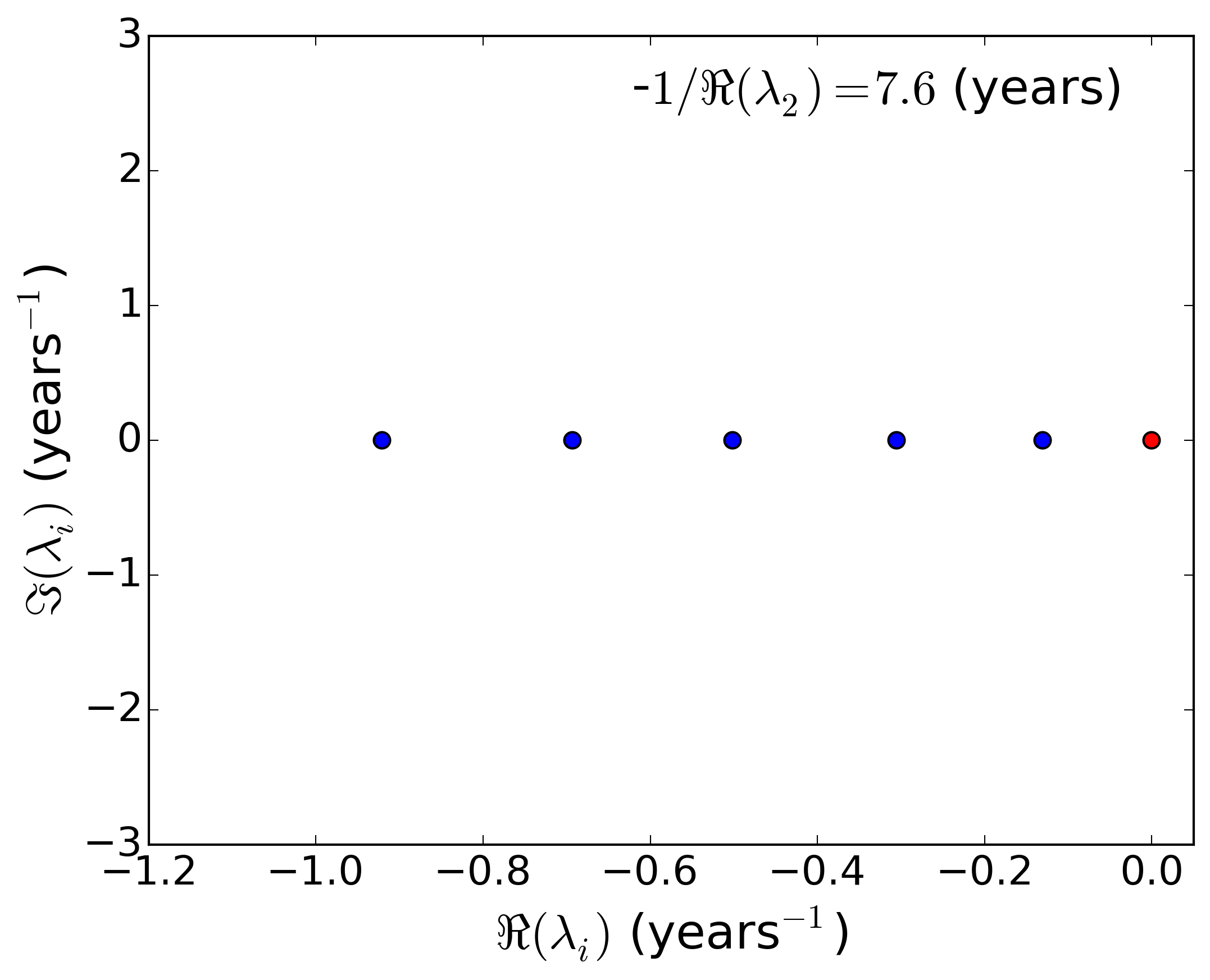}
	\end{subfigure}
        \begin{subfigure}[b]{7cm}
        (f)\\
		\includegraphics[width=\textwidth]{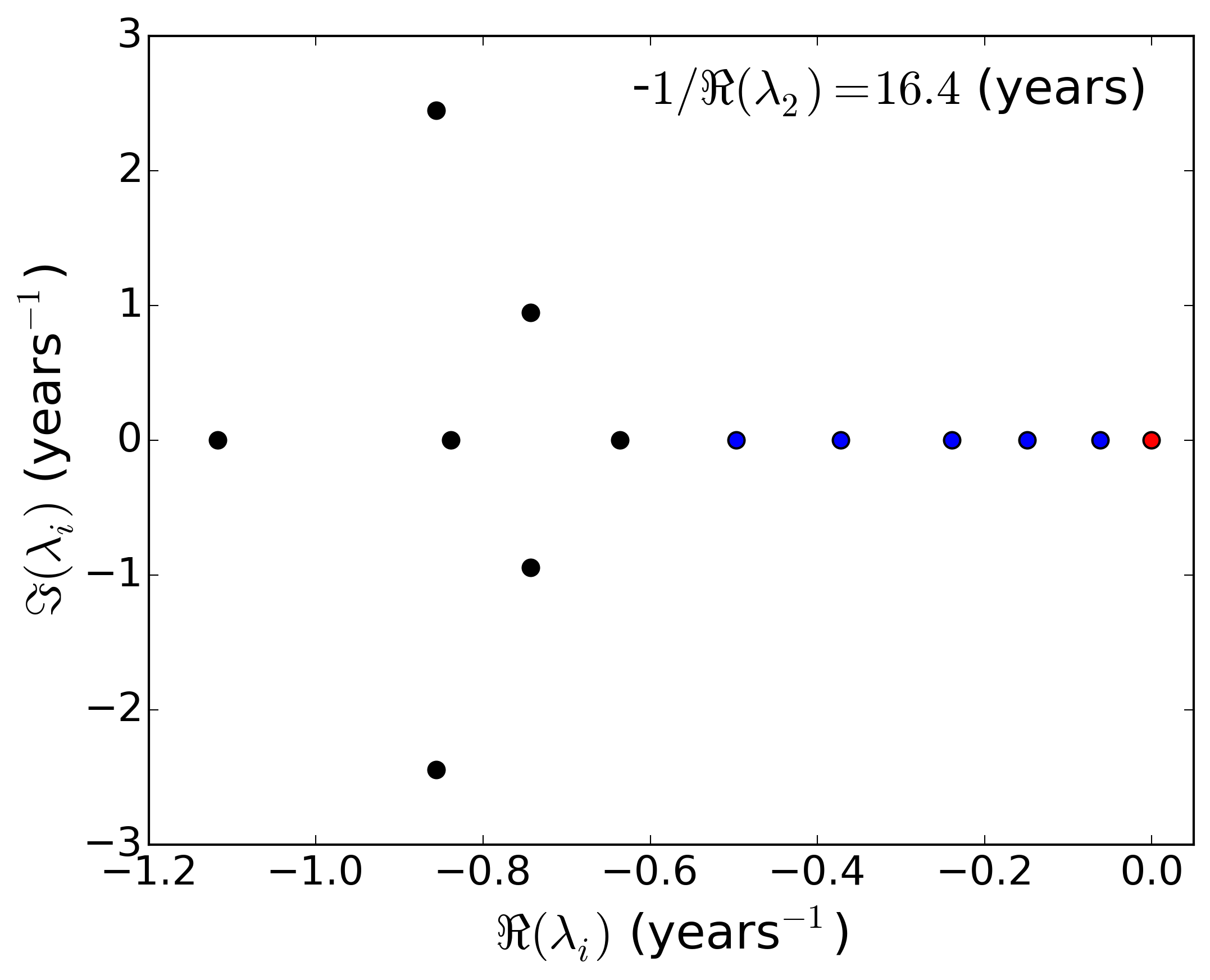}
	\end{subfigure}
	\caption{Rates (\ref{eq:reducedEigenvalues}) corresponding to the reduced unstable resonances estimated from the NH SIC for a solar constant of (a) 1360, (b) 1290, (c) 1280, (d) 1275, (e) 1270 and (f) $1265~W m^{-2}$.
	The first rate, which is zero, is represented in red, 
	while the first 5 secondary rates are represented in blue.
	The relaxation time associated with the leading secondary rate
	and given by minus the inverse of the real part
	of the rate is also given in the upper-right corner of the panels.}
	\label{fig:polesW2SBSICNH}
\end{figure}

The leading reduced unstable resonances,
calculated from the eigenvalues of the transition matrices,
are estimated for different values of the solar constant and the corresponding rates
calculated from (\ref{eq:reducedEigenvalues}) are represented in figure \ref{fig:polesW2SBSICNH}
for the NH SIC.
The rate of the first reduced unstable resonance, represented in red,
is always 0 and is associated with the unit vector,
which is by construction a fixed point for the transition matrices.
%
%
Next, the rates of the first five secondary eigenvalues are represented in blue to help following their evolution.
In addition, the time scale given by minus the inverse of the real part of the rate of the leading reduced
unstable resonance is also given in the upper-right corner of figure \ref{fig:polesW2SBSICNH}. 

The main result here is that, in agreement with the slower decay of correlations of the NH SIC observed in figure \ref{fig:ACF}(a), the leading (secondary) reduced unstable resonances get closer and closer
to the unit circle, as the solar constant nears its critical value. Furthermore, that the leading eigenvalues have vanishing imaginary part agrees with the fact that no oscillations are found in the correlation functions represented figure \ref{fig:ACF}(a).

\begin{figure}[h!]
	\centering
        \begin{subfigure}[b]{7.5cm}
        (a)\\
		\includegraphics[width=\textwidth]{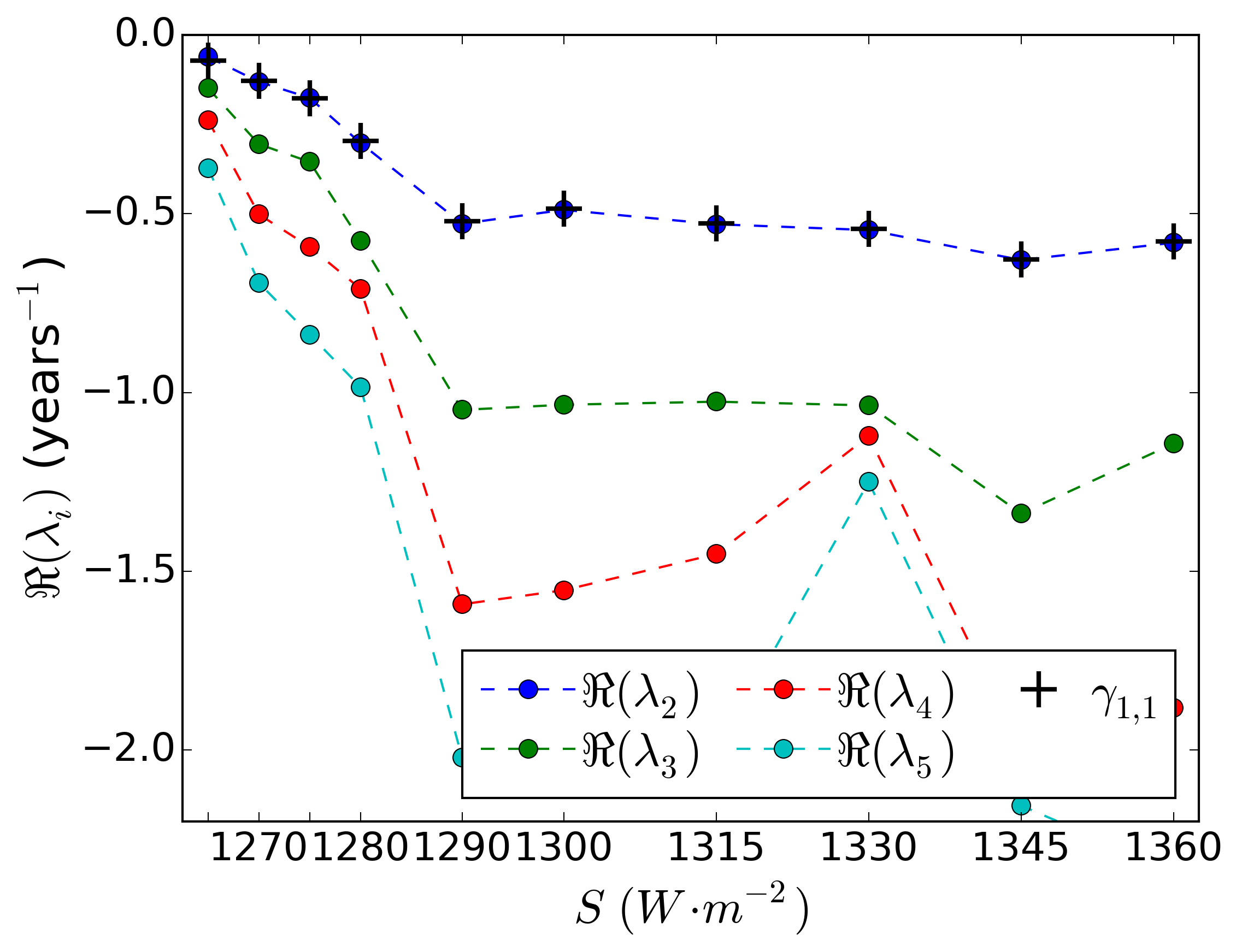}
	\end{subfigure}\\
        	\begin{subfigure}[b]{7.5cm}
 	(b)\\
		\includegraphics[width=\textwidth]{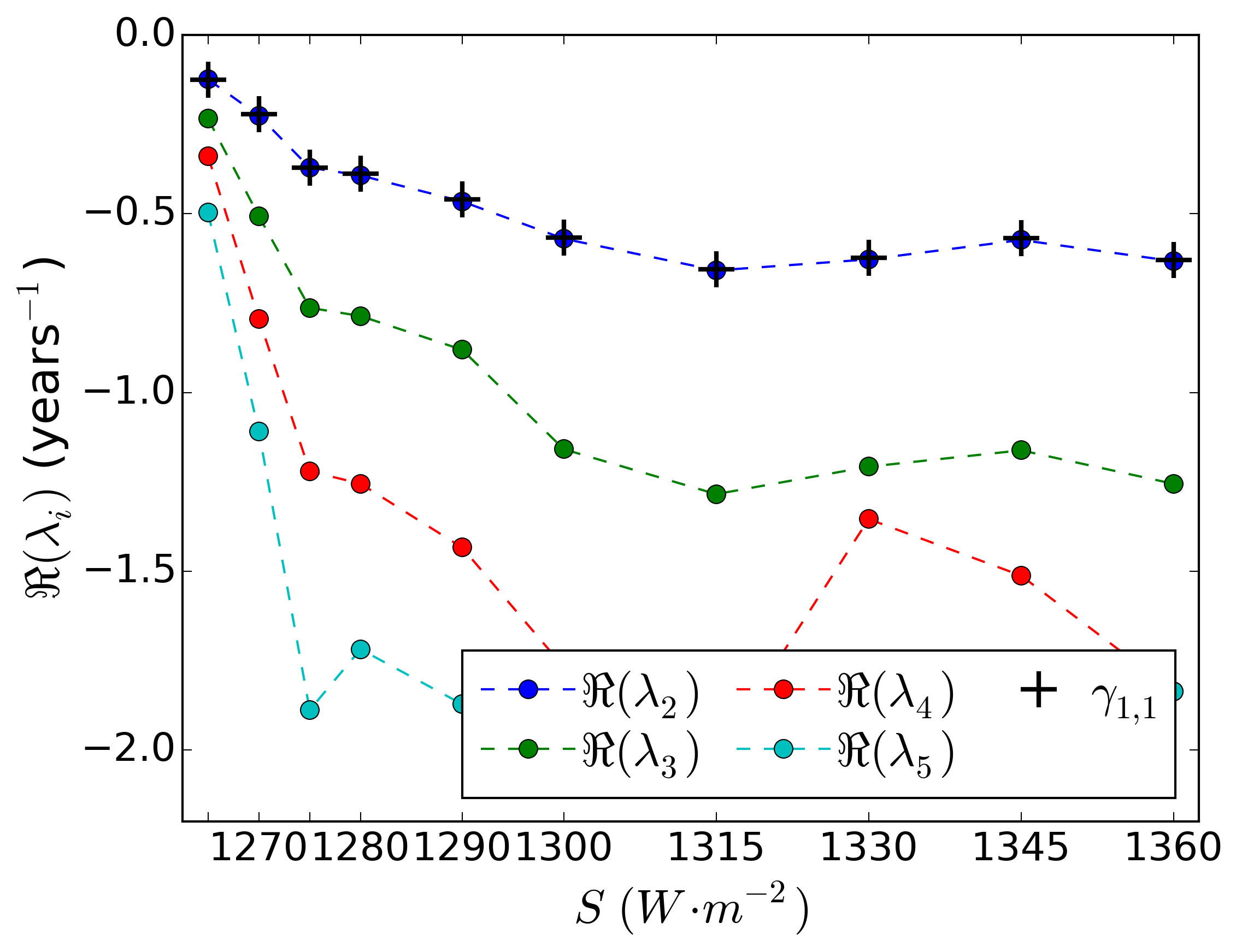}
	\end{subfigure}\\
        	\begin{subfigure}[b]{7.5cm}
 	(c)\\
		\includegraphics[width=\textwidth]{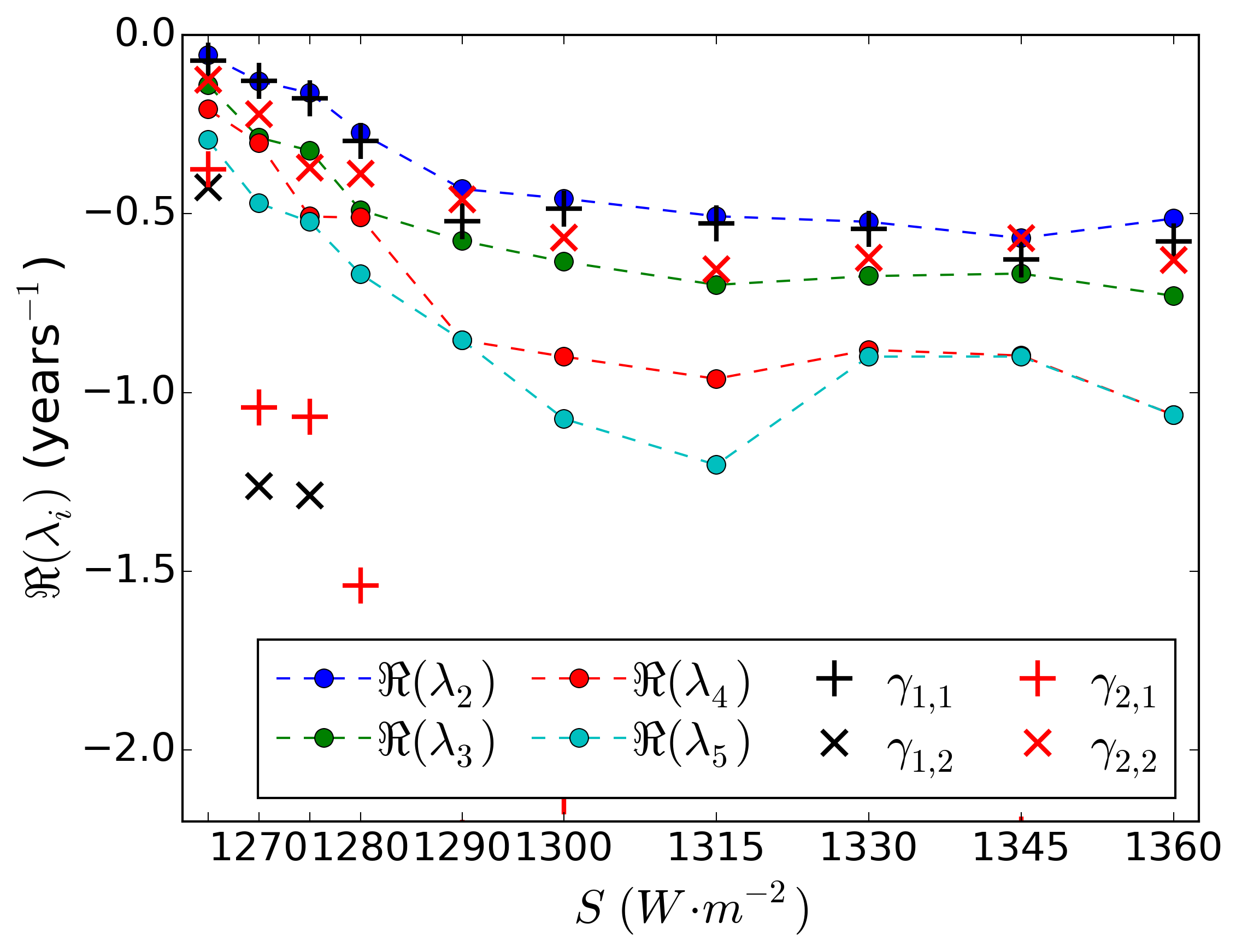}
	\end{subfigure}
	\caption{Real part of the rates for the four four leading reduced unstable resonances for (a) the NH SIC and (b) the Eq MST and (c) the (NH SIC, Eq MST), versus the solar constant.
	The sample decorrelation rates $\gamma_{i, j}$ between the observables are also represented as black and red crosses.}
	\label{fig:ratesW2SB}
\end{figure}

For a more detailed analysis, the real part of the rates (\ref{eq:reducedEigenvalues}) of the four leading reduced unstable resonances for the NH SIC and for the Eq MST are plotted in figure \ref{fig:ratesW2SB}(a) and (b),  respectively. Furthermore, the corresponding plot for the transition matrices estimated using the  two dimensional space composed of both the NH SIC and the Eq MST is also shown  in figure \ref{fig:ratesW2SB}(c). For this two dimensional observation operator $h$, a coarser grid of 25$\times$25 boxes was used, because of the higher dimension of the reduced state space compared to the one dimensional case.
To these plots are added the sample decorrelation rates $\gamma_{i, j}$ of the ACFs and CCFs
to give an indication of what would be the real part of the unstable resonance
if only one of them would dominate.
They are defined here as
$\gamma_{i, j} = \log(|C_{f_i, f_j}(1)|)$, where $C_{f_i, f_j}(1)$ is the sample ACF or CCF
with observable $f_i$ leading observable $f_j$ by one year.
Here the lag of one year is chosen to be the same as  for the estimation of the transition matrices.
For example, $\gamma_{1, 2}$ in figure \ref{fig:ratesW2SB}(c) shows the sample decorrelation rate with the NH SIC leading the Eq MST by one year.

The plots in figure \ref{fig:ratesW2SB} confirm the approach of the unit circle
by the leading reduced unstable resonances for the NH SIC, the Eq MST and the (NH SIC, Eq MST), as the solar constant is decreased to its critical value.
In particular, the time scale associated with the second eigenvalue for the (NH SIC, Eq MST) transition matrices increases from about 2 years to more than 17 years right before the crisis, yielding a quantitative measure of the slowing down of the decay of correlations associated with the loss of stability of the attractor.
Moreover, for $S < 1290~W  m^{-2}$, the increase of the real parts is almost linear and is reminiscent of the results found for bifurcations of low-dimensional systems  \cite{Gaspard1995, Gaspard2001a}. Overall, these changes are in agreement with the slower decay of the ACFs and CCFs shown  in figure \ref{fig:ACF}. In particular, one can see that the maximum sample decorrelation rates coincide very well with the real part of the second leading rate.
Different observables (e.g~highly nonlinear ones) acting on the reduced space may however project differently on the eigenvectors
of the transition matrices, so that different reduced unstable resonances may dominate their decay of correlations.

%
\begin{rmk}
	Let us note that, while showing results only for the NH SIC and the Eq MST for definiteness,
	other observables exhibit a slowing down of the decay of correlations at the approach of the crisis.
	This is not true, however, for every observables as the temporal fluctuations of some
	are not affected by the crisis or are experience too fast a decay of correlations
	to be accessible by the reduction method and the sampling strategy used here.  
\end{rmk}
\begin{rmk}
Note that some reduced unstable resonances for some other values of the solar
constant than those presented here have been found to be associated with an even slower decay of correlations
than for values of the solar constant closer to the critical point of $1265~W  m^{-2}$.
This may happen when meta-stability arises when the sea-ice cover
(giving the \emph{number} of grid boxes covered by sea ice) intermittently switches
between states with slightly different latitudinal extents.
Thus, the appearance of these eigenvalues is due to the coarse representation of this integer valued observable
rather than to the approach to some critical point.
\end{rmk}

To conclude, it is found that the slowing down of the decay of correlations
observed in section \ref{sec:slowing} is explained by
the approach of the reduced eigenvalues to the imaginary axis.
While one cannot prove that the actual unstable resonances approach
the imaginary axis before the boundary crisis in PlaSim,
the results show that important changes in the unstable resonances do occur before the crisis
which are responsible for the approach of the imaginary axis by the reduced eigenvalues.


\section{Summary and Discussion}
\label{sec:discussion}

Motivated by the question of whether critical slowing down of the decay of correrlations
can be observed in a high-dimensional chaotic system undergoing an attractor crisis
and whether such a slowing down of the decay of correlations can be explained in terms of spectrum
of operators governing the dynamics of statistics, we have taken as test bed
a high-dimensional chaotic climate model undergoing a change of attractor as a control parameter is varied.

Correlation functions estimated from long simulations of the model have revealed that a slowing down indeed occurs as the critical value of the control parameter is approached.
Importantly, the slowing down of the decay of correlations
could be seen from the evolution of observables in steady-state,
without having to perturb the system away from the attractor.
The second step of this study was to verify that the slowing down of the decay of correlations
can in this case be explained by the approach of unstable resonances of transfer operators to the unit circle
in the complex plane (i.e. the approach of the corresponding rates to the imaginary axis).
 
The behaviour of the resonances of transfer operators has recently been studied 
for a boundary crisis in the Lorenz flow \cite{Tantet2017}.
It was shown that only the stable resonances associated with contraction towards the attractor
were affected to the crisis, as opposed to the unstable resonances associated with the mixing
or decay of correlations of observables on the attractor.
This result is supported by the fact that these stable resonances 
may be used to characterise the global stability of invariant sets \cite{Mauroy2016}.

Moreover, it has recently been shown for a simplified climate 
model incorporating the dynamical core of PlaSim and a simplified representation 
of the oceanic  transport and of the ice-albedo feedback that the crisis occurring in PlaSim
is indeed a boundary crisis during which a strange saddle (the so-called "edge state")
collides with the attractor \cite{Lucarini2017}.
We are therefore led to think that the edge state exists also in the PlaSim 
configuration adopted here.

We are thus led to ask what mechanism is responsible for the fact that the unstable resonances
and the decay of correlations are affected by the boundary crisis in PlaSim, while this
cannot be expected to be the case in general.
This could be explained in physical terms by the fact that the primary mechanism of instability leading to the destruction of the attractor, the ice-albedo feedback, is not only activated by exogenous perturbations but also by spontaneous fluctuations along the attractor (i.e.~internal variability, in climatic jargon).
This aspect, related to the presence of positive Lyapunov exponents associated with unstable processes, makes the treatment of deterministic chaotic systems intrinsically more interesting than that of simpler systems featuring fixed points or periodic orbits as attractors.

A more mathematical explanation can be suggested in terms of stochastic model reduction.
While in the model considered here the sea ice and the ocean are fully coupled with the atmosphere,
there exists a relatively strong time-scale separation between the dominant variability in the atmosphere,
due to baroclinic instability, and the thermodynamic evolution of the sea-ice and of the ocean.
One is then tempted to take advantage of this time-scale separation to apply homogenisation methods (see \cite{Pavliotis2008} and references therein, in particular \cite{Just2001})
to derive a reduced model of the sea-ice and the ocean where the dynamics
of the atmosphere are, loosely speaking, replaced by noise.
In this reduced model, the stochastic forcing representing the atmosphere could be considered
to drive the ocean-sea-ice system away from its unperturbed statistical steady-state
and thus to excite the stable resonances of this ocean-sea-ice system.
In this case, it has rigorously been shown in \cite{Crommelin2011} that the eigenvalues of the
reduced transfer operator actually converge to the leading unstable resonances,
in the limit of infinite time-scale separation.
This explanation is supported by the fact that the projection of the transfer operator
on the space spanned by the NH SIC and the Eq MST is robust to changes in the lag (see \ref{sec:robust}).
While in the presence of a  strong-time separation, homogenisation methods allow to
give an approximation of the high-dimensional system, it is not yet clear, however,
whether such approximation is able to reflect global stability properties of the original system.

In the case when memory effects are strong in the state space,
the family of reduced transfer operators (for varying number of iterations of the original operator)
cannot be considered as a semigroup
and several operators at different times need to be considered (section \ref{sec:approxSpectrum}).
The choice of the observation operator defining the reduced space is not straightforward
and should eventually be guided by the behaviour of correlation functions
(looking for observables with the slowest decay rate)
and by an understanding of the fundamental physical mechanisms responsible for the slow dynamics.
Approximating this spectrum for systems with competing time-scales can reveal much richer dynamics than by the estimation of the decorrelation rates alone \cite{Tantet2015, Tantet2016}.

Let us now put these results in perspective with Ruelle's response theory  \cite{Ruelle2009}
which has recently been extended to deal with the response of correlations to perturbations \cite{Lucarini2017c}.
The approach of the resonances, both stable and unstable, to the unit circle
is an indicator that the radius of  convergence of the series expansion of the perturbed invariant measure shrinks  
\cite{Kato1995, Lucarini2016, Chekroun2014}.
In other words, the size of the perturbation for which response theory applies \cite{Ruelle2009}
vanishes at the criticality, accordingly to the fact that 
a dramatic change in the statistics occurs during the attractor crisis.
Assessing the sensitivity to perturbation of the invariant measure
is particularly relevant to the problem of the climate response to anthropogenic forcing \cite{Ghil2015, Ragone2016a}.
In particular, the current efforts to calculate the linear concept of \emph{climate sensitivity}
\cite{Knutti2008, Roe2009, Collins2013} may be of little help if higher order nonlinear terms are important \cite{Zaliapin2010, Rugenstein2016b} or if this response has a strong dependence on the steady-state itself \cite{VonderHeydt2014, Kohler2015, Knutti2015}.

Finally, more than just to show that slowing down of the decay of correlations can be 
observed from long time series of a high-dimensional chaotic system
(as anticipated by \cite{Boulton2014}, also in a geophysical context),
this study gathers existing results from the ergodic theory of dynamical systems
to give the theoretical foundations necessary to
better understand the behaviour of statistics near a chaotic attractor crisis
and in particular the potential of early warning indicators such as the lag-1 autocorrelation \cite{Held2004a}
or more advanced ones, able to capture non-Markovian effects \cite{Faranda2014d}.
For example, PlaSim provides an example of a chaotic system undergoing a boundary crisis
for which the correlation functions can provide an indicator of the crisis.
However, such an indicator would not work for any boundary crisis \cite{Tantet2017},
since the unstable resonances are not always affected by the crisis.
Even in the former case, the rather smooth changes in the reduced unstable resonances
observed in section \ref{sec:resultSpectrum} before the crisis (and, a fortiori, in the decorrelation rates)
suggest that autocorrelation-based indicators may not give a strong signal at the approach of an attractor crisis.
This is in agreement with the analytical results obtained by \cite{Gaspard1995, Gaspard2001a, Tantet2016}  for  low-dimensional systems.
Moreover, invoking the chaotic hypothesis \cite{Gallavotti1995}, such numerical results could be 
explained by the differentiability of the unstable resonances of Anosov systems, proved in \cite{Gouezel2006}.
We do not know, however, if such stability of the mixing spectrum is common in high-dimensional systems
and possibly nonuniformly hyperbolic systems such as found in geophysical fluid dynamics.

This analysis thus shows, that even though the methodology presented section \ref{sec:approxSpectrum} is too demanding in terms of data to apply it to observations and use the spectral gap as an early-warning indicator of a crisis, it is very well suited to understand and design such indicators in the more general context of high-dimensional chaotic (and also stochastic) systems than the one of auto-regressive processes for which the lag-1 auto-correlation indicator was originally developed in \cite{Held2004a}.
However, whether the numerical method presented in section \ref{sec:approxSpectrum} is amenable
to state-of-the-art climate models with thiner grids and for which obtaining long time series is computationally costly
can be questioned. The answer to this question will, however, depend on the strength of the resonances associated with the physical processes of interest.

\ack
The authors are grateful to the two reviewers for giving detailed comments,
which allowed to greatly improve this contribution.
AT is particularly thankful to Micka\"el D. Chekroun for the stimulating discussions
on the theory and applications of ergodic theory.
AT and HD would like to acknowledge the support of the LINC project (no. 289447)  funded by EC's 
Marie-Curie ITN program (FP7-PEOPLE-2011-ITN).
VL acknowledges fundings from the Cluster of Excellence for Integrated Climate Science (CLISAP) from the European Research Council under the European Commission's Seventh Framework Programme (FP7/2007-2013)/ERC Grant agreement No. 257106 Starting Investigator Grant NAMASTE - Thermodynamics of the Climate System and from the DFG Project MERCI.

\appendix
\section{Robustness of the numerical estimates of the spectrum of the reduced transition matrices}
\label{sec:robust}

In this section, we test the robustness of the eigenvalues of the reduced transition matrices represented figure \ref{fig:ratesW2SB} to the sampling length, the grid size and the lag. For convenience, all the tests are presented for the NH SIC only. However, the corresponding tests done for the Eq MST and (NH SIC, Eq MST) do not challenge the conclusions taken in this study. Unless specified, all the parameters used in the following to estimate the transition matrices, such as the length of the times series, the grid, or the lag, are taken the same as for the ones used section \ref{sec:resultSpectrum} to produce figure \ref{fig:ratesW2SB}.

\subsection{Robustness to the sampling length}

\begin{figure}[h!]
	\centering
        \begin{subfigure}[b]{7.5cm}
        (a)\\
		\includegraphics[width=\textwidth]{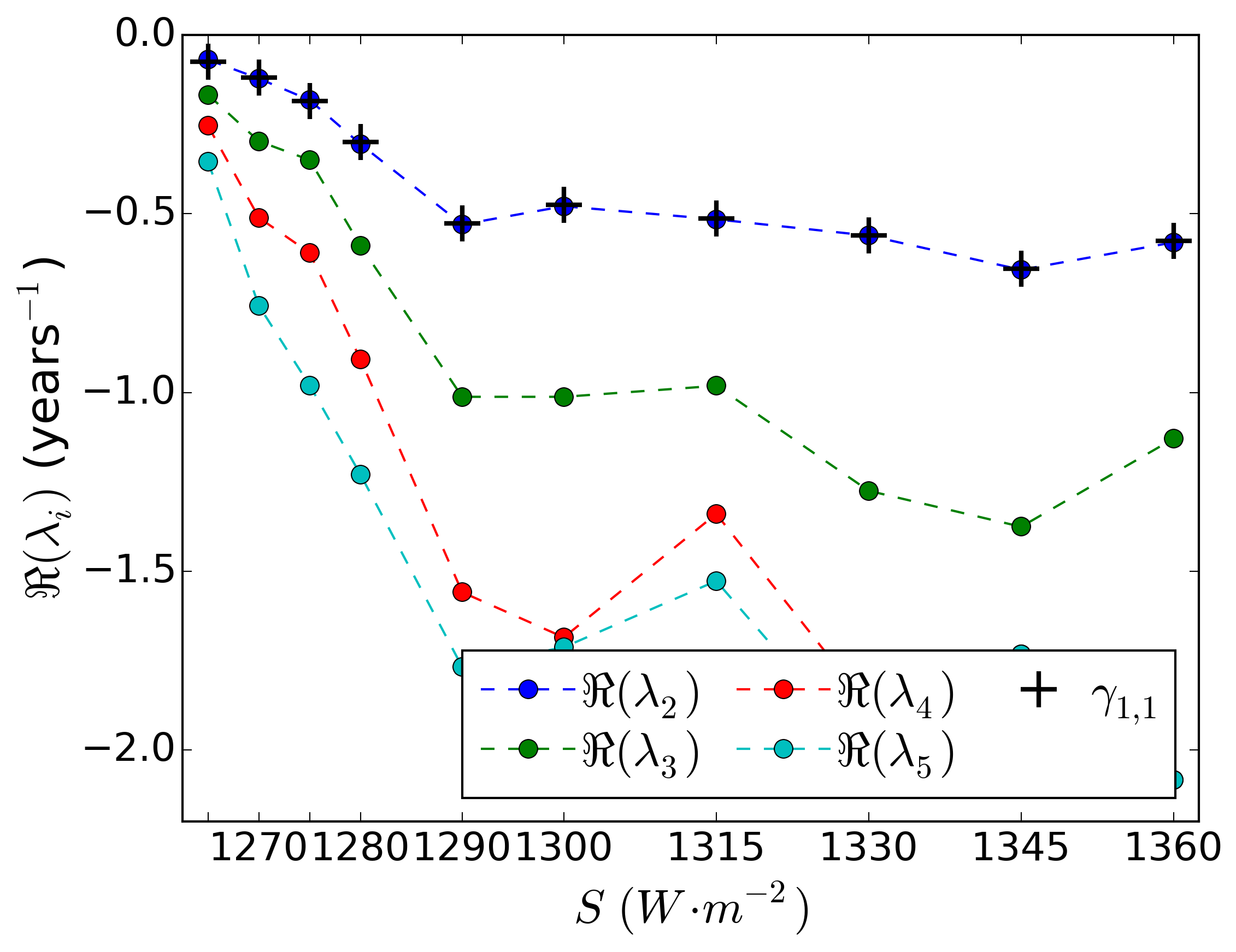}
	\end{subfigure}
        	\begin{subfigure}[b]{7.5cm}
 	(b)\\
		\includegraphics[width=\textwidth]{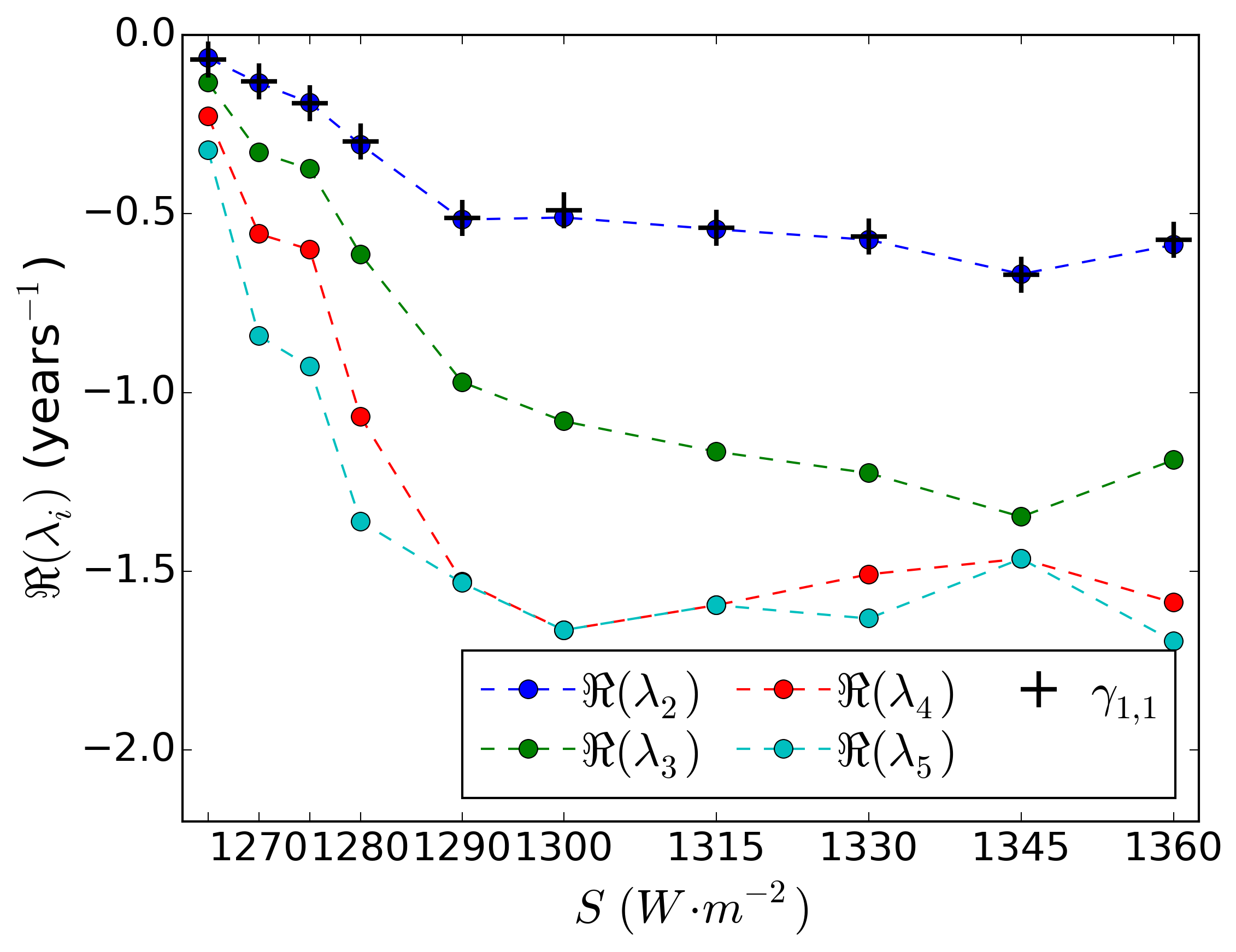}
	\end{subfigure}
	\caption{Real part of rates (\ref{eq:reducedEigenvalues}) for the four leading reduced unstable resonances for the NH SIC, versus the solar constant, using only (a) half and (b) a quarter of the time series used to produce the corresponding figure \ref{fig:ratesW2SB}.}
	\label{fig:robustLength}
\end{figure}

To test the robustness of the approximated spectra to the length of the time series used to estimate the transition probabilities, one could calculate confidence intervals from a version of the non-parametric bootstrap \cite{Efron1982a, Mudelsee2010}, adapted to the estimation of transition matrices \cite{Craig2002a} as was done in \cite{Chekroun2014, Tantet2015}. Instead, we simply look at the changes in the leading eigenvalues of the transition matrices when only the first half ($4850$ years) or quarter ($2425$ years) of the time series is used. The thus obtained real parts of the rates of the reduced unstable resonances calculated for the NH SIC, with a lag of 1 year and on a grid of 50 boxes, are represented figure \ref{fig:robustLength}. Furthermore, the inverse of the real part of the rate of the second eigenvalue is catalogued in the third and fourth column of table \ref{tab:robust}.

It is sufficient for this analysis to see, by comparing figure \ref{fig:ratesW2SB} to figure \ref{fig:robustLength} and from table \ref{tab:robust}, that using time series of NH SIC twice or even four times as short as the original one to estimate the transition probabilities results in errors of less than $10\%$ in the inverse of the real part of the eigenvalues. This shows the convergence of the estimates with the length of the time series, and that the qualitative picture, of the approach of the eigenvalues to the imaginary axis as the attractor crisis is neared, is not affected by the sampling length.

\subsection{Robustness to the grid size}

\begin{figure}[h!]
	\centering
        \begin{subfigure}[b]{5.1cm}
        (a)\\
		\includegraphics[width=\textwidth]{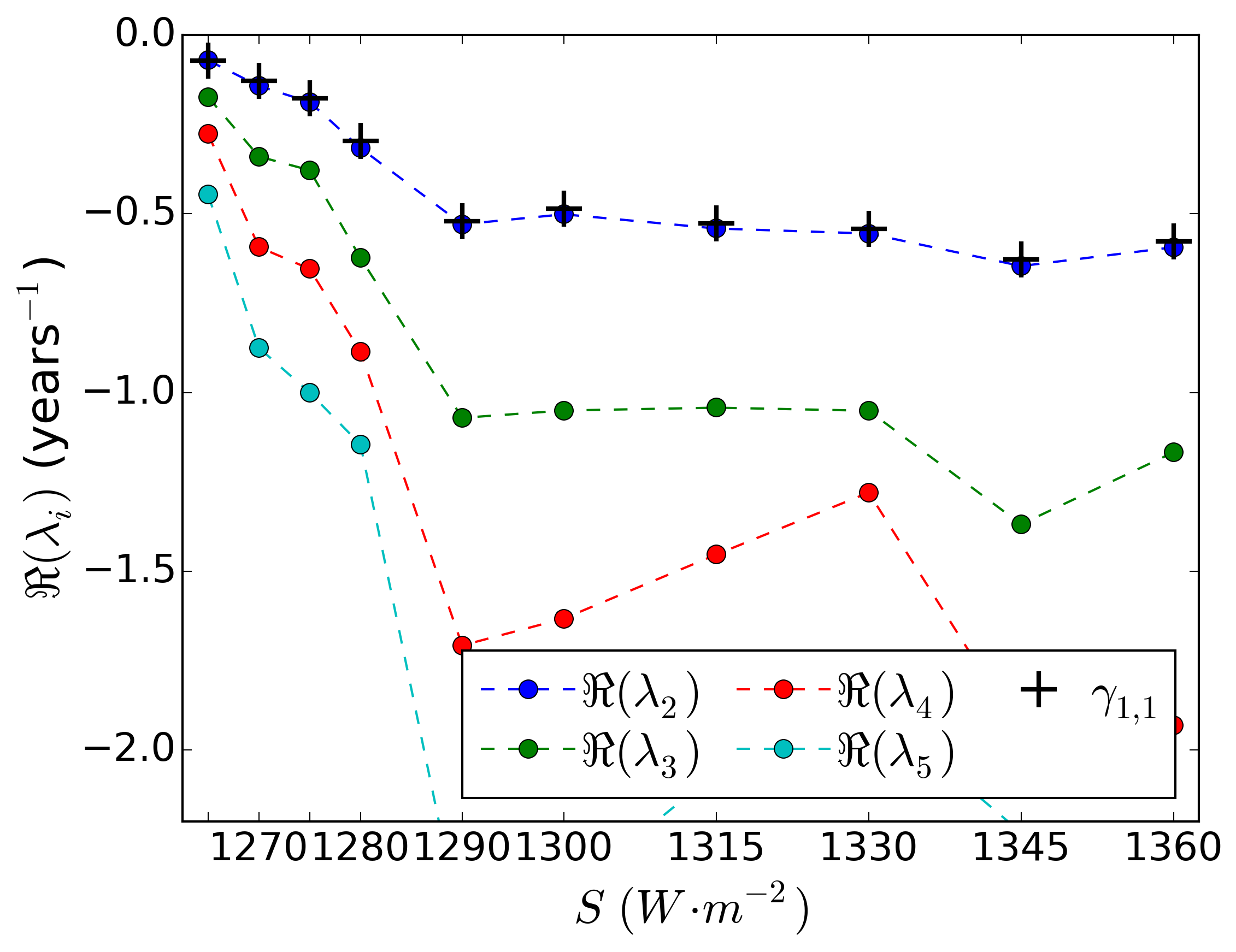}
	\end{subfigure}
        	\begin{subfigure}[b]{5.1cm}
 	(b)\\
		\includegraphics[width=\textwidth]{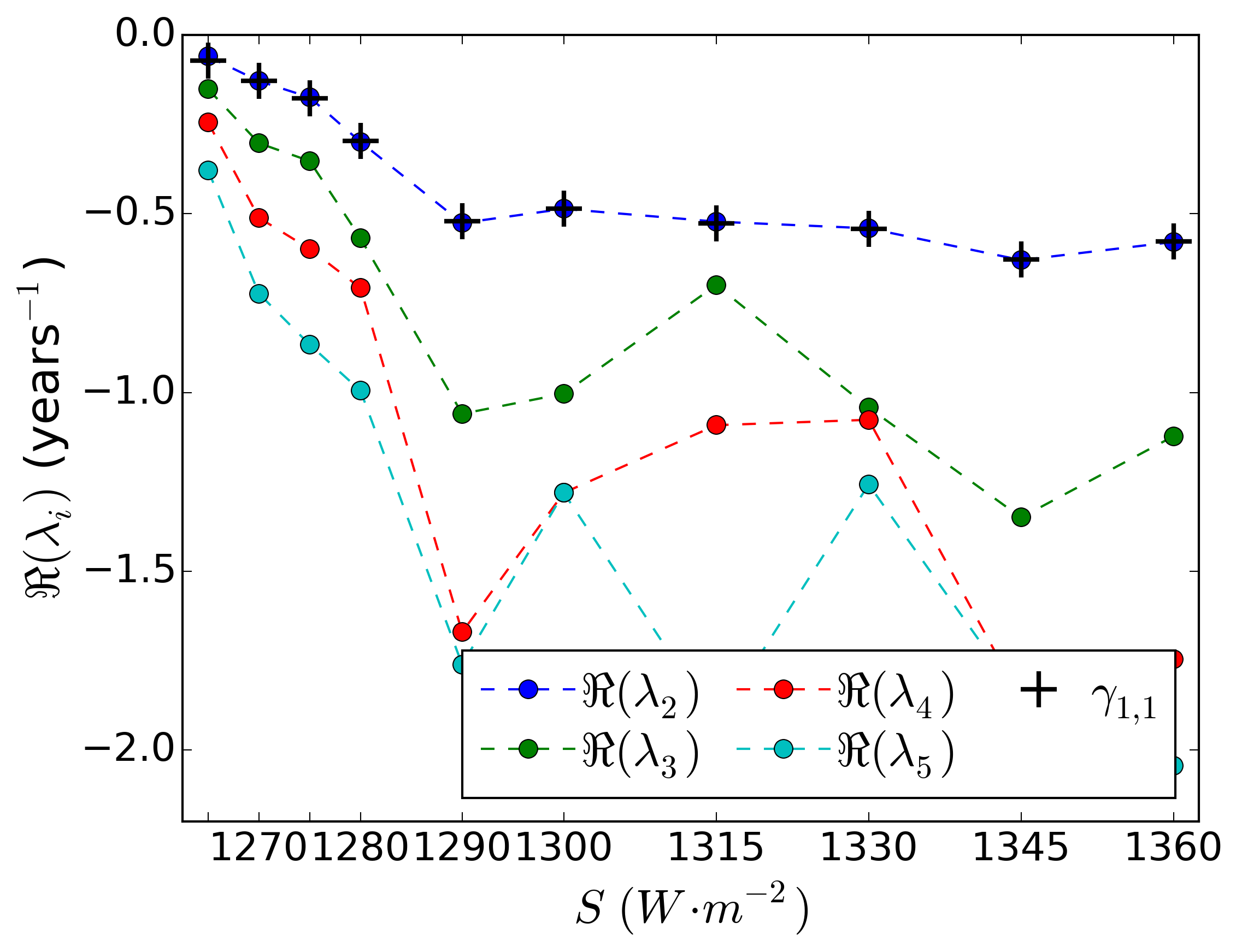}
	\end{subfigure}
        	\begin{subfigure}[b]{5.1cm}
 	(c)\\
		\includegraphics[width=\textwidth]{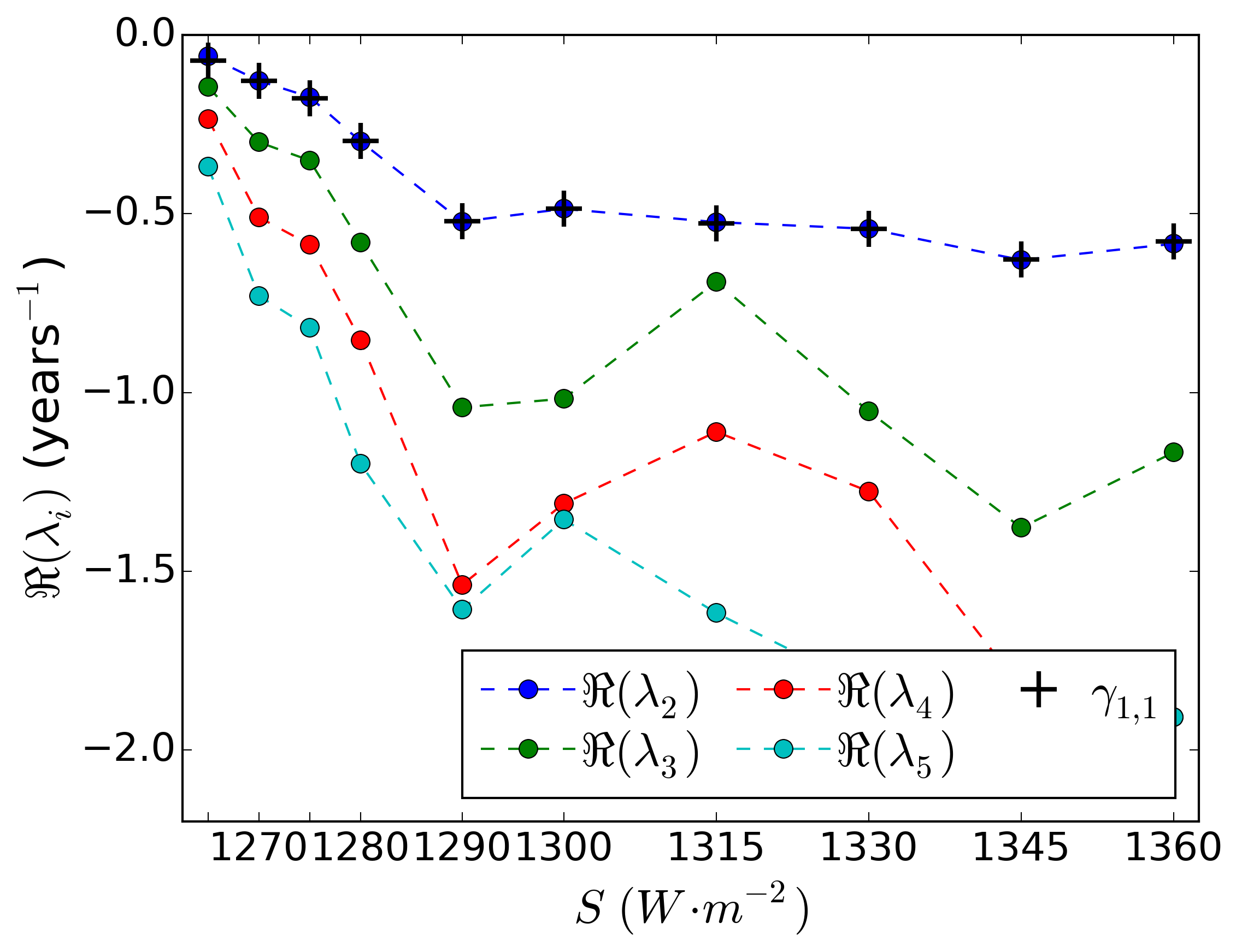}
	\end{subfigure}
	\caption{Real part of the rate  (\ref{eq:reducedEigenvalues}) of the four leading reduced unstable resonances for the NH SIC, versus the solar constant, using a grid of (a) 25 boxes, (b) 75 boxes and (c) a 100 boxes, compared to the 50 boxes grid used to produce figure \ref{fig:ratesW2SB}.}
	\label{fig:robustGrid}
\end{figure}

The robustness of our results to the size of the grid is now addressed. Figure \ref{fig:robustGrid}(a) to (c) represents the real part of the rates  (\ref{eq:reducedEigenvalues}) of the leading reduced unstable resonances estimated from transition matrices on a grid of 25, 75 or a 100 boxes, respectively, instead of the original grid of 50 boxes used to produce figure \ref{fig:ratesW2SB}. Figure \ref{fig:robustGrid}(a-c) and figure \ref{fig:ratesW2SB} are qualitatively similar, with the approach to 0 of the real part of the leading eigenvalues as the solar constant is decreased. The inverse of the real part of the rate of the second eigenvalue is also given in the fifth to the seventh column of table \ref{tab:robust} from which one can see that increasing the resolution results in differences in the inverse of the real part of the rate of the second eigenvalue of less than $5\%$. This suggests that the grid of $50$ boxes used for the NH SIC is sufficiently thin to obtain good estimates of, at least, the second eigenvalue, allowing one to conclude that the outcome of this study is not challenged by the coarse resolution of the grid used to estimate the transition matrices.

\subsection{Robustness to the lag}
\label{ref:robustLag}

\begin{figure}[h!]
	\centering
        \begin{subfigure}[b]{7.5cm}
        (a)\\
		\includegraphics[width=\textwidth]{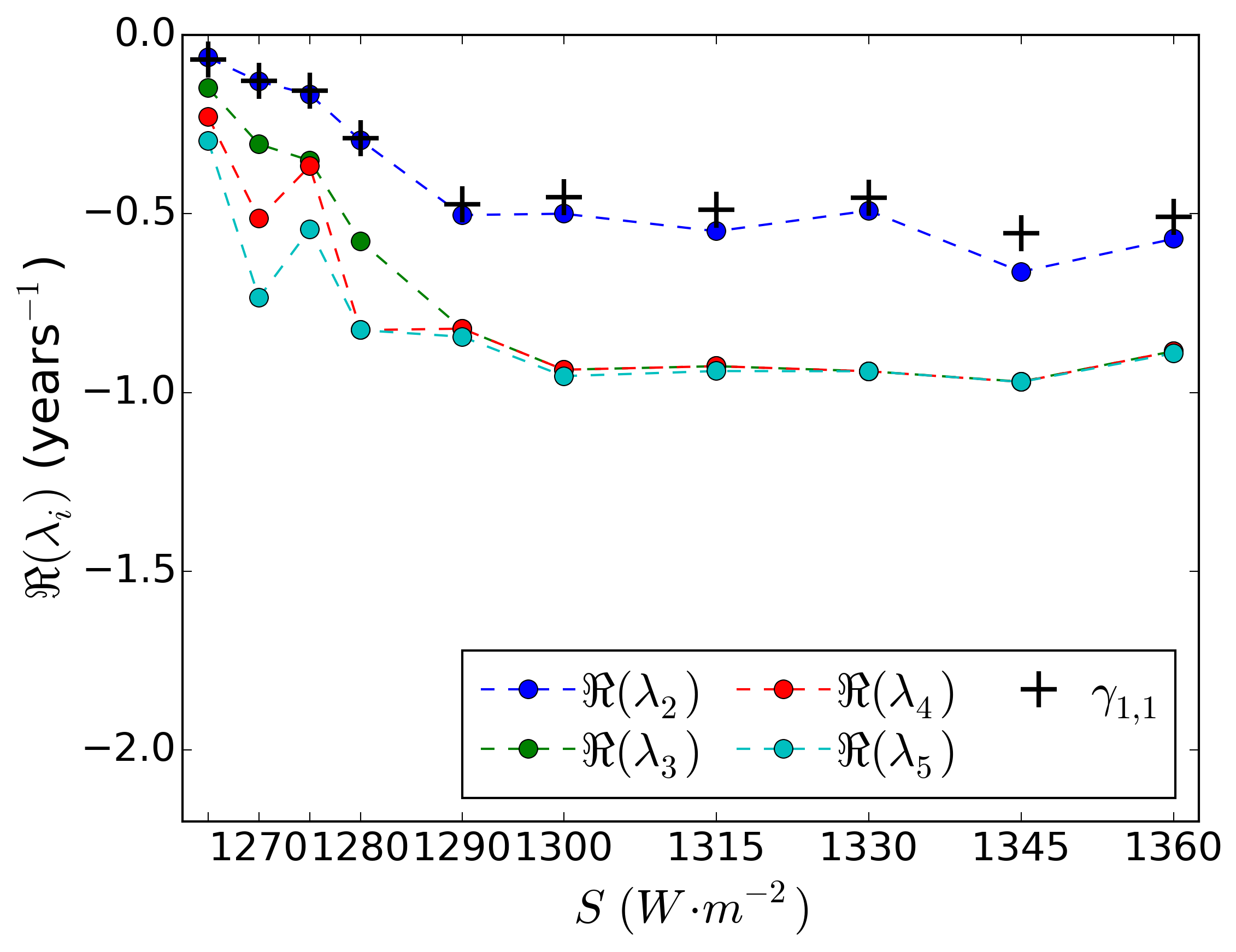}
	\end{subfigure}
        	\begin{subfigure}[b]{7.5cm}
 	(b)\\
		\includegraphics[width=\textwidth]{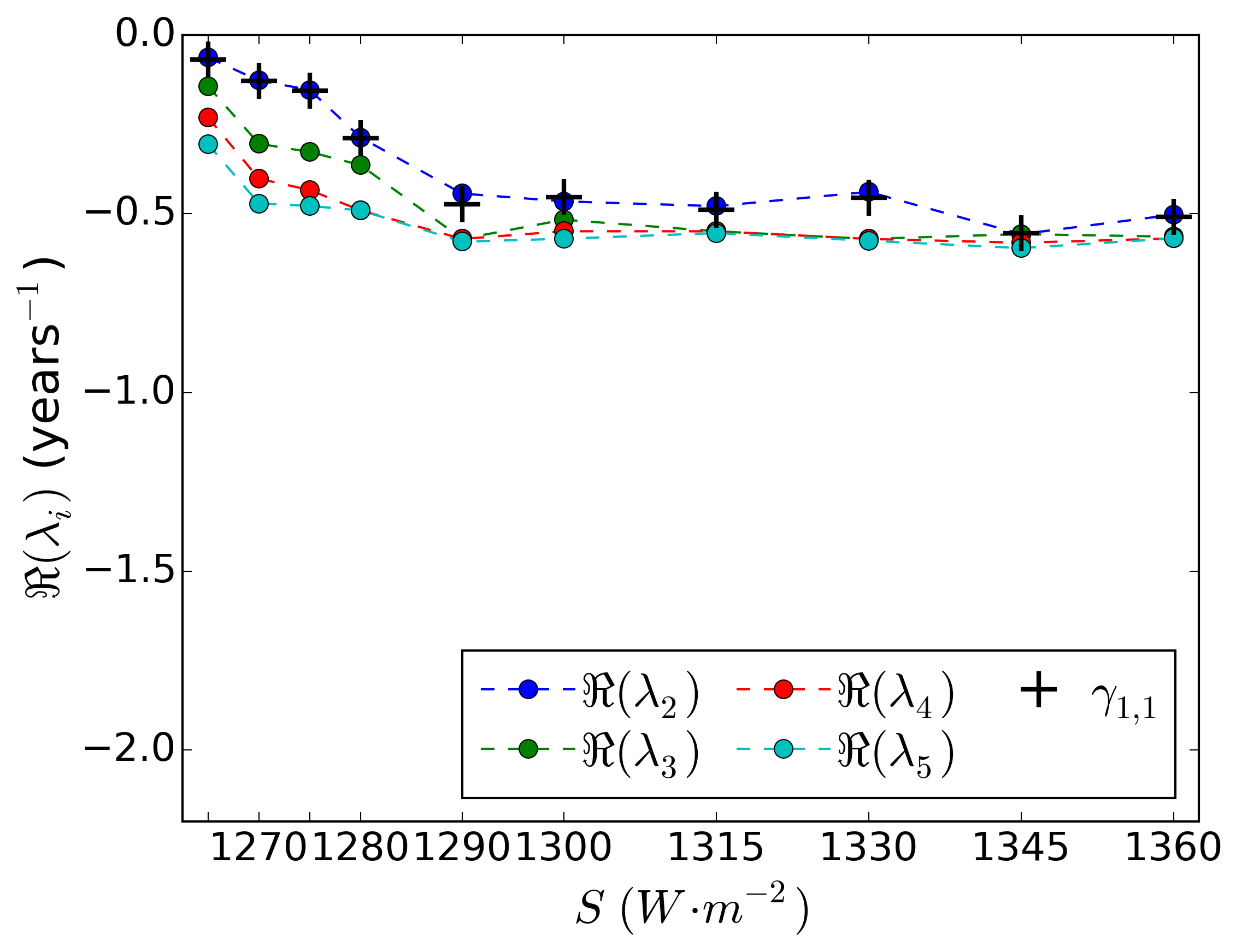}
	\end{subfigure}
	\caption{Real part rates corresponding to the four leading reduced unstable resonances for the NH SIC, versus the solar constant, using a lag of (a) 3 years and (b) 5 years, compared to the lag of 1 year used to produce figure \ref{fig:ratesW2SB}.}
	\label{fig:robustLag}
\end{figure}

All the transition matrices in section \ref{sec:resultSpectrum} were calculated for a lag of 1 year, in order to have access to time scales as short as possible.
However, due to the fact that the transition matrices are calculated on a reduced state space,
the family of operators $\mathcal{P}_\rho^{(n)}, n \ge 0$ reduced from $n$ iterates of the transfer operator does not constitute a semigroup, so that the spectral formula  (\ref{eq:reducedEigenvalues}) does not apply in general and the rates  (\ref{eq:reducedEigenvalues}) computed from the eigenvalues
of the $\mathcal{P}_\rho^{(n)}, n \ge 0$ depend on $n$ (see section \ref{sec:approxSpectrum}).
To verify that the conclusions of the analysis presented section \ref{sec:resultSpectrum} are not questioned by a different choice of the lag $n$, we represent, in figure \ref{fig:robustLag}(a) and (b), the real part of rates corresponding to the leading reduced unstable resonances calculated, as for figure \ref{fig:ratesW2SB}, for the NH SIC and on a grid of 50 boxes but for a longer lag of 3 and 5 years, respectively. Again, the inverse of the real part of the rate of the first secondary eigenvalue thus obtained is catalogued in table \ref{tab:robust}.

It is obvious from figure \ref{fig:robustLag}, that rates far away from the imaginary axis are not well estimated for long lags.
This effect if known \cite{Tantet2016a} and due to the fact that the eigenvectors associated with rates far from the imaginary axis decay fast and quickly become overwhelmed by the different sources of numerical imprecisions. 
However, even when a lag of 5 years is taken, the approach of the second leading rate to the imaginary axis as the control parameter is decreased towards its critical value can still be observed figure \ref{fig:robustLag}(b), since the real part of this rate corresponds to a time scale larger than 5 years for $S < 1280~W  m^{-2}$.

\begin{table}[htb!]
\centering
\caption{Inverse of the real part of the rates corresponding to the second reduced unstable resonance (in years) of transition matrices estimated for the NH SIC using different parameters, for different values of the solar constant.
For the second column, the whole time series was used as well as a grid of 50 boxes and a 1 year lag, as for figure \ref{fig:ratesW2SB}.
For the third and the fourth columns, only half or a quarter (cf. fig. \ref{fig:robustLength}.a and b, respectively) of the time series was used, respectively.
For the fifth to the seventh column, a grid of 25 boxes, 75 boxes and a 100 boxes (cf. fig. \ref{fig:robustGrid}.a to c, respectively) was used.
Finally, for the last two columns, a lag of 3 and 5 (cf. fig. \ref{fig:robustLag}.a to b, respectively) years was used.}
\begin{tabular}{| c | c | c | c | c | c | c | c | c |}
\hline
$S (W  m^{-2})$	& Fig. \ref{fig:ratesW2SB} & $1/2$ length	& $1/4$ length	& 25 boxes	& 75 boxes	& 100 boxes	& 3 years	& 5 years\\ \hline
$ 1265 $ 			& 16.36				& 14.48		& 15.63		& 14.12		& 16.69		& 16.95		& 15.84	& 15.95\\ \hline
$ 1270 $ 			& 7.65				& 8.22		& 7.45		& 7.02		& 7.77		& 7.81		&  7.70	& 7.85 \\ \hline
$ 1275 $ 			& 5.69				& 5.49		& 5.29		& 5.33		& 5.73		& 5.77		& 6.02	& 6.52 \\ \hline
$ 1280 $ 			& 3.31   				& 3.28		& 3.26		& 3.16		& 3.34		& 3.36		& 3.40	& 3.49 \\ \hline
$ 1290 $ 			& 1.90 				& 1.89		& 1.94		& 1.89		& 1.90		& 1.92		& 1.99	& 2.25 \\ \hline
$ 1300 $ 			& 2.04 				& 2.09		& 1.96		& 1.99		& 2.06		& 2.06		& 2.00	& 2.15 \\ \hline
$ 1315 $ 			& 1.89				& 1.94		& 1.84		& 1.85		& 1.92		& 1.91		& 1.82	& 2.09 \\ \hline
$ 1330 $ 			& 1.83 				& 1.78		& 1.75		& 1.80		& 1.85		& 1.84		& 2.04	& 2.28 \\ \hline
$ 1345 $ 			& 1.59				& 1.52		& 1.50		& 1.55		& 1.59		& 1.59		& 1.51	& 1.80 \\ \hline
$ 1360 $ 			& 1.72 				& 1.72		& 1.70		& 1.68		& 1.73		& 1.71		& 1.75	& 1.99 \\ \hline
\end{tabular}
\label{tab:robust}
\end{table}

\newpage 
\section*{References} 
\bibliographystyle{iopart-num}
\bibliography{2015_PlaSim}

\providecommand{\newblock}{}
\begin{thebibliography}{100}
\expandafter\ifx\csname url\endcsname\relax
  \def\url#1{{\tt #1}}\fi
\expandafter\ifx\csname urlprefix\endcsname\relax\def\urlprefix{URL }\fi
\providecommand{\eprint}[2][]{\url{#2}}

\bibitem{Held2004a}
Held H and Kleinen T 2004 {\em Geophys. Res. Lett.\/} {\bf 31} 1--4

\bibitem{Mauroy2016}
Mauroy A and Mezi{\'{c}} I 2016 {\em IEEE Trans. Automat. Contr.\/} {\bf 61}
  3356--3369

\bibitem{Tantet2017}
Tantet A, Lucarini V and Dijkstra H~A 2017 {\em J. Stat. Phys.\/}  1--33

\bibitem{Lorenz1963a}
Lorenz E~N 1963 {\em J. Atmos. Sci.\/} {\bf 20} 130

\bibitem{Lucarini2007}
Lucarini V, Speranza A and Vitolo R 2007 {\em Phys. D Nonlinear Phenom.\/} {\bf
  234} 105--123

\bibitem{Ghil2010}
Ghil M, Read P~L and Smith L~A 2010 {\em A{\&}G\/} {\bf 51} 4.28--4.35

\bibitem{Grebogi1983}
Grebogi C, Ott E and Yorke J~A 1983 {\em Phys. D Nonlinear Phenom.\/} {\bf 7}
  181--200

\bibitem{Guckenheimer1983}
Guckenheimer J~M and Holmes P 1983 {\em {Nonlinear Oscillations, Dynamical
  Systems, and Bifurcation of Vector Fields}\/} (New York: Springer)

\bibitem{Kuznetsov1998}
Kuznetsov Y~A 1998 {\em {Elements of Applied Bifurcation Theory, Second
  Edition}\/} (New York: Springer-Verlag)

\bibitem{Ruelle1999b}
Ruelle D 1999 {\em J. Stat. Phys.\/} {\bf 95} 393--468

\bibitem{Gallavotti2014}
Gallavotti G and Lucarini V 2014 {\em J. Stat. Phys.\/} {\bf 156} 1027--1065

\bibitem{Oseledets1968}
Oseledets V~I 1968 {\em Tr. Mosk. Mat. Obs.\/} {\bf 19} 179--210

\bibitem{Eckmann1985}
Eckmann J~P and Ruelle D 1985 {\em Rev. Mod. Phys.\/} {\bf 57} 617

\bibitem{Halmos1956}
Halmos P~R 1956 {\em {Lectures on Ergodic Theory}\/} (New York: Chelsea
  Publishing Company)

\bibitem{Arnold1968}
Arnold V~I and Avez A 1968 {\em {Ergodic Problems of Classical Mechanics}\/}
  (New York: Advanced Book Classics)

\bibitem{Baladi2001}
Baladi V 2001 {Spectrum and Statistical Properties} {\em Eur. Congr. Math.\/}
  ed Casacuberta C, Mir{\'{o}}-Roig R~M, Verdera J and Xamb{\'{o}}-Descamps S
  (Basel: Birkh{\"{a}}user) pp 203--223

\bibitem{Young2013}
Young L~S 2013 {\em Commun. Pure Appl. Math.\/} {\bf 66} 1439--1463

\bibitem{Pollicott1985}
Pollicott M 1985 {\em Invent. Math.\/} {\bf 81} 413--426

\bibitem{Ruelle1986}
Ruelle D 1986 {\em Phys. Rev. Lett.\/} {\bf 56} 405--407

\bibitem{Liverani1995a}
Liverani C 1995 {\em Ann. Math.\/} {\bf 142} 239--301

\bibitem{Blank2001a}
Blank M, Keller G and Liverani C 2001 {\em Nonlinearity\/} {\bf 15} 58

\bibitem{Gouezel2006}
Gou{\"{e}}zel S and Liverani C 2006 {\em Ergod. Theory Dyn. Syst.\/} {\bf 26}
  26

\bibitem{Butterley2007}
Butterley O and Liverani C 2007 {\em J. Mod. Dyn.\/} {\bf 1} 301--322

\bibitem{Faure2014}
Faur{\'{e}} F and Tsujii M 2014 {Semiclassical Approach for the
  Ruelle-Pollicott Spectrum of Hyperbolic Dynamics} {\em Anal. Probabilistic
  Approaches to Dyn. Negat. Curvature\/} ed Dal'Bo F, Peign{\'{e}} M and
  Sambusetti A (Cham: Springer) chap~2, pp 65--138

\bibitem{Baladi2017}
Baladi V 2017 {\em J. Stat. Phys.\/} {\bf 166} 525--557

\bibitem{Ruelle2009}
Ruelle D 2009 {\em Nonlinearity\/} {\bf 22} 855--870

\bibitem{Cessac2007}
Cessac B and Sepulchre J~A 2007 {\em Phys. D Nonlinear Phenom.\/} {\bf 225}
  13--28

\bibitem{Young2002}
Young L~S 2002 {\em J. Stat. Phys.\/} {\bf 108} 733--754

\bibitem{Vaidya2008}
Vaidya U and Mehta P~G 2008 {\em IEEE Trans. Automat. Contr.\/} {\bf 53}
  307--323

\bibitem{Lucarini2016}
Lucarini V 2016 {\em J. Stat. Phys.\/} {\bf 162} 312--333

\bibitem{Chekroun2014}
Chekroun M~D, Neelin J~D, Kondrashov D, McWilliams J~C and Ghil M 2014 {\em
  Proc. Natl. Acad. Sci. U. S. A.\/} {\bf 111} 1684--1690

\bibitem{Chekroun2016}
Chekroun M~D, Tantet A, Neelin J~D and Dijkstra H~A 2017 {\em Phys. D Nonlinear
  Phenom.\/}

\bibitem{Tantet2016a}
Chekroun M~D, Tantet A, Neelin J~D and Dijkstra H~A 2016 {\em Prep.\/}

\bibitem{Lucarini2014b}
Lucarini V, Blender R, Herbert C, Ragone F, Pascale S and Wouters J 2014 {\em
  Rev. Geophys.\/} {\bf 52} 1--51

\bibitem{Fraedrich2005a}
Fraedrich K, Jansen H, Kirk E and Lunkeit F 2005 {\em Meteorol. Zeitschrift\/}
  {\bf 14} 305--314

\bibitem{Fraedrich2005b}
Fraedrich K, Jansen H, Kirk E, Luksch U and Lunkeit F 2005 {\em Meteorol.
  Zeitschrift\/} {\bf 14} 299--304

\bibitem{Schalge2013}
Schalge B, Blender R, Wouters J, Fraedrich K and Lunkeit F 2013 {\em Phys. Rev.
  E\/} {\bf 87} 052113

\bibitem{Lucarini2010a}
Lucarini V, Fraedrich K and Lunkeit F 2010 {\em Q. J. R. Meteorol. Soc.\/} {\bf
  136} 2--11

\bibitem{Boschi2013}
Boschi R, Lucarini V and Pascale S 2013 {\em Icarus\/} {\bf 226} 1724--1742

\bibitem{Lucarini2017}
Lucarini V and Bodai T 2017 {\em Nonlinearity\/} {\bf 30} R32--R66

\bibitem{Bodai2015}
B{\'{o}}dai T, Lucarini V, Lunkeit F and Boschi R 2015 {\em Clim. Dyn.\/} {\bf
  44} 3361--3381

\bibitem{Budyko1969}
Budyko M~I 1969 {\em Tellus\/} {\bf 21} 611 -- 619

\bibitem{Sellers1968}
Sellers W~D 1968 {\em J. Appl. Meteorol.\/} {\bf 8} 392--400

\bibitem{Ghil1976a}
Ghil M 1976 {\em J. Atmos. Sci.\/} {\bf 33} 3--20

\bibitem{Pierrehumbert2011a}
Pierrehumbert R, Abbot D, Voigt A and Koll D 2011 {\em Annu. Rev. Earth Planet.
  Sci.\/} {\bf 39} 417--460

\bibitem{Voigt2010}
Voigt A and Marotzke J 2010 {\em Clim. Dyn.\/} {\bf 35} 887--905

\bibitem{Zaliapin2010}
Zaliapin I and Ghil M 2010 {\em Nonlinear Process. Geophys.\/} {\bf 17}
  113--122

\bibitem{Lucarini2013a}
Lucarini V, Pascale S, Boschi R, Kirk E and Iro N 2013 {\em Astron.
  Nachrichten\/} {\bf 334} 576--588

\bibitem{Fraedrich1998a}
Fraedrich K, Kirk E and Lunkeit F 1998 {\em DKRZ Tech. Rep.\/} {\bf 16} 24
  pages

\bibitem{Eliasen1970}
Eliasen E, Machenhauer B and Rasmussen E 1970 {On a numerical method for
  integration of the hydrodynamical equations with a spectral representation of
  the horizontal fields} Tech. rep. Inst. of Theor. Met., K{\o}benhavns
  University Copenhagen

\bibitem{Orszag1970}
Orszag S~A 1970 {\em J. Atmos. Sci.\/} {\bf 27} 890--895

\bibitem{Hoskins1975}
Hoskins B~J and Simmons A~J 1975 {\em Q. J. R. Meteorol. Soc.\/} {\bf 101}
  637--655

\bibitem{Sasamori1968}
Sasamori T 1968 {\em J. Appl. Meteorol.\/} {\bf 7} 721--729

\bibitem{Lacis1974}
Lacis A~a and Hansen J 1974 {\em J. Atmos. Sci.\/} {\bf 31} 118--133

\bibitem{Stephens1978}
Stephens G~L, Paltridge G~W and Platt C~M~R 1978 {\em J. Atmos. Sci.\/} {\bf
  35} 2133--2141

\bibitem{Stephens1984}
Stephens G~L, Ackerman S and Smith E~A 1984 {\em J. Atmos. Sci.\/} {\bf 41}
  687--690

\bibitem{Slingo1991}
Slingo A and Slingo J~M 1991 {\em J. Geophys. Res.\/} {\bf 96} 15341

\bibitem{Kuo1965}
Kuo H~L 1965 {\em J. Atmos. Sci.\/} {\bf 22} 40--63

\bibitem{Kuo1974}
Kuo H~L 1974 {\em J. Atmos. Sci.\/} {\bf 31} 1232--1240

\bibitem{Louis1979}
Louis J~F 1979 {\em Boundary-Layer Meteorol.\/} {\bf 17} 187--202

\bibitem{Louis1981}
Louis J~F, Tiedke M and Geleyn M 1981 {A short history of the PBL
  parameterisation at ECMWF} {\em Proc. ECMWF Work. Planet. Bound. Layer
  Parameterization\/} (Reading) pp 59--80

\bibitem{Laursen1989}
Laursen L and Eliasen E 1989 {\em Tellus\/} {\bf 41A} 385--400

\bibitem{Roeckner1992}
Roeckner E, Arpe K, Bengtsson L, Brinkop S, D{\"{u}}menil L, Esch M, Kirk E,
  Lunkeit F, Ponater M, Rockel B, Sausen R, Schlese U, Schubert S, Windelband
  M, Schlese U, Brinkop S, Ponater M, Sausen R and Rockel B 1992 {Simulation of
  the present-day climate with the ECHAM model: Impact of model physics and
  resolution} Tech. rep. Max Planck Institut f{\"{u}}r Meteorologie Hamburg

\bibitem{Semtner1976}
Semtner A~J 1976 {\em J. Phys. Oceanogr.\/} {\bf 6} 379--389

\bibitem{Ragone2016a}
Ragone F, Lucarini V and Lunkeit F 2016 {\em Clim. Dyn.\/} {\bf 46} 1459--1471

\bibitem{Lucarini2017d}
Lucarini V, Ragone F and Lunkeit F 2017 {\em J. Stat. Phys.\/} {\bf 166}
  1036--1064

\bibitem{Lucarini2014a}
Lucarini V and Pascale S 2014 {\em Clim. Dyn.\/} {\bf 43} 981--1000

\bibitem{Fraedrich2008}
Fraedrich K and Lunkeit F 2008 {\em Tellus, Ser. A Dyn. Meteorol. Oceanogr.\/}
  {\bf 60} 921--931

\bibitem{Poulsen2004a}
Poulsen C~J and Jacob R~L 2004 {\em Paleoceanography\/} {\bf 19} 1--11

\bibitem{Wetherald1975}
Wetherald R~T and Manabe S 1975 {\em J. Atmos. Sci.\/} {\bf 32} 2044--2059

\bibitem{Scott1999}
Scott J~R, Marotzke J and Stone P~H 1999 {\em J. Phys. Oceanogr.\/} {\bf 29}
  351

\bibitem{Lucarini2007a}
Lucarini V, Calmanti S and Artale V 2007 {\em Russ. J. Math. Phys.\/} {\bf 14}
  224--231

\bibitem{Leutbecher2008}
Leutbecher M and Palmer T~N 2008 {\em J. Comput. Phys.\/} {\bf 227} 3515--3539

\bibitem{Yosida1980}
Yosida K 1980 {\em {Functional Analysis}\/} vol 123 (Berlin Heidelberg New
  York: Springer-Verlag)

\bibitem{Engel2001}
Engel K~J and Nagel R 2001 {\em {One-parameter semigroups for linear evolution
  equations}\/} (New York: Springer)

\bibitem{Davies2007}
Davies E~B 2007 {\em {Linear Operators and Their Spectra}\/} (Cambridge:
  Cambridge University Press)

\bibitem{Koopman1931}
Koopman B~O 1931 {\em Proc. Natl. Acad. Sci. U. S. A.\/} {\bf 17} 315--318

\bibitem{Neumann1932}
von Neumann J 1932 {\em Proc. Natl. Acad. Sci. U. S. A.\/} {\bf 18} 70--82

\bibitem{Lasota1994}
Lasota A and Mackey M~C 1994 {\em {Chaos, Fractals and Noise}\/} (Berlin:
  Springer)

\bibitem{Katok1996}
Katok A and Hasselblatt B 1996 {\em {Introduction to the Modern Theory of
  Dynamical Systems}\/} (Cambridge: Cambridge University Press)

\bibitem{Gallavotti2014a}
Gallavotti G 2014 {\em {Nonequilibrium and irreversibility}\/} (Cham: Springer)

\bibitem{Misra1979}
Misra B, Prigogine I and Courbage M 1979 {\em Phys. A Stat. Mech. its Appl.\/}
  {\bf 98} 1--26

\bibitem{Gaspard1998}
Gaspard P 1998 {\em {Chaos, Scattering and Statistical Mechanics}\/}
  (Cambridge: Cambridge University Press)

\bibitem{Garbaczewski2002}
Garbaczewski P and Olkiewicz R (eds) 2002 {\em {Dynamics of Dissipation}\/}
  (Berlin: Springer)

\bibitem{Keller1998}
Keller G, Liverani C and Roma T~U~D 1998 {Stability of the spectrum for
  transfer operators} Tech. rep. Scuola Norm. Sup. Pisa Pisa

\bibitem{Baladi1999}
Baladi V and Holschneider M 1999 {\em Nonlinearity\/} {\bf 12} 525

\bibitem{Froyland2007a}
Froyland G 2007 {\em Discret. Contin. Dyn. Syst.\/} {\bf 17} 671--689

\bibitem{Baladi2007b}
Baladi V 2007 {\em Commun. Math. Phys.\/} {\bf 275} 839--859

\bibitem{Hasegawa1992b}
Hasegawa H and Saphir W 1992 {\em Phys. Rev. A\/} {\bf 46} 7401--7423

\bibitem{Gaspard1992}
Gaspard P and Ramirez D~A 1992 {\em Phys. Rev. A\/} {\bf 45} 8383--8397

\bibitem{Gaspard1995}
Gaspard P, Nicolis G, Provata A and Tasaki S 1995 {\em Phys. Rev. E\/} {\bf 51}
  74--94

\bibitem{Gaspard2001a}
Gaspard P and Tasaki S 2001 {\em Phys. Rev. E\/} {\bf 64} 056232

\bibitem{ulam1964collection}
Ulam S~M 1964 {\em {Problems in Modern Mathematics}\/} science ed (New York:
  Wiley)

\bibitem{Dellnitz1999}
Dellnitz M and Junge O 1999 {\em SIAM J. Numer. Anal.\/} {\bf 36} 491--515

\bibitem{Klus2015a}
Klus S, Koltai P and Sch{\"{u}}tte C 2015 {\em arXiv\/}  1--19

\bibitem{Williams2015}
Williams M~O, Kevrekidis I~G and Rowley C~W 2015 {\em J. Nonlinear Sci.\/} {\bf
  25} 1307--1346

\bibitem{Korda2017}
Korda M and Mezi{\'{c}} I 2017 {\em arXiv Prepr.\/}  1--18

\bibitem{Fishman2002}
Fishman S and Rahav S 2002 {Relaxation and Noise in Chaotic Systems} {\em Dyn.
  Dissipation\/} (Berlin: Springer) chap~6, pp 165--192

\bibitem{Shutte1999}
Sch{\"{u}}tte C 1999 {Conformational Dynamics: Modelling, Theory, Algorithm and
  Application to Biomolecules} Tech. Rep. July Konrad-Zuse-Zentrum f{\"{u}}r
  Informationstechnik Berlin

\bibitem{Dudley2004}
Dudley R~M 2004 {\em {Real Analysis and Probability}\/} (Cambridge: Cambridge
  University Press)

\bibitem{Kallenberg2002}
Kallenberg O 2002 {\em {Foundations of Modern Probability}\/} (New York:
  Springer)

\bibitem{Tantet2015}
Tantet A, van~der Burgt F~R and Dijkstra H~A 2015 {\em Chaos An Interdiscip. J.
  Nonlinear Sci.\/} {\bf 25} 036406

\bibitem{Zwanzig2001}
Zwanzig R 2001 {\em {Nonequilibrium Statistical Mechanics}\/} (Oxford: Oxford
  University Press)

\bibitem{Chorin2009}
Chorin A~J and Hald O~H 2009 {\em {Stochastic tools in mathematics and
  science}\/} (New York: Springer)

\bibitem{Kondrashov2014}
Kondrashov D, Chekroun M~D and Ghil M 2015 {\em Phys. D Nonlinear Phenom.\/}
  {\bf 297} 33--55

\bibitem{Crommelin2011}
Crommelin D and Vanden-Eijnden E 2011 {\em Multiscale Model. Simul.\/} {\bf 9}
  1588--1623

\bibitem{Pavliotis2008}
Pavliotis G~A and Stuart A~M 2008 {\em {Multiscale Methods Averaging and
  Homogenization}\/} vol~53 (New York: Springer)

\bibitem{Billingsley1961}
Billingsley P 1961 {\em {Statistical Inference for Markov process}\/} (Chicago:
  University of Chicago Press)

\bibitem{Lehoucq1997}
Lehoucq R~B, Sorensen D~C and Yang C 1997 {ARPACK Users' Guide: Solution of
  Large Scale Eigenvalue Problems with Implicitly Restarted Arnoldi Methods}

\bibitem{Crommelin2009}
Crommelin D and Vanden-Eijnden E 2009 {\em Multiscale Model. Simul.\/} {\bf 7}
  1--28

\bibitem{Just2001}
Just W, Kantz H, R{\"{o}}denbeck C and Helm M 2001 {\em J. Phys. A. Math.
  Gen.\/} {\bf 34} 3199--3213

\bibitem{Tantet2016}
Tantet A, Chekroun M~D, Neelin J~D and Dijkstra H~A 2017 {\em Phys. D Nonlinear
  Phenom.\/}

\bibitem{Lucarini2017c}
Lucarini V and Wouters J 2017 {\em J. Phys. A Math. Theor.\/} {\bf 50} 355003

\bibitem{Kato1995}
Kato T 1995 {\em {Perturbation Theory for Linear Operators}\/} (Berlin:
  Springer)

\bibitem{Ghil2015}
Ghil M 2015 {A Mathematical Theory of Climate Sensitivity or, How to Deal With
  Both Anthropogenic Forcing and Natural Variability?} {\em Clim. Chang.
  Multidecadal Beyond\/} (WORLD SCIENTIFIC) chap~2, pp 31--51

\bibitem{Knutti2008}
Knutti R and Hegerl G~C 2008 {\em Nat. Geosci.\/} {\bf 1} 735--743

\bibitem{Roe2009}
Roe G 2009 {\em Annu. Rev. Earth Planet. Sci.\/} {\bf 37} 93--115

\bibitem{Collins2013}
Collins M, Knutti R, Arblaster J, Dufresne J~L, Fichefet T, Friedlingstein P,
  Gao X, Gutowski W~J, Johns T, Krinner G, Shongwe M, Tebaldi C, Weaver A~J and
  Wehner M 2013 {Long-term Climate Change: Projections, Commitments and
  Irreversibility} {\em Clim. Chang. 2013 Phys. Sci. Basis. Contrib. Work. Gr.
  I to Fifth Assess. Rep. Intergov. Panel Clim. Chang.\/} (Cambridge, United
  Kingdom and New York, NY, USA: Cambridge University Press) pp 1029--1136 ISBN
  9781107415324

\bibitem{Rugenstein2016b}
Rugenstein M~A~A, Gregory J~M, Schaller N, Sedl{\'{a}}{\v{c}}ek J and Knutti R
  2016 {\em J. Clim.\/} {\bf 29} 5643--5659

\bibitem{VonderHeydt2014}
von~der Heydt A~S, Dijkstra H~A, K{\"{o}}hler P and Wal R~V~D 2014 {\em
  Geophys. Res. Lett.\/} {\bf 41} 6484--6492

\bibitem{Kohler2015}
K{\"{o}}hler P, de~Boer B, von~der Heydt A~S, Stap L~B and van~de Wal R~S~W
  2015 {\em Clim. Past Discuss.\/} {\bf 11} 3019--3069

\bibitem{Knutti2015}
Knutti R and Rugenstein M~A~A 2015 {\em Phil. Trans. R. Soc. A\/} {\bf 373}
  20150146

\bibitem{Boulton2014}
Boulton C~A, Allison L~C and Lenton T~M 2014 {\em Nat. Commun.\/} {\bf 5} 5752

\bibitem{Faranda2014d}
Faranda D, Pons F~M~E, Dubrulle B, Daviaud F, Saint-Michel B, Herbert {\'{E}}
  and Cortet P~P 2014 {\em Phys. Fluids\/} {\bf 26} 105101

\bibitem{Gallavotti1995}
Gallavotti G and Cohen E~G~D 1995 {\em J. Stat. Phys.\/} {\bf 80} 931--970

\bibitem{Efron1982a}
Efron B 1980 {The Jacknife, the bootstrap, and other resampling plans} Tech.
  rep. Stanford University Stanford

\bibitem{Mudelsee2010}
Mudelsee M 2010 {\em {Climate Time Series Analysis: Classical Statistical and
  Bootstrap Methods}\/} (Dordrecht: Springer)

\bibitem{Craig2002a}
Craig B~A and Sendi P~P 2002 {\em Health Econ.\/} {\bf 11} 33--42

\end{thebibliography}

\end{document}